\documentclass[aps,rmp,notitlepage,tightenlines,12pt]{revtex4-1}

\usepackage{natbib}

\usepackage{amssymb,amsmath,amsfonts,amstext,graphics,graphicx}
\usepackage{color}

\graphicspath{{figs/}{figs_lo/}}

\newcommand{\be}{\begin{equation}}
\newcommand{\bq}{\begin{equation}}
\newcommand{\ee}{\end{equation}}
\newcommand{\eq}{\end{equation}}
\newcommand{\beq}{\begin{equation}}
\newcommand{\eeq}{\end{equation}}
\def\sgn{{\;{\rm sgn}}}
\def\ii{{\rm i}}

\def\p{\partial}

\def\half{{\mbox{$\frac{1}{2}$}}}
 
\def\bqy{\begin{eqnarray}}
\def\eqy{\end{eqnarray}}
\def\bqyn{\begin{eqnarray*}}
\def\eqyn{\end{eqnarray*}}

\def\cI{{\cal I}}

\def\ve{{\varepsilon}}

\def\calc{{\cal C}}
\def\cald{{\cal D}}
\def\cale{{\cal E}}

\def\cali{{\cal I}}
\def\calj{{\cal J}}
\def\calk{{\cal K}}
\def\call{{\cal L}}

\def\calq{{\cal Q}}

\def\calz{{\cal Z}}

\def\cD{{\cal D}}
\def\cK{{\cal K}}

\def\Ome{{\Omega}}
\def\ome{{\varsigma}}

\def\tswm{{t}}
\def\xswm{{x}}
\def\yswm{{y}}

\def\cC{{\cal C}}

\def\be{\beta}

\def\de{\delta}

\def\ze{\zeta}
\def\zep{\zeta}
\def\th{\theta}
\def\om{\omega}

\def\R{\mathbb{R}}

\newcommand{\comment}[1]{}
\newcommand{\mathnotation}[2]{\newcommand{#1}{\ensuremath{#2}}}

\mathnotation{\pd}{\partial}
\mathnotation{\ldef}{\mathrel{\raisebox{.069ex}{:}\!\!=}}	
\mathnotation{\rdef}{\mathrel{=\!\!\raisebox{.069ex}{:}}}	
\mathnotation{\dint}{\,{\mathrm{d}}}		
\renewcommand{\l}{\left}
\renewcommand{\r}{\right}
\mathnotation{\grad}{\nabla}
\mathnotation{\compconj}{\mathrm{c.c.}}
\mathnotation{\pvint}{{\int\!\!\!\!\!\!-}}
\mathnotation{\onehalf}{\tfrac{1}{2}}		
\mathnotation{\qq}{\theta}		
\mathnotation{\pp}{I}		
\mathnotation{\pdt}{\pd_t}	
\mathnotation{\pdT}{\pd_T}	
\mathnotation{\pdq}{\pd_\qq}	
\mathnotation{\pdp}{\pd_\pp}	
\mathnotation{\pdy}{\pd_Y}
\mathnotation{\imi}{\mathrm{i}}			
\mathnotation{\exe}{\mathrm{e}}			
\mathnotation{\ampl}{{\mathcal{A}}}
\mathnotation{\phimode}{\phi}
\mathnotation{\phisw}{\varphi}
\mathnotation{\amplsw}{A}

\begin{document}

\title{Pattern formation in Hamiltonian systems
with continuous spectra; a normal-form single-wave model}



\author{N. J. Balmforth}
\email{njb@math.ubc.ca}
\affiliation{Department of Mathematics, University
  of British Columbia, Vancouver, Canada}
\author{P. J. Morrison}
\email{morrison@physics.utexas.edu}
\affiliation{Department of Physics and Institute for Fusion Studies, University
  of Texas at Austin, Austin, TX 78712, USA}
\author{J.-L. Thiffeault}
\email{jeanluc@math.wisc.edu}
\affiliation{Department of Mathematics, University
  of Wisconsin, Madison, WI 53706, USA}

\begin{abstract}

 Pattern formation in biological, chemical and physical
  problems has received considerable attention, with much attention paid to
  dissipative systems. For example, the Ginzburg--Landau equation is a normal
  form that describes pattern formation due to the appearance of a single mode
of instability in a wide variety of dissipative problems.
In a similar vein, a certain ``single-wave model''
arises in many physical contexts that share common
  pattern forming behavior.  These systems have Hamiltonian 
structure, and the single-wave model is a kind of Hamiltonian
  mean-field theory describing the patterns that form in phase space. 
The single-wave model was originally
  derived in the context of nonlinear plasma theory, where it
  describes the behavior near threshold and subsequent nonlinear
  evolution of unstable plasma waves.  However, the single-wave model
  also arises in fluid mechanics, specifically shear-flow and vortex dynamics,
  galactic dynamics, the XY and Potts models of condensed matter
physics, and other Hamiltonian theories characterized by mean field 
interaction.  We demonstrate, by a
  suitable asymptotic analysis, how the single-wave model emerges from
  a large class of nonlinear advection-transport
theories. An essential ingredient for the reduction is 
that the Hamiltonian system has a continuous
spectrum in the linear stability problem, 
arising not from an infinite spatial domain but 
from singular resonances along curves in phase space
whereat wavespeeds match material speeds
(wave-particle resonances in the plasma problem, or critical levels
in fluid problems). The dynamics of the continuous spectrum 
is manifest as the phenomenon of Landau damping when the system is stable,
and when the system becomes unstable, an embedded neutral mode appears
within the continuum. The single-wave model captures the dynamics
of the embedded mode if the system is perturbed so as to make the mode
unstable, or if perturbations destroy that mode 
and the system remains stable. As a consequence, the model describes
a range of universal phenomena: for a bifurcation to instability,
the model features the so-called ``trapping scaling'' dictating
the saturation amplitude, and the ``cats-eye'' structures that
characterize the resulting phase-space patterns. In the stable situation,
the model offers one of the simplest descriptions of Landau damping,
and illustrates how this can be arrested by nonlinearity.
Such dynamical phenomena have been rediscovered in different contexts,
which is unsurprising in view of the normal-form character of the single-wave
model.
\end{abstract}

\maketitle

\clearpage

\tableofcontents

\section{Introduction}
\label{sec:intro}

\subsection{Perspective}
\label{ssec:perspective}

The dynamics of forming patterns in biological, chemical, 
and physical systems has received considerable attention 
over recent years (see e.g.\ \citealp{cross}).
In the simplest settings, one deals with spatial  patterns 
that form as a result of the appearance of
a small number of ``modes'' of instability, such 
as a convection cell  in a fluid contained between differentially 
heated plates.
These are the eigenmodes of the linear stability problem
and provide a natural set of tools to describe the forming spatial patterns.
In particular, by decomposing perturbations about 
the unpatterned equilibrium into the
eigenmodes and exploiting the center-manifold reduction,
the full system of governing equations can be approximated by
a low-order set of amplitude equations near the onset of linear instability  
(e.g.\  \citealp{crawford91a}).

The center-manifold reduction works because, in spatially
bounded, dissipative systems, the eigenvalues of the normal modes
compose a set of distinct points on the spectral plane,
and instability arises when one of these modes becomes unstable.
The purpose of the present article is to summarize extensions of  
 some of these ideas to non-dissipative systems that suffer instabilities
that cannot be described in low-dimensional terms, and for which the patterns 
that form occur naturally in the associated phase space.   
These  systems are Hamiltonian systems that have a neutrally stable
continuous spectrum of eigenvalues in the linear stability problem. 
Moreover, this spectrum is not the result of 
the problem being couched in an infinite spatial
domain, but due to singularities in the linear eigenvalue
problem that arise due to a resonant interaction between
wave-like perturbations and the background equilibrium state;
the singular resonances occur along curves in phase space whereat
wavespeeds match material speeds.
Such systems are also commonplace, and include ideal plasmas, inviscid fluids,
self-gravitating stellar systems,
and various other Hamiltonian models characterized by mean-field interactions.
The plasma problem is that of electrostatic
instability in an ideal, single-species plasma \cite{vankampen55a}, 
which was the original
building block in understanding a multitude of more complicated instabilities
in plasma theory and fusion science. The fluid problem is the instability
of inviscid shear flow or vortices \cite{case60a}, which are also key problems
in fluid mechanics.  Another mean-field Hamiltonian model of interest 
is that with the XY interaction 
(see \citealp{chaikin}), which models spin-spin interaction 
and has been used to describe  
e.g.\ superfluid helium and hexatic liquid crystals (see  the 
`Hamiltonian Mean Field' model of \citealp{campa} for an extensive 
treatment).  

The mathematical issue which underlies our discussion is that,
although instability is described by an exponentially growing discrete
eigenmode, the  mode detaches from the continuous spectrum when it becomes 
unstable.   Consequently, at onset, the distinguished mode cannot 
be isolated from an infinite number of other, singular eigensolutions 
representing
the continuum (the center-manifold reduction relies on such a spectral gap).
This aspect of the problem fundamentally affects the weakly nonlinear
description, which, as a consequence, has to proceed down a relatively
novel and much more tortuous pathway. 

This pathway leads to the single-wave model of
O'Neil, Malmberg \& Winfree (1971), following terminology in plasma physics,
or simply the ``single-wave model'', for short.
This model is a Hamiltonian normal form
that describes the transition to instability described above in a wide
variety of physical systems of fluid,  plasma, and other disciplines. 
The model captures the dynamics of the patterns that form in the phase space
of the system due to the growth and nonlinear saturation of the
unstable mode detaching from the continuous spectrum.  
All of the physical models for which the single-wave model is appropriate 
are hyperbolic Hamiltonian systems with Hamiltonian characteristic equations.
Thus the patterns occur in a conventional phase space with canonically 
conjugate coordinates, and consequently have a Liouville theorem and other 
Hamiltonian properties, all of which help to recognize the circumstances
that lead to the single-wave model.

For the physical systems we treat as examples, the patterns
have two characteristic features: there is a relatively smooth, 
wave-like pattern ocupying most of the domain, coupled to a
more complicated, finely scaled structure appearing over a localized
region surrounding the resonance of linear theory. 
The larger-scale wave pattern reflects the shape of the embedded neutral
mode. The finely scaled structure is generated by the interaction
of that mode with a locally resonant packet of the continuous spectrum, and
takes the form of either localized ``holes'' in the plasma 
distribution function (see, e.g., \citealp{shukla}),  ``cat's-eye'' 
vortices in the fluid shear flow, and `magnetization' zones in the XY model.
Explicating the universal nature of these patterns as described by the 
single-wave model is the main goal of this article.

\subsection{Single-wave model preview}
\label{ssec:preview}

The single-wave model is a Hamiltonian normal form that 
captures the dynamics of Landau damping in stable situations, and the
nonlinear saturation of unstable modes.  
In light of its origin, the model must be of Hamiltonian and transport form,
and be defined in a phase space that we take, for simplicity, to be
two-dimensional.
The basic dynamical variables are a density-like variable $Q(x,y,t)$,
and an amplitude variable $A(t)$. The density $Q$ depends
on the coordinates $(x,y)$, describing the phase space over the
resonant, finely scaled region. The amplitude $A(t)$ depends purely
on time, and represents the amplitude of the embedded neutral mode;
given that mode's eigenfunction, one can reconstruct the large-scale
wave pattern outside the resonant region.
In addition, the model contains parameters that must be adjusted to 
characterize the physical system of interest.  A special case of the 
single-wave model is compactly given by following:
\begin{align}
&Q_\tswm + [Q,\mathcal{E}]=0\,,
\label{eq:swm1pre}\\
& \phisw = \amplsw \exe^{\imi \xswm} + \amplsw^* \exe^{-\imi \xswm}\,, 
  \label{eq:swm2pre}\\
&\imi  \amplsw_\tswm =
         \l\langle  Q \, \exe^{-\imi \xswm}\r\rangle \,,
 \label{eq:amppre}
\end{align}
where $\mathcal{E}=y^2/2 -\varphi$, 
\bq
 [f,g]=f_xg_y-f_yg_x\,, \ \  {\rm and} \ \  \l\langle \ \cdot\  \r\rangle 
 =\frac1{2\pi}\int_{-\infty}^{\infty}\!\! dy\int_0^{2\pi}\!\!\! dx \, \ \cdot  
\eq
and $f_x:=\p f/\p x$,  etc.  Equation (\ref{eq:swm1pre}) is the transport 
equation for $Q$ that resembles the Vlasov or Euler equation,  with 
dependence   on the variable $\varphi$ through $\mathcal{E}$,  an energy 
for the characteristic `particle orbit' equations.   From 
the second of Eqs.~(\ref{eq:amppre}) it is clear that $\varphi$ represents
 a wave of amplitude 
$A$,  with its single-wave spatial dependence upon the variable $x$ being 
obvious, while equation  (\ref{eq:swm2pre}) gives the temporal evolution 
of $A$, which is driven in an integral manner by the rearrangements of $Q$ 
throughout the entire resonant region.
For simplicity, the full parameter dependence of the single-wave model has 
been chosen in (\ref{eq:swm1pre})--(\ref{eq:amppre}) to furnish
the simplest possible form of the equations;
thus no parameters appear explicitly.

\begin{figure}[t]
\centering
\includegraphics[width=15 cm]{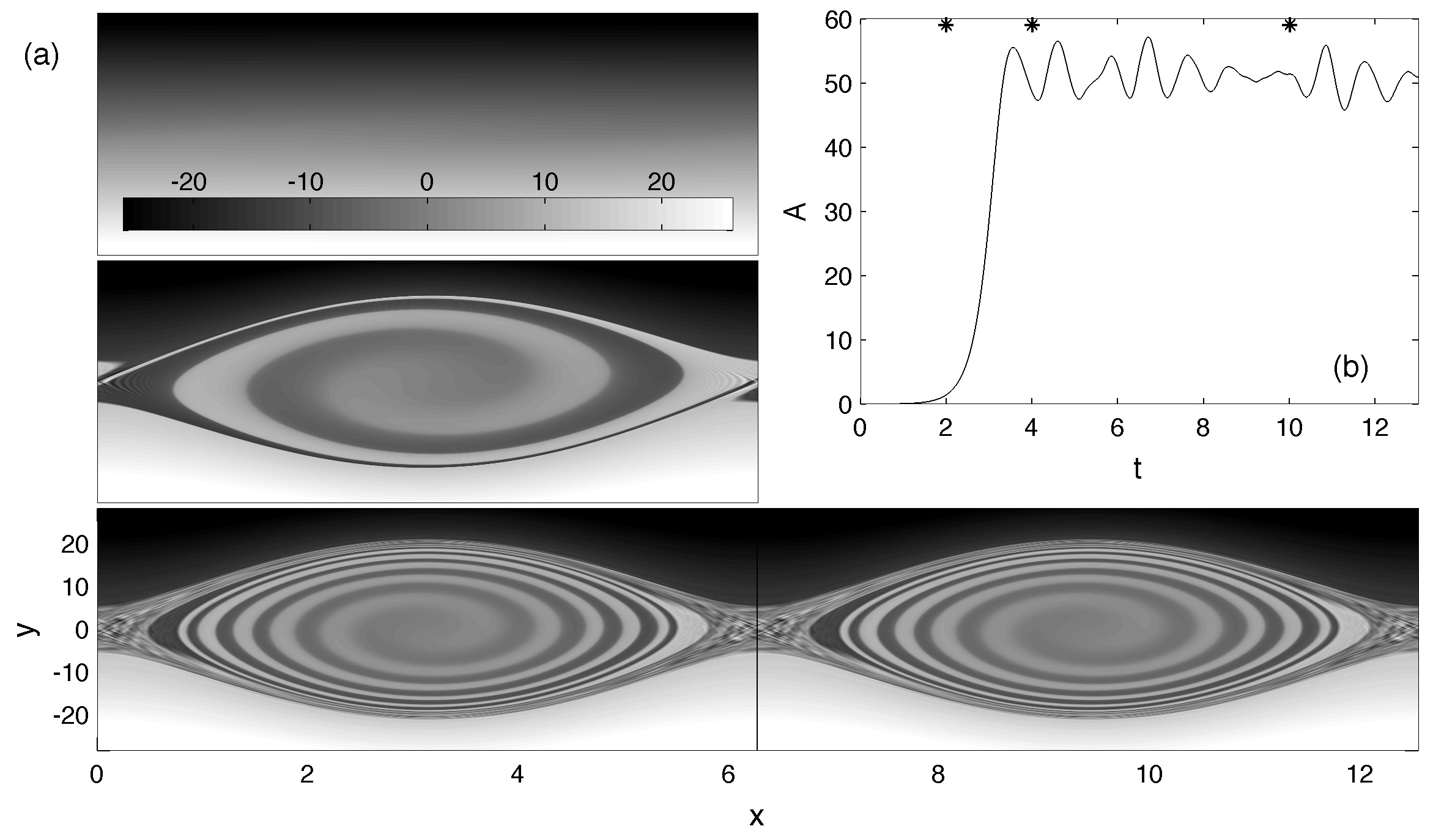}
\caption{
A solution of the single-wave model in its normal form.  Parameter values 
are $\kappa=0$ and $\gamma=-1$ (cf.\ Sec.~\ref{sec:derivation}). 
The system is initialized with $A(0)=0$ and $Q(x,y,0)=-y$  and  then kicked 
into action  by adding a forcing term  $0.1 t \exp (-200t^2) $ to the mode 
amplitude equation (see \citealp{balmforth2001b} for further details of the 
computational scheme).  Panel (a) shows three snapshots of $Q(,x,y,t)$ as 
densities on the $(x,y)$-plane.
Two wavelengths are shown for the final snapshot to emphasize the
periodic geometry. Panel (b) shows the time evolution of the
mode amplitude, $A(t)$; the stars indicates the times of the snapshots in (a).
}
\label{mum}
\end{figure}

The solution of (\ref{eq:swm1pre})--(\ref{eq:amppre})
is presented in Fig.~\ref{mum}, which shows three
snapshots of the density $Q(x,y,t)$ on
the  $(x,y)$-phase plane (panel a), and the mode 
amplitude $A(t)$ as a function of time (panel b). 
For the earliest snapshot, $Q(x,y,t)$ has not been significantly
rearranged, but 
at later times the $Q$ distribution twists up into a cats-eye pattern.
This is an example of the universal pattern seen in many physical systems 
described by the single-wave model, and is typical of vortex 
formation in fluid mechanics and electron hole formation in plasma physics. 

The system of Eq.~(\ref{eq:swm2pre}) is a Hamiltonian field theory given in Poisson bracket form as 
\bq
Q_t=\{Q, H\} \qquad {\rm and}\qquad A_t=\{A, H\}
\eq
where the Hamiltonian is given by 
\bq
H=\int_{-\infty}^{\infty}\!\! dy\int_0^{2\pi}\!\!\! dx \, \mathcal{E}  Q 
\eq
 and the Poisson bracket $\{\ ,\ \}$ is
\bqy
\{F,G\}&=&  \frac{i}{2\pi}\left(
\frac{\p F}{\p A} \frac{\p G}{\p A^*}-\frac{\p F}{\p A^*}\frac{\p G}{\p A} 
\right)
\nonumber\\
&{\ }& \hspace{.5 cm} + \int_{-\infty}^{\infty}\!\! dy\int_0^{2\pi}\!\!\! dx \, 
Q\left[\frac{\delta F}{\delta Q},\frac{\delta G}{\delta Q}\right] \, ;
\label{swpb}
\eqy
functional derivatives  such as $\delta F/\delta Q$ are defined in Sec.~\ref{sec:theories}.  This Hamiltonian structure is inherited from  that of the various `parent' models for which the single-wave model  is an encompassing  normal form. 

\subsection{Overview}
\label{ssec:overview}

In Sec.~\ref{sec:theories} we describe a large class of Hamiltonian  
2+1 field theories that contain the basic ingredients for the phenomena described by the single-wave model. In this section, we also establish our notation and 
provide examples of specific physical systems, including shear flow from 
fluid mechanics, Vlasov theory from plasma physics, and the XY model from 
condensed matter physics.   We use, mainly, these three examples to help 
make our general points throughout the paper.  We note that the mean field 
model with the XY interaction  has been called the `Hamiltonian Mean Field
model' by \textcite{campa} and collaborators.  We refrain from this terminology 
since all of the models treated in this paper are in fact examples of 
Hamiltonian mean field models. Instead, we will refer to this 
example as the XY model. 

As mentioned above, an essential feature of single-wave dynamics lies in the nature of the linear theory of the parent models.  In particular, to understand single-wave phenomena it is necessary to understand the CHH  transition to instability that it describes.  This  transition has  been developed and rediscovered independently in many fields, and since this linear dynamics is crucial to our development, we review this in an historical context in Sec.~\ref{sec:linear}.  We  first describe the linear theory for our general class of Hamiltonian models and then consider the early history due to Rayleigh and Kelvin, followed by the developments in plasma physics, stellar dynamics, and in the XY model.   A main purpose of producing a normal form theory is  to unify discoveries made in disparate fields;  the nature and considerable amount of rediscovery and interchange in several areas emphasizes the need for a review of the single-wave model. 


As is common in normal form theory, asymptotic calculations must be performed 
and the situation  for the single-wave model is particularly intricate.  
Thus, in Sec.~\ref{sec:nonlinear} we outline the nonlinear dynamics that 
one must understand to appreciate the single-wave model. In particular,
we first demonstrate how conventional weakly nonlinear theory is unable to
describe the transition to instability and fails to capture the scale
of the nonlinear saturation level. That level is referred to as
trapping scaling in plasma physics, and we provide a historical commentary 
on the origin of this scaling, as well as parallel development in fluid 
mechanics regarding what is termed critical layer theory.  These descriptions 
motivate  the asymptotic 
sequences used in the derivation of the single-wave model. 
 
 With the intuition gained from  Secs.~\ref{sec:linear} and \ref{sec:nonlinear}, the stage is set for the derivation of the singe-wave model, which we do in Sec.~\ref{sec:derivation}.   The derivation involves an asymptotic expansion featuring both the discrete embedded mode and the continuous spectrum. We provide a general derivation that produces the Hamiltonian normal form embodied in the single-wave model.  We also observe that, in some specific physical problems, a certain degeneracy occurs that necessitates a degenerate form of the single-wave model, a form shared by top-hat profiles in the XY model and some  waterbag distributions for the plasma and stellar dynamics problems.   Thus we delineate a norm form and a degenerate form of the single-wave model.  Next, we discuss the Hamiltonian nature and conservation laws of the single-wave model that are inherited from the Hamiltonian parent model.  
Finally, for later use, we outline the leading-order forms 
expected for different types of dissipation.  

The linearized version of the single-wave model is arguably the simplest
system capturing the transition
to instability of a discrete mode embedded in the continuous spectrum,
as well as the disappearance of such a mode and Landau damping.
In Sec.~\ref{sec:linearsinglewave} we describe this linear theory in detail, including the effect of dissipation, so as to address issues of structural stability of our model. 
Various  patterns emerging from  the single-wave model are described in Secs.~\ref{sec:normalpatterns} and \ref{sec:degeneratepatterns}.  
In Sec.~\ref{sec:normalpatterns} we discuss patterns of  the normal form: first where the pattern  arises out of  instability and second where the pattern arises from nonlinearly forcing a stable situation. In Sec.~\ref{sec:degeneratepatterns} we provide a similar analysis for the degenerate form, considering
in detail the XY model.
In Sec.~\ref{sec:deviants} several variants of the single-wave theme 
are discussed.   
Finally, in Sec.~\ref{conclusion}, we conclude and discuss the 
universality and limitations of the single-wave model.

\section{A class of Hamiltonian theories and specific examples}
\label{sec:theories}

We now describe a general class of 2+1 Hamiltonian mean-field theories
and outline several specific examples that we discuss in more detail
in later sections.

\subsection{A class of Hamiltonian field theories}
\label{ssec:badass}

The class of 2+1 Hamiltonian field theories \cite{morrison03} 
possesses a single independent variable,  $\zep(q,p,t)$, which is a density-like 
variable that depends on the independent  phase space  variables $z:=(q,p)$, coordinates for some phase space $\calz$,  as  well as time.    We suppose the density satisfies a  transport  equation of the following form:
\bq
\frac{\partial \zep}{\partial t} + [\zep, \cale]=0\,,
\label{eq:den.eq}
\eq 
where the `particle' Poisson bracket is defined by the usual expression
$[f,g]= f_{q} g_{p} - g_{q} f_{p}$, where $f_{q}:=\p f/\p q$ etc.,
and the quantity $\cale$ is an energy-like quantity that we call the particle energy.
If (\ref{eq:den.eq}) were a Liouville equation, then $\cale$ would be
a given function of $z$,  and we would have a linear theory.  However,
we are concerned with mean field theories where $\cale$ depends on $\zep$ in a global sense.
Thus, our systems are governed by  nonlinear (quasilinear) partial integrodifferential equations.   Such equations arise, for example, by
truncation of BBGKY-like hierarchies, which results in  particular functional dependencies of the particle energy on the density.  We do not pursue such an approach, but postulate that the particle energy arises from a total energy (Hamiltonian) functional of the form $H[\zep]= H_1 + H_2 + H_3 +\dots$, where the one-point energy, $H_{1}$, the two-point, $H_{2}$, etc.\  are given as follows:
\bqy
H_1[\zep] &=&\int_{\calz}\!\! d^2z\,  h_1(z) \, \zep(z,t) \,,
\nonumber\\
H_2[\zep] &=&\frac1{2} \int_{\calz} \!\! d^2z \int_{\calz} \!\! d^2z' \, 
\zep(z,t)\, h_2(z,z') \, \zep(z',t)\,,
\label{eq:1.2.ptle}
\eqy
with the generalizations to $H_3$, $H_4$, etc.\  being obvious. 
The quantities $h_{1}$ and $h_{2}$, the interaction kernels, are left unspecified for now.  But, we do 
suppose the two-point interaction possesses the symmetry $h_2(z,z') =
h_2(z',z)$. The particle energy is obtained from the field energy by functional
differentiation:
\bq
\cale :=\frac{\de H}{\de \zep} = h_1 +  \int_{\calz}\!\! d^{2}z'\,  h_2(z,z') \,
\zep(z',t)\,,
\label{eq:ptle.field}
\eq
where the functional derivative is defined as usual by
$\de H= \int_{\calz} d^2z \,  \de\zep\,  \de H/\de \zep$.
It is not difficult to show that $H[\zep]$ is a constant 
of motion under the dynamics of  (\ref{eq:den.eq}).

Equation (\ref{eq:den.eq}) with $\cale = \de H/\de \zep$ is a Hamiltonian field theory \cite{morrison03}.  However, since  there is only one field variable  $\zep$  the theory is not of a canonical form, but possesses a noncanonical Lie-Poisson bracket description (for review see, for
example,  \citealp{morrison98} and  \citealp{marsden}) given by
\bq 
\{F,G\}=\int_{\calz}\!\!d^2z\,  \zep 
\left[\frac{\de F}{\de \zep},\frac{\de G}{\de \zep}\right]\,.
\label{eq:pb}
\eq
Noncanonical brackets of this form, which  depend explicitly upon the variable $\zep$ unlike conventional
Poisson brackets that only depend on (functional) derivatives of the
canonical variables,  express the Hamiltonian form of matter in terms of Eulerian variables.   Such noncanonical Poisson brackets were introduced in the context of magnetohydrodynamics in \textcite{MG80},  and  the specific form of (\ref{eq:pb}) was given for the   Vlasov-Poisson system in \textcite{morrison80} and  the two-dimensional Euler equation in \textcite{morrison82}.
 
Using (\ref{eq:pb}) the equations of motion are obtained in the form 
\bq
\frac{\partial \zep}{\partial t}=\{\zep, H\}
= - \left[\zep, \frac{\de H}{\de \zep}\right]
= - [\zep, \cale]\,.
\label{eq:eom0}
\eq

Associated with the preceding Hamiltonian form are various constants of 
motion. The Hamiltonian $H[\zep]$ itself is one such constant.
In addition, there are the Casimir invariants,
\bq
C[\zep]= \int_{\calz}\!\! d^2z \,  \calc(\zep) \,,
\label{eq:casimir}
\eq
where $\calc(\zep)$ is an arbitrary function, which arise from
degeneracies in the Poisson bracket.

Other important invariants are momenta, $P[\zep]$, that are 
Hamiltonian dependent.   
Momentum invariants generally arise from translation symmetries that in the
present context would be determined by the form of the interaction kernels  $h_{1}, h_{2}, \dots$.  This is how the
strong version of Newton's third law is built into the $n$-body problem.  This idea can be generalized in various ways, 
but we do so by observing that  our system conserves 
momentum if there exists a canonical transformation of the phase space $\calz$ 
$$
{z=(q,p) \longleftrightarrow \bar{z}:=(\qq,\pp)}
$$
such that in the new particle coordinates $\bar{z}:=(\qq,\pp)$, the
interactions $h_{1}$,  $h_{2}$, etc.\  have upon composition with $z(\bar z)$ one of the following two forms:
 \bq 
 h_1\circ z= \bar h_1(\pp)\,,\quad\quad h_2\circ (z, z') =\bar h_2(\pp,\pp',|\qq -\qq'|) 
 \label{vpform}
\eq 
or
 \bq 
 h_1\circ z= 0\,,\quad\quad h_2\circ (z,z') =\bar h_2(|\pp-\pp'|,|\qq-\qq'|) \,.
 \label{euform}
\eq  
In the first case  
\bq
P[\zep]= \int_{\calz}\!\!d^2z  \, \pp\, \zep(z)\,.
\label{eq:mom}
\eq
is conserved, while in the second case we have two kinds of translation invariance and thus two components of the momentum
\bq
P_1[\zep]= \int_{\calz}\!\!d^2z  \, \pp\, \zep(z) \, \quad {\rm and} \quad P_2[\zep]= \int_{\calz}\!\!d^2z \, \qq\, \zep(z)\,.
\label{eq:mom2}
\eq
All examples treated in this review will be of one of these types 
and three-point and higher interactions will not be considered.

Associated with each mean-field model is a Hamiltonian n-body problem. 
This follows by assuming a Klimontovich type of distribution 
\bq
\zep=\sum_{i=1}^n\de(z-z^i(t))\,,
\label{klim}
\eq
with $z^i=(q^i,p_i),\quad i=1,2,\dots,n$, and   substituting (\ref{klim}) 
into (\ref{eq:den.eq}) and seeking a weak solution.  Alternatively, the 
n-body problem arises upon effecting the functional chain rule on 
$F[\zep]=f(z^1,z^2,\dots, z^n)$, where $f$ is defined upon inserting 
of (\ref{klim}) into $F$, giving
\bq
\frac{\p }{\p z^i}\left.\frac{\de F}{\de f}\right|_{z^i}=\frac{\p f}{\p z^i}\,,
\eq
to  establish  a mapping between the canonical and noncanonical  Poisson 
brackets.  Both methods yield the Hamiltonian n-body problem
\bq
\dot{z}^i= [z^i, H]\,,
\eq
where self-energy terms are removed from $H$.  The existence of this 
Hamiltonian n-body problem reinforces why our patterns occur in a phase 
space in the mean-field theory, rather than a configuration space.  One 
consequence of this is that the characteristic equations of the single-wave model 
possess Liouville's theorem on conservation of phase space volume (here area). 
 
All of the examples treated here arise from equilibria that only depend 
on $\pp$.  For this reason we set $\zep(\qq,\pp,t) = R(\pp) + \rho(\qq,\pp,t)$ 
and then when a choice of $R$ is made,  $\rho(\qq,\pp,t)$ represents  the 
main dynamical variable.  We further assume that $\qq$ is an angle-like
variable, so that the phase space is periodic in 
$\qq\in  [0,2\pi]$, and
$\pp\in\cD$ with $\cD$ equal to $[-1,1]$ or $(-\infty,\infty)$, depending 
on the example.  Upon substitution of $\zep = R + \rho$ into $\cale$, both 
of the forms of (\ref{vpform}) and (\ref{euform}) can be written as follows:
\bq
\cale[R+\rho]=\cale[R] + \cale[\rho]=:h(\pp) + \Phi(\qq,\pp)\,, 
\eq
with
\bq
\Phi(\qq,\pp) = \calk \rho 
\ldef \int_\cD\!\! \dint\pp'\int_{0}^{2\pi}\!\!\! \dint  \qq'  \, K(\pp,\pp',|\qq-\qq'|)
         \,\rho(\qq',\pp',t)\,,
 \label{Phi}
\eq
where $h$ and $K$ are determined by $h_1$ and $h_2$. 
Thus the governing equations become 
\bq
  \rho_t + [R,\Phi] + [\rho,h+\Phi]=0
\label{eq:eomA}
\eq
or
\bq
  \rho_t + {\Ome}{\rho_\qq} - R' \Phi_\qq
       + [\rho,\Phi] = 0,
         \label{eq:eom}
\eq
where $[f,g]=f_{\th} g_{\pp} - g_{\th} f_{\pp}$ and $\Ome(\pp)=h'$.    The quantity $h(\pp)$ corresponds to a Hamiltonian for an integrable system and the coordinates $(\qq,\pp)$ are action-angle coordinates for the characteristics of this system. 

Equation (\ref{eq:eom}) will serve as our starting point in subsequent analyses of our examples. 
In addition for all the examples we consider, the interaction is assumed to have zero mean, i.e.
\beq
\int_0^{2\pi}\!\!\dint\theta' \,  K(I,I',\theta-\theta') = 0.
\eeq
Of note is the case where $K$ is independent of $I$ and $I'$; such a  $K$  will be referred  to as a {\it spatial kernel}.

\subsection{Notation}

By tradition, different independent and dependent variables  
have been used in different contexts. For example, in plasma physics,
the phase space coordinates, representing position and velocity,
are often written as $(x,v)$. In fluid mechanics, both coordinates denote
position, just as the example of Sec.~\ref{ssec:preview} used $(x,y)$.
To be a little more definite,
we will always describe the first independent variable as coordinate-like
or angular, 
while the second will be momentum-like or an action, as for
the general independent variables, $(q,p)$ and
$(\theta, I)$, defined above. Moreover,
throughout, the following two shorthands will be used:
\bqy
[f,g]&=&f_{q}\, g_{p}- f_{q}\, g_{p} \\
\l\langle f\r\rangle &=&\frac1{2\pi}\int_\mathcal{D} \!\!d p \int_0^{2\pi}\!\!\!d q \, f(q,p)\,, 
\label{bkts}
\eqy
where $(q,p)$ denotes any `conjugate'  pair and, as above,  
subscript denotes partial differentiation.

When there is a proliferation of subscripts, we will use e.g. $\p_x$ to denote $\p/\p x$ etc.  In all cases our coordinate-like 
variable will be periodic and eventually scaled to be $2\pi$-periodic, while the momentum domain $\mathcal{D}$ is example specific. When  $\mathcal{D}=(-\infty,\infty)$ we will denote it by $\R$.

\subsection{Examples}

\subsubsection{Shear flow}
\label{ssec:shear_flow}

\begin{figure}[t]
\begin{center}
(a)
\includegraphics[width=6.5 cm]{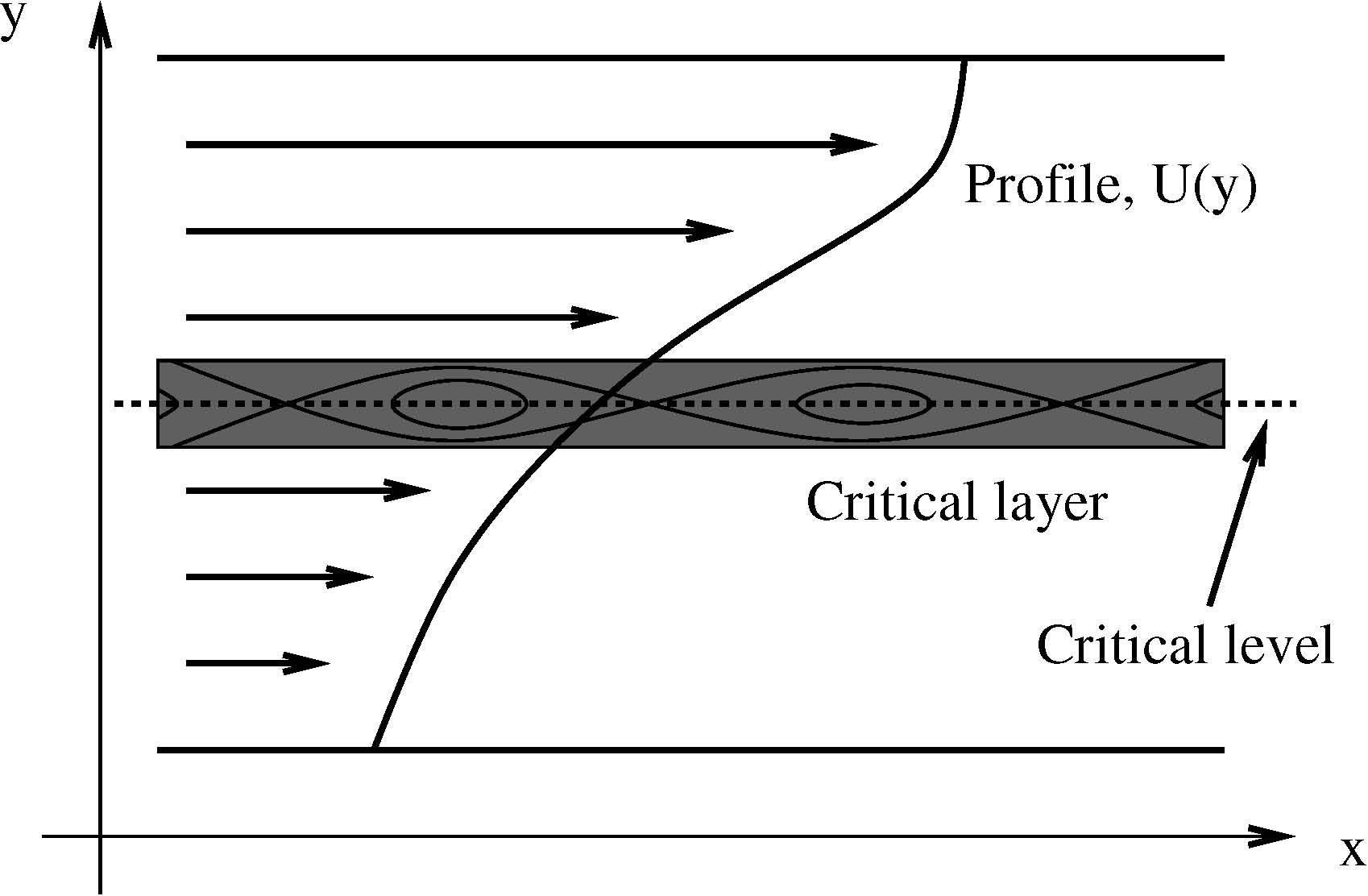}
\hspace{0.5cm}
(b)
\includegraphics[width=6.5 cm]{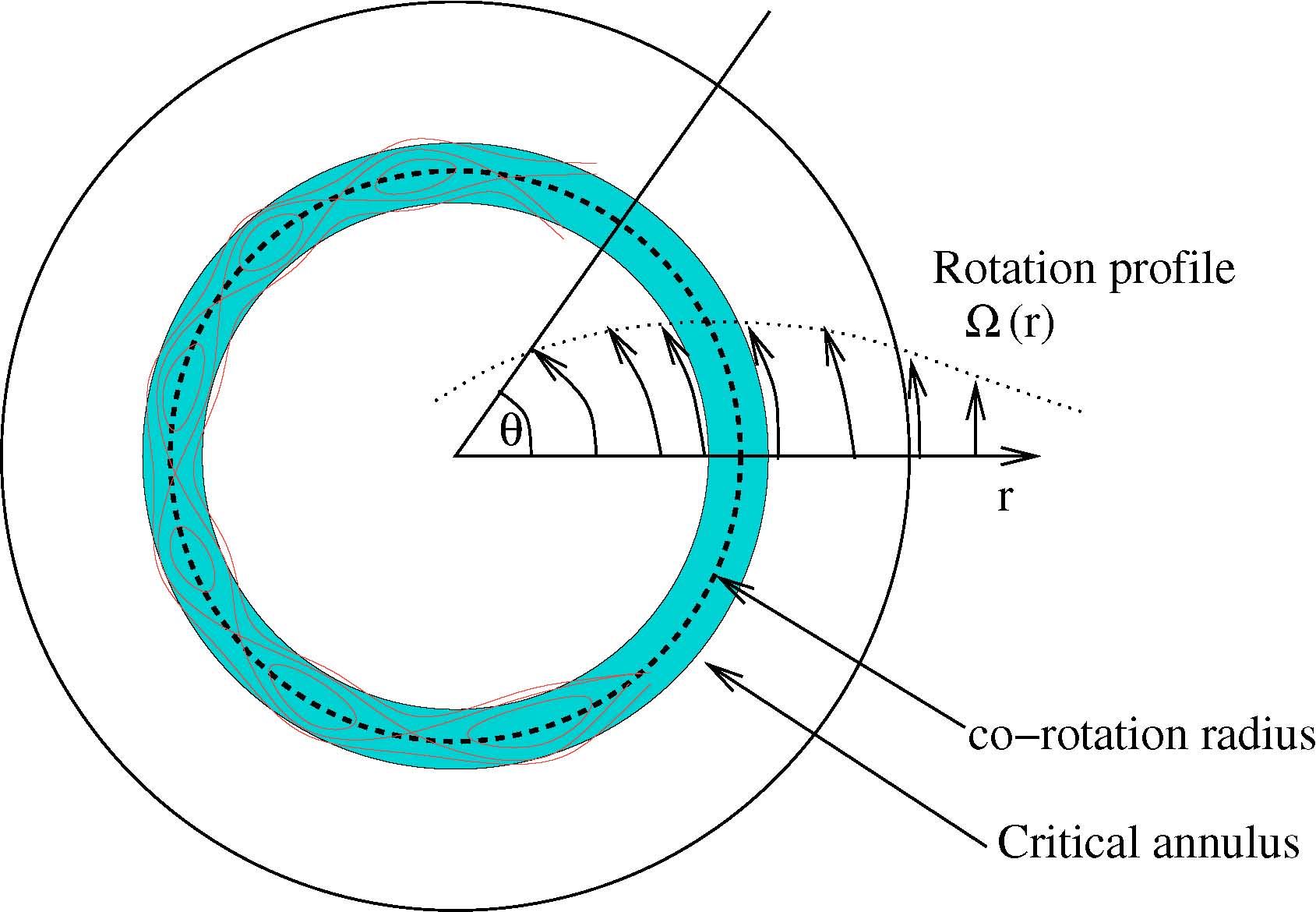}
\end{center}
\begin{center}
(c)
\includegraphics[width=6.5 cm]{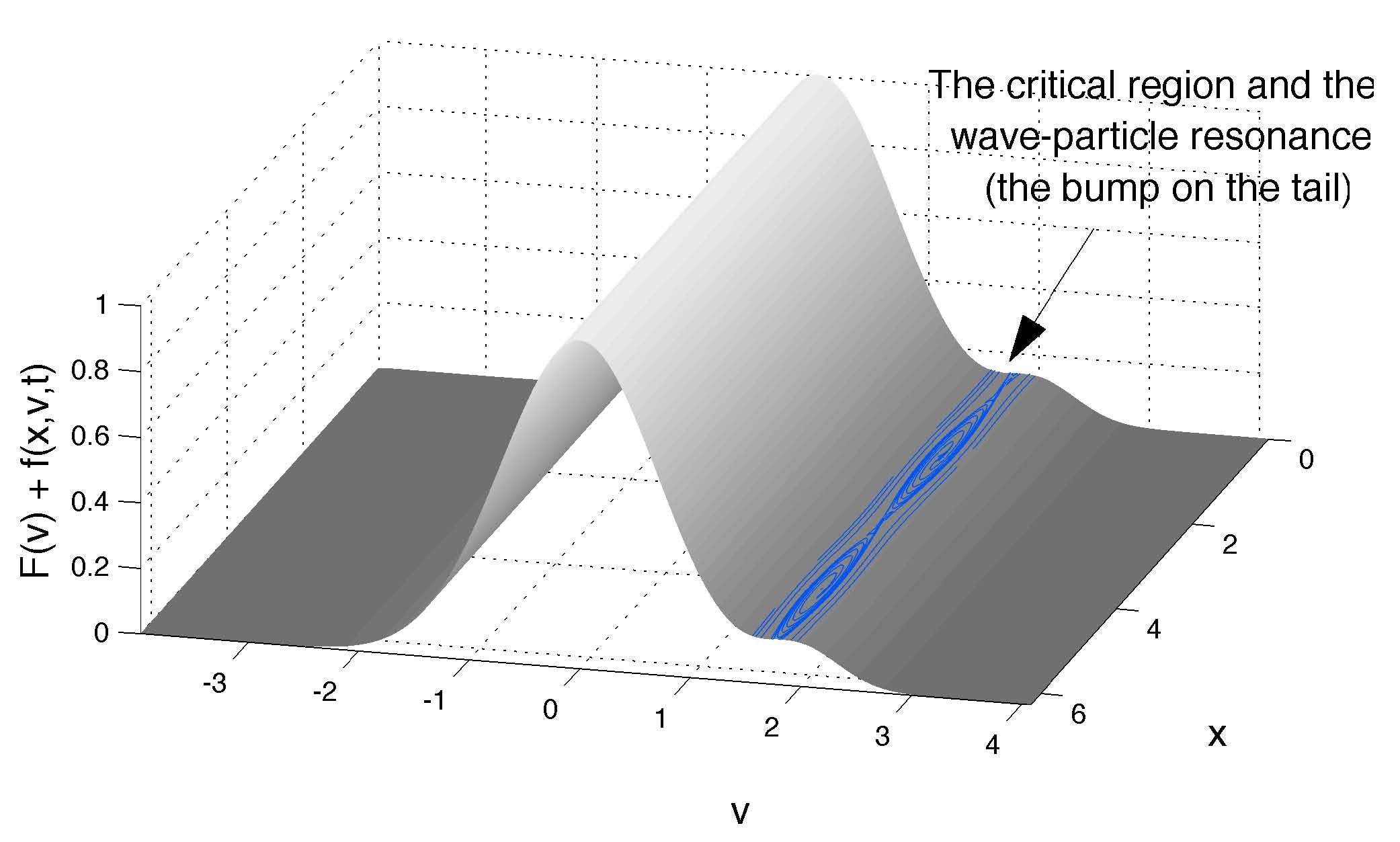}
(d)
\includegraphics[width=6.5 cm]{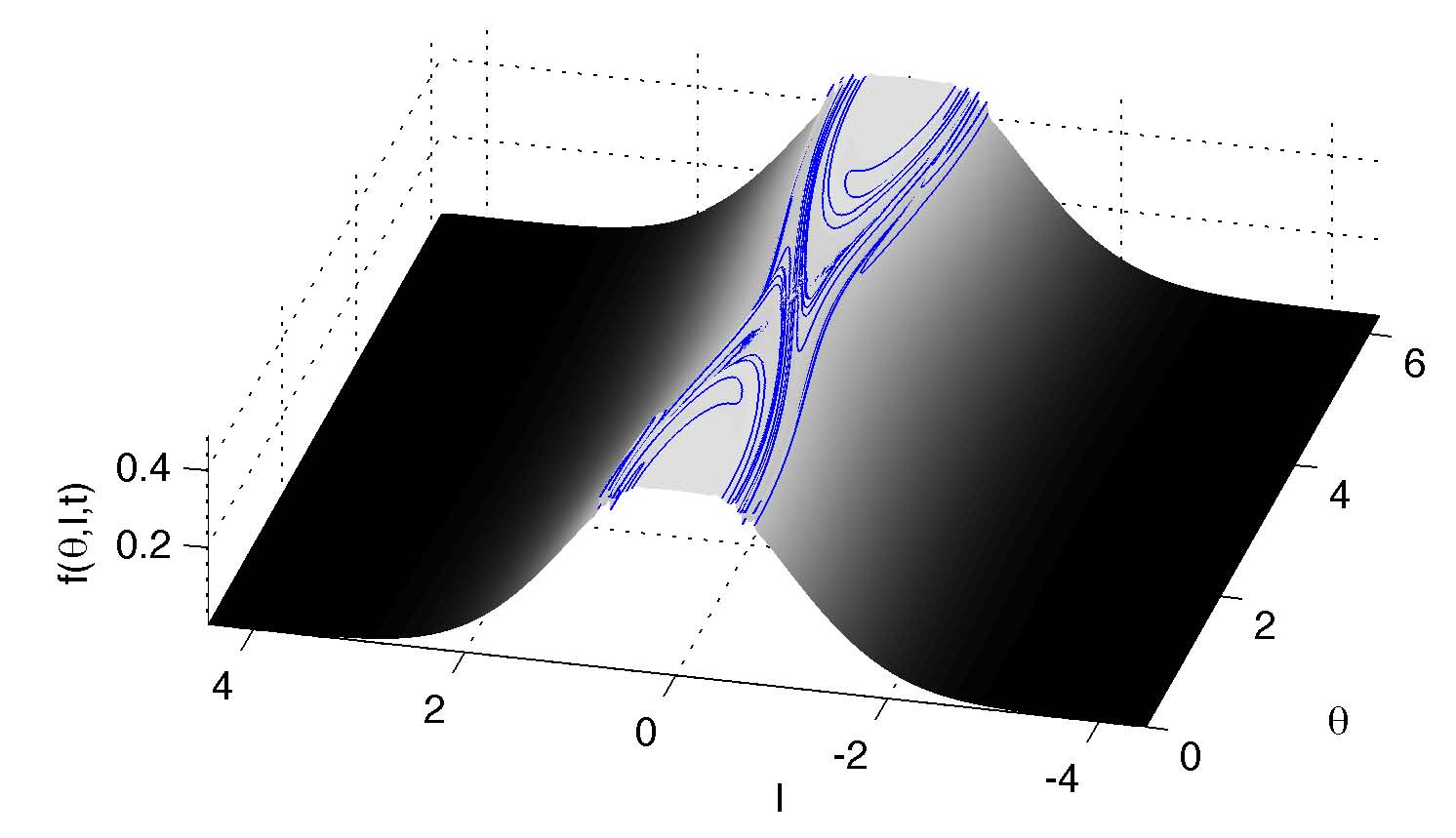}
\end{center}
\caption{
Sketches of the geometry of our inviscid shear flows and ideal plasma.
In the channel flow in (a), the flow proceeds in the $x$-direction, and shears in the $y$-direction
according the equilibrium profile, $U(y)$.  The unstable disturbances
considered in this article take the form of weakly unstable modes
that have a pronounced effect within a localized ``critical layer''
which surrounds a line, the ``critical level''
along which a neutrally stable wave propagates at the same speed
as the mean flow. In (b), the circular variant of the problem is
shown: a differentially rotating vortex with equilibrium
rotation rate, $\Omega_0(r)$. The critical ring, or co-rotation radius,
now locates the critical region. The sketch for the plasma problem in (c) displays the total
distribution function, $F(v)+f(x,v,t)$, as a surface over the
$(x,v)$-plane; the location of the wave-particle resonance
and the surrounding critical region are indicated.
For the XY model example, shown in (d), the distribution function,
$f(\qq,\pp,t)$,
develops a cat's eye pattern around the resonance at $\pp=0$.
}
\label{shearfig}
\end{figure}

Our example of shear flow considers the evolution of disturbances to a channel flow or to perturbations of differentially rotating vortices, as illustrated in Fig.~\ref{shearfig}.  Thus, we  begin with Euler's equation in two-dimensions, in which case the field $\zep$ corresponds to the scalar  vorticity field.  In cylindrical geometry the coordinates for $\calz$ would be $(\qq,\pp)= (\qq,r^2/2)$, the usual polar coordinates; however, we are principally interested in channel flow with a periodic boundary condition and so we adopt $(\qq,\pp)= (x,y)$, where  the spatial coordinate $x$ is equivalent, after a simple scaling, to the ``angle''  $\theta$.  Thus, for shear flow, the action-like coordinate $\pp$ is equivalent to $y$, the cross stream coordinate,  the equilibrium channel flow speed in the $x$-direction is $U(y)$, and $\calk$  is the inverse of the two-dimensional Laplacian for this geometry.  We then set  $\zep=R+\rho= U'(y) + \om(x,y)$, which here physically corresponds  to   divorcing the  background equilibrium flow from an evolving perturbation  described by a relative 
vorticity field $\omega(x,y,t)$.   Thus, (\ref{eq:eom}) for this case is
\begin{equation}
\omega_t + U \omega_x - U''\psi_x 
+ [\psi,\omega]
 = 0\, ,  
\label{f.1}
\end{equation}
where 
(\ref{Phi}) is equivalent to 
\begin{equation}
\omega=\psi_{xx}+\psi_{yy}\,,
\label{f.2}
\end{equation}
and $\psi(x,y,t)$,  the  streamfunction, is identified with $\Phi$. 
Equations (\ref{f.1})--(\ref{f.2}) are dimensionless, the channel width
and maximum equilibrium flow speed acting as the relevant units.
The flow is bounded by impermeable walls at $y=\pm1$,
and so $\psi(x,\pm1,t)=0$, and periodic in $0\leq x<L$,
where the domain length is $L$.

Thus, to summarize, the shear-flow problem is recovered from the system of Sec.~\ref{ssec:badass} by the following identifications: 
\bqy
&&\qq\leftrightarrow x,\  \pp\leftrightarrow y,\  \Ome(\pp)\leftrightarrow U(y),\  R(\pp)\leftrightarrow U'(y),\ \nonumber\\
&&   \rho \leftrightarrow \om,\   {\rm and\  (formally)}\ \    K \leftrightarrow \Delta^{-1}\,.
\nonumber
\eqy

\subsubsection{The Vlasov-Poisson plasma}
\label{sssec:vp}

In the Vlasov-Poisson system,
the plasma is described by an electron distribution function (phase space density) 
depending on a spatial coordinate, $x$, and speed, $v$, with a stationary neutralizing ion background.
The relevant equilibrium distribution function is spatially homogeneous
and determined by a profile, $F(v)$. Disturbances
about this equilibrium are described by
the perturbation distribution function, $f(x,v,t)$, 
and electrostatic potential, $\varphi(x,t)$, that satisfy
the Vlasov equation,
\begin{equation}
f_t + v f_x + \varphi_x F_v +  [\varphi, f] 
= 0,
\label{e.1}
\end{equation}
and Poisson equation,
\begin{equation}
\varphi_{xx} = \int_{\R}\!\!\dint v\,  f(x,v,t) - N_B
\,,
\label{e.2}
\end{equation}
where $N_B$ is the neutralizing background (i.e. $N_B = \langle f \rangle$). 
Here, the electron mass, charge, and an assumed thermal velocity scale 
are used to nondimensionalize the equations.  Thus, lengths are measured 
in units of  Debye length and time in units of the  electron plasma frequency. 
We assume periodic boundary conditions in $x$, and we demand that $f$ vanish
as $v\rightarrow\pm \infty$. 

Thus, the  plasma problem is recovered from the system of Sec.~\ref{ssec:badass} 
by the following identifications: 
\bqy
&&\qq\leftrightarrow x,\  
\pp\leftrightarrow v,\  \Ome(\pp)\leftrightarrow v,\  
R(\pp)\leftrightarrow F(v),\  \rho \leftrightarrow f,\   \nonumber\\
&&
\varphi \leftrightarrow -\Phi,\  {\rm and\  (formally)}\ 
K \leftrightarrow -\partial_x^{-2}=|x-x'|\,.
\nonumber
\eqy

\subsubsection{XY model}
\label{sssec:xyinteraction}

Our XY model captures
essential features of  long-range spin-spin interaction and 
has been studied in a variety of contexts \cite{chaikin,campa}. 
The phase  space coordinates $(\qq,\pp)$ and
basic dynamical variable $f(\theta,I,t)$ are equivalent to
those for the Vlasov-Poisson system of Sec.~\ref{sssec:vp},
and the model differs only by a change in the interaction:
\bq
f_t + I f_\theta - \Phi_\theta R_I - \Phi_\theta f_I
= 0,
\label{hmf.1}
\eq
\begin{equation}
\Phi = - \frac1{\pi} \int_{\R}\!\!  \dint  I  \int_{0}^{2\pi}\!\! \! \dint \theta'  \,   f(\theta',I,t)\; \cos(\theta-\theta')\,.
\label{hmf.2}
\end{equation}
Thus, the  XY  model  is the system of Sec.~\ref{ssec:badass} with 
the following identifications: 
\bqy
&&
\qq\leftrightarrow \theta,\  
\pp\leftrightarrow,\  \Ome(\pp)\leftrightarrow I,\  
R(\pp)\leftrightarrow R(I),\nonumber\\
&&  \rho \leftrightarrow f,\   {\rm and \ }\  
K \leftrightarrow- \pi^{-1}\cos(\theta-\theta')
\,.
\eqy

\subsubsection{Other models}
\label{sssec:other}

There are many models with physical content that  fit into the same general Hamiltonian framework 
 described here.   One example is the Jeans equation for stellar dynamics  (obtained by changing the sign of the interaction in the Vlasov equation and removing the zero net charge condition).  
Other examples include various quasi-geostrophic models for describing ocean and atmospheric dynamics \cite{pedlosky}.  
In shear  flow, the vorticity
defect model of  \textcite{balmforth97a} (obtained by simplifying the relationship between 
the vorticity and the streamfunction) is also applicable. Interesting 
related examples where the underlying characteristics correspond to  integrable $n$-body problems are the Cologero-Moser system
(\citealp{Moser75,Illner00}) and  a model introduced by \textcite{smerk}.

\section{Linear theory and  historical perspective}
\label{sec:linear}

The linear theory of fluid and plasma  systems with continuous spectra has 
a long history.  Here we present a general theory in 
Sec.~\ref{ssec:generaltheory} for the bifurcation to instability, 
followed by discussions of the  early developments in fluid mechanics in Sec.~\ref{ssec:raleigh}, plasma physics in Sec.~\ref{ssec:plasmalandau}, and other systems  in Sec.~\ref{ssec:xylinear}.

\subsection{General eigenspectra; bifurcation to instability through
an embedded neutral mode}
\label{ssec:generaltheory}

We begin by substituting  $\rho = \hat\rho(\pp,t) \exe^{\imi k\qq}$ 
(with wavenumber $k$)  into Eqs.~(\ref{Phi}) and (\ref{eq:eom})  
of the general formulation of Sec.~\ref{ssec:badass}.
After linearizing the latter, we arrive at
\begin{equation}
\hat\rho_t = - \imi k \Ome \hat\rho + \imi k R' \hat \cK_k \hat\rho ,
\qquad
\hat \cK_k \hat\rho = \int_\cD \!\! \dint\pp' \, \hat K_k(\pp,\pp') \hat\rho(\pp',t)
\,,
\label{njb.0}  
\end{equation}
and
\begin{equation}
\hat K_k(\pp,\pp') = \int_{0}^{2\pi}\! \!\! \dint\qq\, K(\qq,\pp,\pp') \exe^{\imi k\qq}
\,.
\end{equation}
Special cases of $\hat K_k$ of interest are the following:
\begin{description}
\item[\rm \underline{shear  flow}]
\[
 \hat{K}_k= 
\begin{cases}
 \frac{\sinh[k(I+1)] \sinh [k(I'-1)]}{ k\sinh[2k]}\,,& \text{if }\  -1\leq I\leq I'\\
 \frac{\sinh [k(I-1)] \sinh [k(I'+1)]}{ k\sinh[2k]}\,,& \text{if }\  \quad   I'\leq I\leq 1
\end{cases}
\]
\item[\rm \underline{Vlasov  plasma}]
\[
\hat{K}_k=-k^2\,,
\]
\item[\rm \underline{$XY$ - interaction}]
\[
 \hat{K}_k=-\delta_{k,1}-\delta_{k,-1}\,,
\]
\end{description}
with  $\delta_{i,j}$ being the Kronecker delta symbol.

The first term on the right of (\ref{njb.0})
corresponds to  free-streaming motion of fluid elements or particles
along each level of $\pp$.   In the Fourier representation, 
the free-streaming dynamics
is analogous to a continuum of uncoupled oscillators
with a range of frequencies $\Omega(I)$ indexed by $\pp$. The shearing action
of the basic flow corresponds to phase mixing due to the frequency spread.
This can be seen particularly clearly for cases in which the
second term on the right of (\ref{njb.0}) is absent, leading to
\bqy
\rho &=& \hat \rho(\pp,t) e^{\ii k\qq}
= \hat \rho(\pp,0) e^{\ii k[\qq-\Ome(\pp)t]}
\nonumber
\\
&\equiv& 
 \hat\rho (\pp,0) \int_\cC\!\! \dint c\,  \exe^{\ii k(\qq-ct)}\,  \delta(\Ome(\pp)-c) \,,
\label{njb.00}
\eqy
where $\cC$ denotes the range of values of $\Ome(\pp)$ that correspond
to $\cD$. This solution models the tilting over of the initial perturbation
by differential advection; the lines
of constant phase are given by $\qq-\Ome(\pp)t=const$.
The final piece of (\ref{njb.00}) further emphasizes how the solution
cannot be written in a form that is
separable in $\pp$ and $t$ unless one formulates it
as an integral superposition of singular modes of the continuous
spectrum with delta-functions for eigenfunctions \cite{dirac27,vankampen55a,eliassen}.
In other words, one can identify the free-streaming operator
in the term, $\Ome(\pp) \rho_\qq$, as creating the continuous spectrum.

The second, interaction term 
in (\ref{njb.0}) acts as a perturbation of the
free-streaming operator. However, with certain restrictions
on the form of $\cK_k(\pp,\pp')$, this perturbation of the operator
does not change the existence of the continuous spectrum (for rigorous treatments see 
\citealp{kato,degond,morrison03,hagstrom}), and only
introduces the possibility of additional discrete eigenmodes.
This was shown by explicit construction of  the solution of the initial-value
problem by \textcite{case60a} and \textcite{rosencrans66a}.

To find the discrete normal modes introduced by the interaction 
term, we set $\hat\rho(\pp,t)=\tilde\rho(\pp) \exe^{-ikct}$, 
furnishing
\begin{equation}
\tilde\rho = {R' \eta \over \Ome-c}, \qquad
\eta = \hat \cK_k\tilde\rho
\label{njb.1a}
\end{equation}
and
\begin{equation}
\eta(\pp) = \int_\cD\!\!\dint\pp' \, 
\hat K_k(\pp,\pp') \frac{R'(\pp')\eta(\pp')}{\Ome(\pp')-c}
\,.
\label{njb.1}
\end{equation}
If $c$ is complex, this is a regular integral equation. But
in the problems of interest, the integral on the right-hand side of
(\ref{njb.1}) becomes singular when $c$ is real,
which is symptomatic of the continuous spectrum.
Moreover, for $c=c_r+i c_i$ and $c_i\rightarrow 0^\pm$,
\bqy
\eta(\pp) &=& \pvint_{\!\! \cD} \!\!\dint\pp'\, \hat K_k(\pp,\pp')\, \frac {R'(\pp)\eta(\pp')}{\Ome(\pp')-c_r}
\nonumber\\
&{\  }& \hspace{.5 cm} \pm i\pi\sum_j \hat K_k(\pp,\pp_j) {\eta(\pp_j) R'(\pp_j) \over |\Ome'(\pp_j)|}
\,,
\label{njb.2}
\eqy
where $\pvint$ means the singularity of the integral is evaluated by taking the 
Cauchy  principal value, and the $\pp_j$'s denote all values of $\pp$ for which $\Ome(\pp)=c_r$ \cite{gakhov}.
We may satisfy the real and imaginary parts of   (\ref{njb.2}) if we choose  
$R'(\pp_j)=0$ for all the $\pp_j$'s,  and take
$\eta(\pp)$ to be real. This choice is straightforward if $\Ome(\pp)$
is a monotonic function, as we assume  later when we derive the
single-wave model. Then, $\pp_j\rightarrow \pp_*$ and
$R'(\pp_*)=\Ome(\pp_*)-c_r=0$.
Non-monotonic equilibrium with certain symmetries
can also satisfy this condition (such as the jet portrayed below, 
in Fig.~\ref{fig:jet}, which is symmetrical about the midline).
In either circumstance, the special mode that is embedded within the
continuum and which is also the limit of the complex eigenmodes
as $c_i\rightarrow0^\pm$ is the smooth solution of (\ref{njb.2}).
The distinguished modes are therefore associated with the
extrema of the equilibrium profile, $R(\pp)$ (inflection  points
of the velocity profile, $\Ome=U(y)$, in the fluid problem).

Note that, if the kernel is independent of the action coordinates, what we  call  
a  {\it spatial kernel}, then  
 $\cK\to\cK(\theta)$ and the $\cK_k$'s and therefore the
eigenfunctions $\eta$ are constants. Hence, (\ref{njb.1}) reduces
to an explicit dispersion relation,
\beq
D(c,k) = 1 - \cK_k \int_{\cD} \!  \dint I\, \frac{R'(I)}{\Ome(I)-c} = 0\,.
\label{nx.1}
\eeq
For the plasma problem, $D(c,k)$ is the plasma dispersion function,
and its relatively simple form allows one to extract a number
of general criteria for determining the transition to instability,
results such as those of  \textcite{penrose} (see Sec.~\ref{ssec:plasmalandau}). 
Similar results have recently been outlined for the XY model (Hamiltonian mean field model) by 
\textcite{chavanis} as well as vorticity defects  by \textcite{balmforth97a}, 
although they amount to a minor generalization of those
provided by Penrose, which are  standard in plasma physics \cite{krall}.

\subsection{Rayleigh's problem}
\label{ssec:raleigh}

In the late 1800s, Rayleigh explored a variety of fluid stability
problems motivated by experiments with smoke jets and flames. 
As vividly illustrated by cigarettes and chimneys,
smoke jets often rise initially uniformly, but then begin tortuous
undulations and meanders. Rayleigh's theoretical explanation for the
breakdown of the unidirectional shear flow in the jet amounted to
a linear stability analysis of
(\ref{f.1}) and (\ref{f.2}). For
exponentially growing, wave-like, normal-mode solutions with
$\psi \propto \exe^{ik(x-ct)}$, the system reduces to the
so-called ``Rayleigh equation,'' 
\begin{equation}
\psi_{yy}-k^2 \psi = \frac{U'' \psi}{U-c}
\,.
\label{f.3}
\end{equation}
This equation follows from (\ref{njb.1}) on recalling that the
kernel is  Green's  function for the fluid problem, and then recasting
this equation in its differential form. 

Rayleigh found explicit examples of unstable modes for inviscid shear flows
in which $U(y)$ took the form of a sequence of broken lines
(corresponding to a stacked set of layers with constant mean vorticity).
For such profiles, $U''(y)=0$ everywhere except at the line breakages,
which allows for an explicit closed form  solution of (\ref{f.3}).
His  results for these broken-line profiles motivated Rayleigh
to continue further to look for a more useful
stability criterion for general flow profiles, and he formulated
what is now called the ``inflection-point theorem'':
we multiply (\ref{f.1}) by $\psi^*$ and integrate
across the channel. After a little algebra, involving integration
by parts and separation of real and imaginary parts, we arrive
at
\begin{equation}
c_i \int_{-1}^1\!\! dy\,  \frac{U''(y) |\psi|^2}{|U-c|^2}= 0.
\end{equation}
Thus either $c_i=0$ (and there is no exponentially growing mode),
or the integral must vanish, which demands that $U''$ change
sign somewhere within the channel. In other words,
a sufficient condition for {\sl stability} is that $U(y)$ have an
inflection point. Further results that extended Rayleigh's
theorem were presented later by 
\textcite{fjortoft50a,howard64b,rosenbluth64a,balmforth1999}.
These criteria can be argued to be equivalent to energy stability criteria 
\cite{balmforth2002a}, akin to Dirichlet's theorem of classical mechanics 
(see,  e.g, \citealp{morrison98}).

Rayleigh's equation  (\ref{f.3}) has singular points
for real wavespeeds where $U(y)=c_r$. These singularities
typically prohibit the construction of neutral waves 
(e.g.\  \citealp{balmforth1999}), and occur at the 
so-called ``critical levels''
for which neutral waves propagate at the same speed
as the background flow. These locations play a special
role in the single-wave theory described later, and are examples of the
resonances mentioned in Sec.~\ref{sec:intro}. Note that
the critical-level singularities appear only in the normal-mode
problem: as is clear from our discussion in Sec. \ref{ssec:generaltheory},
the linear initial-value problem itself is completely regular,
and it is only by forcing solutions into the normal-mode form that 
one produces the singularity. 

The singularities in Rayleigh's equation, and their prohibition of
neutral waves led Kelvin to object quite strongly to Rayleigh's theory:
``This disturbing singularity vitiates the seeming proof of
stability contained in Lord Rayleigh's equations.''
Kelvin's view was that Rayleigh's singularities reflected a
breakdown of linear inviscid theory. At the critical levels,
``the motion has a startlingly peculiar character'', nonlinearity
could not be ignored, and any slight disturbance created ``a cat's-eye
pattern of elliptic whirls''.
In fact, Kelvin was convinced that viscous shear flows were linearly
stable, as illustrated by his analysis of Couette flow ($U(y)=y$).
Kelvin was also probably prejudiced by Reynolds' experiments
in pipes, which had been undertaken at roughly the same time, and
suggested that there was a finite threshold
in the amplitude of perturbations required to generate turbulent motion.
Kelvin argued further that inviscid flows were strongly
unstable in the sense that the slightest perturbation would generate cat's
eye patterns along any critical levels (``let one or both bounding-surfaces
be infinitesimally dimpled in any place and left free to become plane
again... Hence the interior disturbance essentially involves
elliptic whirls. Thus we see that the given steady laminar motion
is {\sl thoroughly} unstable, being ready to break up into eddies
in every place, on the occasion of the slightest shock or bump
on either plastic plane boundary''). Kelvin  evidently felt that there
were some fundamental problems with the inviscid formulation.

Despite Kelvin's criticisms, Rayleigh remained convinced
that inviscid fluid theory was a useful guide, and that shear flows could
be unstable, perhaps guided by his experiments with flames and smoke
jets.  Rayleigh contended Kelvin's objections by pointing out that the
inviscid linear analysis was perfectly valid for exponentially growing
modes --  in this case $c$ is complex and
there are no critical level singularities. However, he did admit
that his proof of stability for flows without inflection points
only concerned exponentially growing instabilities and did not exclude
algebraically growing instability
(``Perhaps I went too far in asserting that the motion was thoroughly
stable; but it is to be observed that if $c$ be complex, there is
no disturbing singularity. The argument, therefore, does not
fail, regarded as one for excluding complex values.'').
In fact, it was not until the 1960s that
algebraic instability was to a large degree ruled out for shear flow
when researchers followed Landau's Laplace transform 
treatment of the plasma problem  (cf.\ Sec.~ \ref{ssec:plasmalandau}) 
and solved the linear problem as an initial-value 
problem \cite{case60a,dikki60a,engevik,rosencrans66a}.

Rayleigh also criticized Kelvin's assumption that
viscous flows were stable: ``Lord Kelvin arrives at the conclusion
that the flow ... is fully stable for infinitesimal disturbances....  
Naturally it is with diffidence that
I hesitate to follow so great an authority, but I must confess
that the argument does not appear to me demonstrative.'' Nevertheless,
some of the inviscid profiles that Rayleigh aimed to establish were unstable
to explain unsteady motions in experiments were actually shown to be stable by
the inflection-point theorem.
Rayleigh's impression was that the theoretical problem lay in taking the 
inviscid limit (``it is possible that the investigation in which viscosity
is altogether ignored is inapplicable to the limiting case
of a viscous fluid when the viscosity is supposed infinitely small'').
Indeed, as it is now known, the viscous version of the
linear stability theory yields the celebrated Orr--Sommerfeld equation,
rather than (\ref{f.3}), which can have unstable modal solutions even for
flow profiles that are stable in inviscid theory \cite{landahl86}.

\subsection{Plasma oscillations and Landau damping}
\label{ssec:plasmalandau}

In the middle of the 20th century   the analog  of the disturbing singularity discussed 
by  Rayleigh and Kelvin's arose in plasma physics in the context of linear analysis 
of the Vlasov equation; the singularity now appears
at the wave-particle resonance where the particle velocity matches
the speed of a neutral plasma wave (oscillation).
To observe this directly, let $f=\hat f(v) \exe^{\imi k(x-ct)}+\compconj$,
and then linearize equations (\ref{e.1}) and (\ref{e.2}) to obtain
\begin{equation}
 f = -{\varphi F' \over v-c},
\qquad
\varphi = -{1\over k^2}  \int_{\R} \! dv\,  f \,,
\label{e.3}
\end{equation}
which clearly presents the singularity at the
wave-particle resonance, $v=c$ for $c=c_r$ real.
Unlike for Rayleigh's equation the Vlasov problem has a spatial kernel and, thus, 
we can proceed much further analytically and derive
a dispersion relation for the wavespeed $c$: eliminating $\hat f(v)$
yields
\begin{equation}
D(c,k) = 1 - {1\over k^2} \int_{\R} \!\ \!dv \,{F'(v) \over v-c} = 0
\label{e.4}
\end{equation}
(cf.\ Eq.~(\ref{nx.1})).
Complex conjugate solutions for $c$ can be straightforwardly found
from this dispersion relation for certain equilibrium distribution
functions, $F(v)$; again, the mode with $c_i>0$ is unstable.
Even without specific choices of $F(v)$, Nyquist methods can be brought
to bear on the dispersion relation (\ref{e.4}) to construct explicit
stability criteria \cite{penrose}.

The singularity of plasma oscillations is seen in a different way in
(\ref{e.4}): this dispersion relation
contains a singular integral with a branch cut along the
real axis of the complex $c$ plane, which again identifies 
the continuous spectrum (for rigorous spectral theory see 
\citealp{degond}, and \citealp{hagstrom}). 
On using the Plemelj relation \cite{gakhov} 
with $c\rightarrow c_* + \imi 0^\pm$, we find that
\begin{equation}
F'(c_*)=0 \qquad {\rm and} \qquad
1 - {1\over k^2} \pvint_{\R}\! dv\,  {F'(v) \over v-c_*} = 0 \, .
\label{e.10}
\end{equation}
In other words, as the equilibrium profile $F(v)$ is adjusted so that
the complex conjugate pairs move towards the real axis of the
complex $c$-plane, they limit to a special neutral mode
with $c=c_*$. By virtue of the first 
condition in (\ref{e.10}), the special neutral mode is
a smooth solution (the singularity
of $1/(v-c_*)$ is cancelled by the zero of $F'(v)$ in the numerator
of the eigenfunction), even though it is embedded in 
the continuous spectrum. The embedded mode can be interpreted to 
lie at a location within the continuous spectrum where the modal
``signature''  changes sign, consistent with a generalization of Kre\u{i}n's  
theorem \cite{morrison00a,hagstrom} of finite-dimensional Hamiltonian 
dynamics \cite{krein}. 

\begin{figure}[t]
\centering
\includegraphics[width=15 cm]{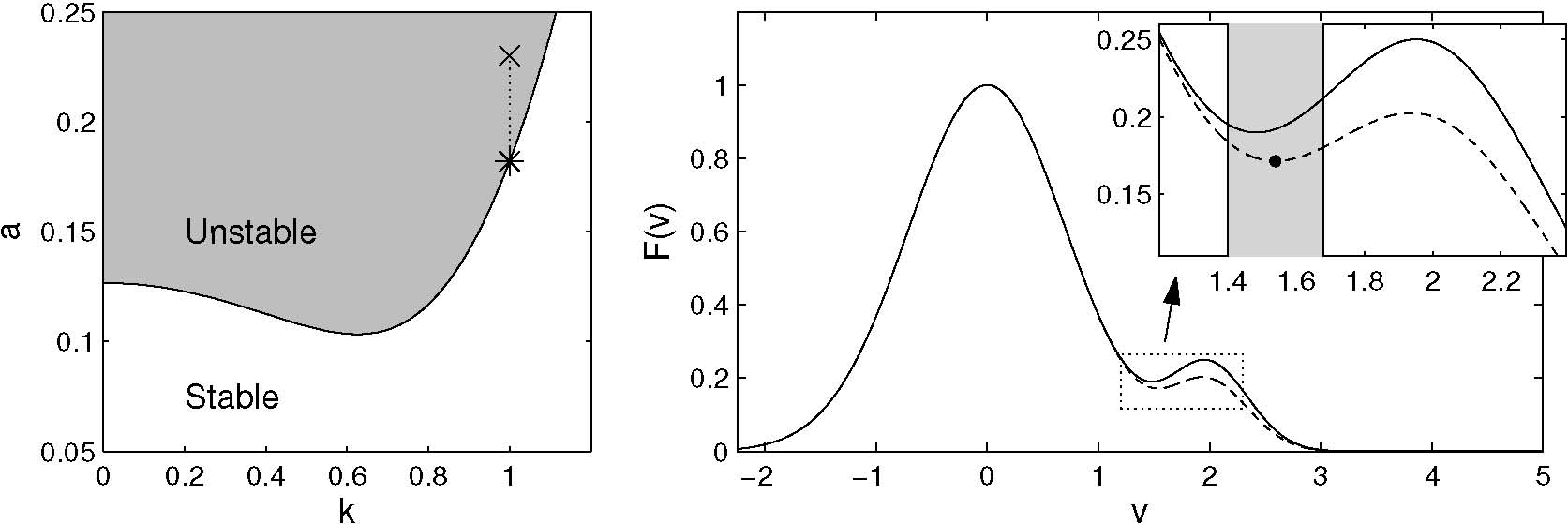}
\caption{
(a) Stability boundary on the $(k,a)$ plane for parameters values of 
(\ref{e.15}),  $\alpha=4$ and $v_0=2$.
The star and cross mark the values of $a$ for which the profile
is plotted in (b). The star denotes the marginally stable profile
for $k=1$ ($a\approx0.182$); the cross marks an unstable profile
with a slightly larger bump ($a=0.23$).
The inset in (b) shows a magnification of the bump-on-tail
region, with the shaded area illustrating the critical region
surrounding the minimum of the marginally stable profile over which
the distribution function is wound up by the unstable mode. This region is 
narrower than the velocity spread of the bump.
}
\label{stab}
\end{figure}

For a family of equilibria indexed by a control parameter $a$,
the embedded mode determines the stability boundary on the $(k,a)-$plane
because it is the limit of an unstable complex eigensolution.
By way of an example, consider the family of equilibria given by
\begin{equation}
F(v) = \exe^{-v^2} + a \,\exe^{\alpha(v-v_0)^2} \,,
\label{e.15}
\end{equation}
where $\alpha$ and $v_0$ are additional parameters.
This parameterized family consists of a primary Maxwellian
with a superposed ``bump-on-tail''. Via Penrose's criterion
one can establish that there are no distinguished neutral modes
associated with the main peak of the distribution near $v=0$;
instability can only arise when the bump on the tail generates further extrema.
If the bump amplitude, $a$, is the main control parameter,
the stability boundary is denoted by the curves $a=a(k)$, as
illustrated in Fig.~\ref{stab}. The minimum of this boundary
($a\approx0.103$ for $k\approx0.62$, $\alpha=4$ and $v_0=2$) occurs
when an inflection point first appears in $F(v)$ in the vicinity
of the bump, and thereafter splits
into two extrema with a further increase of $a$. 
However, if the domain is periodic in $x$, as considered here,
there is a minimum value of $k$ determined by the longest spatial wavelength,
and so the transition to instability can take place elsewhere
on the stability boundary, as illustrated in Fig.~\ref{stab}.

For stable equilibria below the stability boundary, the preceding arguments
rule out all regular eigenmodes, and in particular neutral plasma oscillations.
This physically unsatisfying conclusion led  \textcite{Landau1946} 
to attack the linear initial-value problem 
using Laplace transforms in order to uncover unambiguously how waves 
evolved once
introduced into the plasma.  A direct consequence of Landau's solution
was the prediction of the phenomenon that is now called {\sl Landau damping}:
the exponential decay of the electric field (or potential
$\varphi(x,t)$) that arises due to the phase mixing of particles.

The exponentially decaying contribution to the electic field
is often refered to as a {\sl quasi-mode} (or {\sl Landau pole}),
as it is not a true normal mode of the linear eigenproblem.
To uncover these objects, and the general
long-time behavior of the solution, one inverts the 
Laplace transform solution, which takes the form of a complex integral
over the usual Bromwich contour ({\it e.g.} \citealp{krall}).
The Landau poles can be picked up as the leading-order long-time
contributions by deforming the integration contour through the
branch cut of the continuous spectrum. One thereby emerges on a different
Riemann sheet of the non-analytical dispersion function $D(c,k)$
(which appears in the denominator of the Laplace transform solution), 
allowing one to encircle and pick up residue contributions from
singularities due to zeros of the dispersion relation on a non-physical 
Riemann sheet. These ``fake eigenvalues'' are the quasi-modes.

Despite Landau's reputation as one of the eminent theoretical physicists of his day, 
 the relatively mathematical nature of Landau's predictions
was not convincing to many plasma physicists.
Several questions arose. For example, 
Landau damping required that the energy contained in the electric field
must be removed, which appeared to some to violate energy conservation
(but see \citealp{MP92,morrison00a,morrison94} where it is shown explicitly how energy is conserved and energy is transferred).
Others argued that the damping effect was a mathematical artifact
arising from inconsistencies in the formulation  \cite{All59,All63,Eck63,weitzner63a}. On the other hand,  in three celebrated papers \textcite{Boh49a,Boh49b,Boh50}  argued for the existence of Landau damping
outside of Vlasov theory.   The issue was not  settled until experimental verification of Landau damping was reported by
 \textcite{Mal64} (see also \citealp{Mal66}), nearly twenty years after Landau's original paper. These authors demonstrated experimentally that  the damping vanished when the resonant particles were removed.  They did this by removing the tail of the distribution for electrons streaming in one direction but not the other, and observed Landau damping only for the waves propagating in the direction with resonant particles. 

From the mathematical perspective the story of Landau damping has continued.  Steps toward an early rigorous justification of Landau damping by \textcite{Bac60}, who considered the linear limit, were recently extended to the nonlinear regime by  \textcite{cedric}.  Their proof requires a condition on the initial amplitude in order for the damping to behave essentially like the linear theory.  However, as we will see later, the linear results fail for finite amplitude disturbances  (see Fig.~\ref{quasimodo} of 
Sec.~\ref{ssec:stablepatterns}).  The nonlinear arrest of linear Landau damping remains an open problem from a formal mathematical
perspective.

\subsection{Violent relaxation, XY model,  and phase transitions}
\label{ssec:xylinear}

Closely connected to Vlasov-Poisson dynamics is the relaxation of stellar
systems, which can be described by the Jeans equation \cite{binney}, 
which is identical to the Vlasov-Poisson system but for an attracting
interaction (the sign of $\mathcal{K}$ is switched) 
and the removal of the charge neutrality condition.    
In a famous paper \textcite{lyndenbell} prompted a discussion of this problem
by arguing that such relaxation proceeded in two phases: in the first,
the ensemble of stars behaves like a continuum, relaxing rapidly from
an arbitrary initial condition towards some quasi-steady state (the ``violent
relaxation''). That state
consisted of a structure, or pattern, in phase space characterized by
increasingly wound-up filaments. Eventually, the ever decreasing scales
precipitate the failure of the continuum approximation, and herald the
onset of a second phase in which the discreteness of the system plays an
essential role.

More recently, Lynden Bell's violent relaxation was revisited in the
context of n-body Hamiltonian dynamics 
(e.g.\  \citealp{campa}) with the XY interaction. Like
gravity, this interaction is attractive, and the stellar and XY
models has many common features. In addition to
the consideration of far-from-equilibrium initial conditions,
recent explorations with the XY model also studied
equilibrium distributions with low-amplitude perturbations.
In particular, stability theories were presented to identify
equilibria with linear instabilities 
\cite{inagaki,antoni95,campa,yamaguchi04}, 
and numerical simulations explored the resulting phase-space patterns.
Specific attention was focussed on 
the nature of the phase transition occuring when the linear
instability appeared, and whether statistical techniques
could be exploited to predict integral measures of 
the emerging patterns (such as the magnetization;
see \textcite{yamaguchi04,campa,antoniazzi07,buyl09}).
Unsurprisingly, the phenomenon of Landau damping
was also rediscovered \cite{campa}.

The search for instabilities in the XY model closely follows
the Vlasov-Poisson problem: after introducing the normal-mode form,
$f(\qq,\pp,t)=\hat f(\pp) \exe^{\imi (\qq-ct)}+\compconj$,
into (\ref{hmf.1}) and (\ref{hmf.2}) and linearizing,
we arrive at the explicit dispersion relation,
\beq
D(c) = 1 + \int_{\R} \! dI\,  \frac{R'(I)}{I-c} = 0\,, 
\label{hmf.3}
\eeq
which is the current version of (\ref{nx.1}) since this problem has a spatial 
kernel.  All  Fourier modes 
with $k\ne \pm1$ are stable.

Two simple examples for the equilibrium distribution, $R(I)$,
that allow further analytical progress
are provided by the ``top-hat'' profile,
\beq
R(I) = \frac{a}{2}\left[\Theta(I+1) - \Theta(I-1)\right],
\label{hmfx.1}
\eeq
where $\Theta(x)$ represents a Heaviside step function,
and the Gaussian,
\beq
R(I) = \frac{a}{\sqrt{2\pi}} e^{\frac{-I^2}{2}}
\label{hmf.4}
\eeq
(cf.\  \citealp{yamaguchi04,campa}).
For the top-hat profile, $R'$ consists of a pair of delta-functions
and so the integral in (\ref{hmf.3}) is performed immediately
to furnish
\beq
D(c) = 1 - \frac{a}{1-c^2} = 0 \qquad
{\rm or} \qquad
c = \pm \sqrt{\frac{1-a}{a}}.
\eeq
Thus, when $a>1$, there is an unstable mode. This example
is analogous to the  ``waterbag'' model of  the plasma problem 
(see e.g.\  \citealp{berk}). Moreover, 
both are equivalent to Rayleigh's broken-line profiles.
For the Gaussian, the dispersion relation  (\ref{hmf.3}) is closely 
analogous to the plasma dispersion
relation with a Maxwellian equilibrium distribution function \cite{fried}.
The neutral stability condition can be computed on observing that 
marginally stability occurs for $c=0$, in which case $a=a_c=1$.
Moreover, once again, for $a>1$, there is an unstable mode.

Note that the instability of
both distributions in (\ref{hmf.1}) and (\ref{hmf.4}) 
for $a>1$ is quite different from
Vlasov-Poisson theory, for which both equilibrium
profiles would be stable.
The key difference between the dispersion functions
in (\ref{e.4}) and (\ref{hmf.3}) is the reversal of the sign of the integral,  
which arises because the XY interaction is attractive. 
The sign difference permits the distributions 
to become unstable in the XY model, 
a situation that is forbidden in Vlasov-Poisson theory
because of the positivity of the distribution function. 
Likewise, the gravitational attraction of the stellar dynamics
problem also reverses the sign of the integral in (\ref{e.4});
the instability of a singly peaked distribution that results
is a form of the classical Jeans instability, as explored by 
\textcite{fujiwara}.

\section{Nonlinear dynamics: scaling, saturation and critical layers}
\label{sec:nonlinear}

Before proceeding to our derivation of the single-wave in 
Sec.~\ref{sec:derivation}, we here motivate the necessity and features 
of the derivation.  In particular, we show why conventional weakly 
nonlinear analysis fails, and review both the plasma physics and fluid mechanics literature.

\subsection{Conventional weakly nonlinear analysis}
\label{ssec:weakanalysis}

The power of bifurcation theory lies in its ability to construct nonlinear 
states analytically.  Conventionally,
the center manifold reduction (e.g.~\citealp{crawford91a})
is often the method of choice for dissipative systems, 
whereas Birkhoff normal form
theory is applied to Hamiltonian systems (e.g., \citealp{meer,bryuno}).

\subsubsection{Generalities}
\label{ssec:generalities}

Conventional weakly nonlinear theories aim for an amplitude 
equation that captures the dynamics of the normal mode that bifurcates
to instability at a critical threshold in a control parameter.
The derivation of the equation can be cast in the form of a regular
perturbation expansion,  in which one opens with a neutrally 
stable normal mode balanced on the 
stability boundary, and then kicks that mode into an unstable action
by suitably adjusting the control.
The relevant amplitude equation is dictated by the symmetries of the
system in question; for the problems of interest here,
the equation is a certain Hamiltonian normal form.
This bifurcation is characterized by  asymptotic scalings for the
mode amplitude, the perturbation to the control parameter, and a
redefinition of time.

More specifically, and in the general scheme of the full model 
given at the end of Sec.~\ref{ssec:badass},
the idea is that the equilibrium profile, $R(I)$, is one of a family
parameterized by a constant parameter, $a$. 
After a suitable choice of domain size and scaling, 
we focus on the situation in which the normal mode with
a wavenumber $k=1$ is unstable if $a>a_0$, and all the other normal
modes are stable. By switching the angular frame of reference
we can zero the frequency of that mode
(i.e.\  by transforming coordinates to a frame rotating at the
mode's frequency). We then set $a=a_0+\varepsilon^2a_2$, where
$\varepsilon\ll1$ is the small parameter that we use to organize
the perturbation expansion.
The modification to $a$ embodied in $\varepsilon^2 a_2$
creates a change in the equilibrium profile 
that destabilizes the otherwise neutral mode. Correspondingly,
the relatively slow growth of the mode is captured by scaling time.  Thus, 
we set
\beq
T = \varepsilon t, \quad
\frac{\partial}{\partial t} \to \varepsilon
\frac{\partial}{\partial T} ,
\quad \rho(\qq,\pp,t)\to\rho(\qq,\pp,T) \quad {\rm etc.}
\eeq
The remaining asymptotic scalings are encoded in the following sequences:
\begin{align}
       R(\pp)&= R_0(\pp) + \varepsilon^2 R_2(\pp) + \ldots\,,
		\label{eq:wne1}\\
	\rho &= \varepsilon\rho_1(\qq,\pp,T) 
 + \varepsilon^2\rho_2(\qq,\pp,T)
  + \ldots\,,
		\label{eq:wne2}\\
\Phi={\mathcal{K}}\rho &= \varepsilon\psi_1(\qq,\pp,T) +
		 \varepsilon^2\psi_2(\qq,\pp,T)
		 + \ldots\,,
		\label{eq:wne3}
\end{align}
where $\psi_m = \mathcal{K} \rho_m$ for~$m=1,2,\dots$.
The expansion of $\rho$ opens at order $\varepsilon$,
even though the equilibrium is modified only at $O(\varepsilon^2)$,
which corresponds to the ``Hopf scaling'' in the terminology of
Crawford (see, e.g., \citealp{crawford95a}).

If one substitutes the sequences above into the system 
(\ref{Phi}) and (\ref{eq:eom}), and  gathers terms of equal order
in $\varepsilon$, one arrives at a hierarchy of equations to solve
sequentially for $(\rho_m,\psi_m)$.
The leading-order equations determining $(\rho_1,\psi_1)$ turn out
to be the relations satisfied by the distinguished neutral mode.
That solution has an undetermined amplitude, $A(T)$. To fix this
quantity, we proceed to the next orders. At $O(\varepsilon^2)$
the pair $(\rho_2,\psi_2)$ capture the nonlinear generation of the first
harmonic of the neutral mode (with wavenumber $k=2$)
and a correction to the angular average (with wavenumber $k=0$), but
$A(T)$ yet remains unknown. Finally, at $O(\varepsilon^3)$, 
the requirement that the pair $(\rho_3,\psi_3)$ be bounded and
periodic in $\theta$ demands that $A(T)$ satisfy a solvability
condition, which amounts to the desired amplitude equation.
The construction is quite standard, although for the current
problems it contains a key flaw. We bring out the nature of this flaw  and 
illustrate the expansion in a slightly simpler setting in 
Sec.~\ref{sssec:spatialkernels}.
 
\subsubsection{Weakly nonlinear expansion for spatial kernels}
\label{sssec:spatialkernels}

By way of illustration,
we perform a weakly nonlinear analysis of
the special case of our general model for spatial kernels for which
the kernel is independent of the actions, $I$ and $I'$. As remarked earlier, 
explicit dispersion relations are available for such kernels,
and both the Vlasov and XY models furnish  examples.

For further brevity, we consider the specific situation in which 
the domain is symmetrical in $I$, the equilibrium profile $R(I)$
is an even function, and $\Omega(I):=h'(I)$ is an odd function.
One can then explore the situation in which an 
instability arises due to the appearance of a (zero-frequency)
normal mode with the form of a standing wave, eliminating
any need to transform into the rotating frame of the
mode.  After the rescaling of time,
$T = \varepsilon t$, the governing equations become
\beq
\varepsilon \rho_T + \Omega \rho_\theta - \Phi_\theta R_I 
- \Phi_\theta \rho_I = 0
\label{B.1}
\eeq
\beq
\Phi = \int_0^{2\pi}\!\! \dint\theta' \int_\cD \!\!\dint I'\, K(\theta-\theta')\rho(\theta',I',T)\,,
\label{B.2}
\eeq
where recall $\cD$ denotes the range of $I$.
It is also convenient to take the angular average,
\beq
\overline{f}\ldef \frac1{2\pi}\int_0^{2\pi}\!\!\dint \theta\, f \,,
\eeq
 of the first of
these equations:
\beq
\ve \overline{\rho}_T = (\overline{\Phi \rho_\theta})_I.
\label{B.3}
\eeq

Next, we pose the asymptotic sequences,
\bqy
R &=& R_0 + \ve^2 R_2 \,, \nonumber\\
\rho &=& \ve \rho_1 + \ve^2 \rho_2 + \cdots\,, \nonumber\\
\Phi &=& \ve \Phi_1 + \ve^2 \Phi_2 + \cdots\,, 
\label{B.4}
\eqy
(cf.\ (\ref{eq:wne1})--(\ref{eq:wne3})).
Introducing these into (\ref{B.1}) leads, at order $\ve$ to
\beq
\rho_1 = \frac{R_{0I} \Phi_1}{\Omega},
\qquad
\Phi_1=A(T) \exe^{\ii \theta} + c.c.,
\label{B.5}
\eeq
after adopting the marginally unstable mode (with wavenumber $k=1$)
as the leading-order solution.
The relation in (\ref{B.2}) then gives the marginal stability
condition,
\beq
1 - \hat K_1 \int_\cD\!\!\dint I\,  
\frac{R_{0I}}{\Omega} = 0
\eeq
(cf.\  the dispersion relation (\ref{nx.1}) with $c=0$).

At order $\ve^2$, we find
\beq
\Ome\rho_{2\theta} = R_{0I} \Phi_{2\theta}
- \rho_{1T} - \Phi_{1\theta} \rho_{1I}
.
\label{B.6}
\eeq
Hence, on introducing (\ref{B.5}),
\bqy
\rho_{2} &=& \overline{\rho_2}+\frac{R_{0I} }{\Ome} \Phi_{2}
+ \frac{R_{0I}}{\Ome^2} \left(iA_T\exe^{\ii\theta}+c.c.\right)
\nonumber\\
&-& 
\frac{1}{2\Ome}\left(\frac{R_{0I}}{\Ome}\right)_{\!I}
\left(A^2\exe^{2\ii\theta}+c.c.\right)\, 
.
\label{B.7}
\eqy
Substitution of (\ref{B.7})  into (\ref{B.2}) yields 
\beq
\Phi_2 = 
\frac{\hat K_1\hat K_2 \cI
}{2(\hat K_1-\hat K_2)}
\left(A^2\exe^{2\ii\theta}+c.c.\right)
, 
\eeq
where
\beq
\cI
= \int_\cD\!\!\dint I\,\,  \frac{1}{\Ome} \left(\frac{R_{0I}}{\Ome}\right)_{\!I}\,.
\eeq
Here we have used  $\hat K_0=0$ and that 
the symmetry of $R_0(I)$ and $\Omega(I)$ implies
$\int_\cD\!\dint I \,  R_0' /\Ome^2=0$. In principle, we should also
add a second term to this solution with the form,
$A_2(T)\exe^{\ii\theta}+c.c$, which
denotes a correction to the $k=1$ neutral mode, but
is not needed to determine the amplitude equation for $A$.


At $O(\ve^2)$ we also arrive at
the first non-trivial result from
(\ref{B.3}):
\beq
 \overline{\rho_2}_T = 
(\overline{\Phi_{1\theta} \rho_2 +\Phi_{2\theta}\rho_1})_I
=  \left(\frac{R_{0I}}{\Ome^2}\right)_I (|A|^2)_T
.
\label{B.10}
\eeq
To integrate this relation, we must specify an initial condition.
If $A(0)=A_0$ and $\overline{f_2}(I,0)=0$, corresponding to initializing
the system with the equilibrium plus a low-amplitude normal mode, then
\beq
\overline{\rho_2} = \left(\frac{R_{0I}}{\Ome^2}\right)_I (|A|^2-|A_0|^2),
\eeq

Finally, we arrive at the $O(\ve^3)$ equation,
\beq
\rho_{3\theta} = \frac{1}{\Ome}
( R_{0I} \Phi_{3\theta} + R_{2I} \Phi_{1\theta}  
- \rho_{2T} + \Phi_{2\theta} \rho_{1I} + \Phi_{1\theta} \rho_{2I})
.
\label{B.11}
\eeq
Isolating the contribution of the right-hand side to the
first Fourier mode, $\exe^{\ii \theta}$, and then applying
the second relation in (\ref{B.1}) leads to the amplitude equation,
\beq
A_{TT} \int_\cD\!\!\dint I \,  \frac{R_{0I} }{\Ome^3} 
=  J A  + \Gamma |A|^2A ,
\label{Bx.1}
\eeq
with
\beq
J = \int_\cD\!\!\dint I\,   \frac{R_{2I}}{\Ome} +  |A_0|^2
\int_\cD\!\!  \dint I  \,  \frac{1}{\Omega} \left(\frac{R_{0I}}{\Omega^2}\right)_{II}
\eeq
and
\beq
\Gamma = \frac{\hat K_1\hat K_2\, \cI^2}{2(\hat K_1-\hat K_2)}
+ \int_\cD \!\!\dint I \, \frac{1}{\Ome} \left[\left(\frac{R_{0I}}{\Omega^2}\right)_{I}
-\frac{1}{2\Ome}\left(\frac{R_{0I}}{\Ome}\right)_I\right]_I
\, .
\eeq
Note that the integrals that appear in the coefficients of
the amplitude equation do not exist unless a number
of the derivatives of $R_0(I)$ and $R_2(I)$ vanish 
at $I=0$; this is reflective of critical-level
singularities in the weakly nonlinear expansion. In fact, as shown by
\textcite{crawford99a}, the singularities appear at all orders of the expansion,
becoming progressively more severe as one descends through the asymptotic
hierarchy. The only instance in which one can justify the asymptotics
is if all derivatives of $R_0$ and $R_2$ vanish at the wave-particle
resonance $I=0$, which places severe limitations on the equilibrium profile.
Notable examples of such situations, however, 
include Rayleigh's broken-line shear-flow
equilibria, the waterbag models of plasma physics, and the top-hat
profile of the XY model.

\subsubsection{Top-hat profile of  the XY model}
\label{sssec:tophat}

For the XY model with the top-hat equilibrium profile, $\Ome:=I$, $\cD:=\R$, 
$\hat K_1=-1$, $\hat K_2=0$ and
\beq
R_{jI} = \frac{a_j}{2}[\delta(I+1) - \delta(I-1)],
\eeq
with $j=0$ and 2. If we also assume that $|A_0|\ll1$,  the integrals
can be evaluated to furnish
\beq
A_{TT} = a_2 A + \half |A|^2 A . 
\label{hmfx.2}
\eeq
Note that the sign of the
second term ensures that nonlinearity is destabilizing 
in this example. That is, the bifurcation
is {\sl subcritical}, and an
unstable steady solution branch given by $|A|=\sqrt{-2a_2}$
exists for $a<a_0=1$ (where the unpatterned equilibrium state is stable).
A similar type of bifurcation arises for broken-line shear-flow profiles
in the vorticity defect model of \cite{balmforth12} (where 
$\hat K_k:=(2k)^{-1}$).

\subsection{Trapping scaling versus Hopf scaling in the Vlasov problem}
\label{ssec:traphopf}

In the 1960s and 70s, several attempts were
made to continue the analysis of linear electrostatic instability
into the nonlinear regime.  Some of the key
results were obtained by \textcite{oneil71a,onishchenko71a,frieman63a,rosenbluth76a}.
One of the striking features of these articles is a disagreement regarding the level
of saturation of the instability: the standard way to
estimate the saturation level is in terms of
the amplitude of the electric field disturbance, $A_{\rm sat}$,
relative to the distance to the stability
boundary, as measured by a suitable control parameter, $\epsilon$
(such as $\epsilon=a-a_c$, where $a_c$ denotes a point on the stability
boundary, for the example of Sec. \ref{ssec:plasmalandau}). Two characteristic
scalings were hypothesized: ``trapping scaling''  
\cite{oneil71a,onishchenko71a}, for  which $A_{\rm sat}\sim\epsilon^2$,
and Hopf scaling \cite{frieman63a,rosenbluth76a} with
$A_{\rm sat}\sim\epsilon^{1/2}$, which is the usual saturation level for
strongly dissipative instabilities and equivalent to that used
above in Sec. \ref{ssec:weakanalysis}.

Trapping scaling was the essential  ingredient in the ``single-wave model''
of \textcite{drummond70a,oneil71a,onishchenko71a}, which
was a phenomenologically based theory written down partly
using physical arguments. \textcite{frieman63a,rosenbluth76a}, 
on the other hand, championed the Hopf scaling,
and attempted amplitude expansions of the kind summarized above.
Unfortunately, such expansions are plagued by 
critical-level singularities which unavoidably
haunt the higher-order equations of the expansion, 
as noted in Sec.~\ref{ssec:weakanalysis}. To avoid this difficulty
\textcite{frieman63a,rosenbluth76a} removed the singularities by
displacing the poles from the real axis in a regularization procedure
that was meant to be equivalent to Landau's solution of the initial-value
computation. But the procedure has questionable mathematical validity
(strictly speaking, the asymptotics introduces multiple timescales
to account for a relatively fast adjustment of the system to
the slower growth of the unstable mode; using Landau damping ideas
to remove the singularities in the expansion obscures the
separation of timescales).

\begin{figure}
\begin{center}
\leavevmode
\includegraphics[width=9cm]{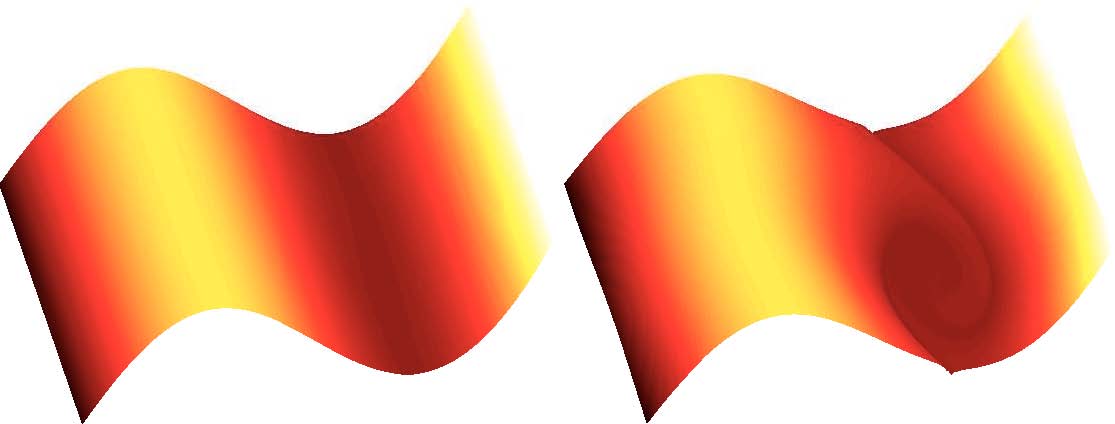}
\includegraphics[width=9 cm]{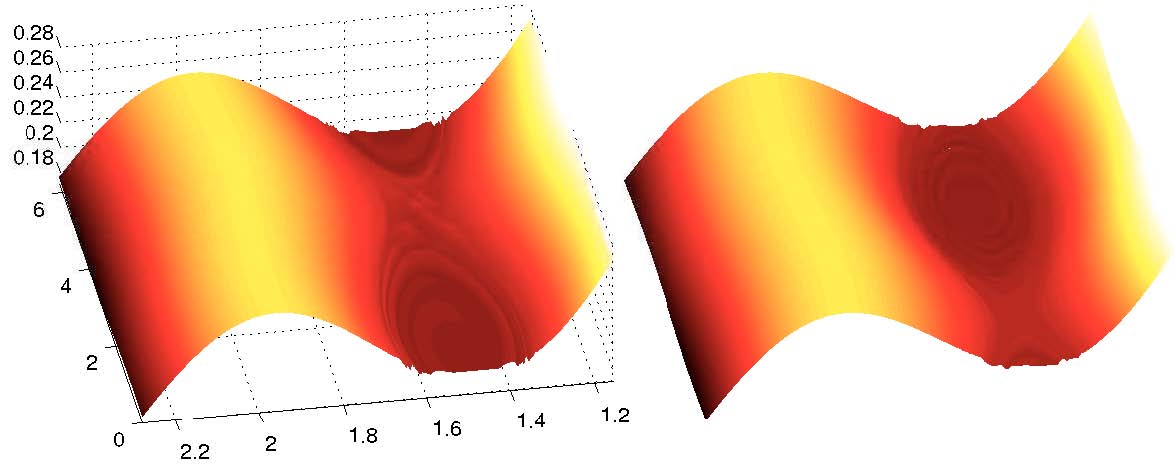}
\includegraphics[width=9 cm]{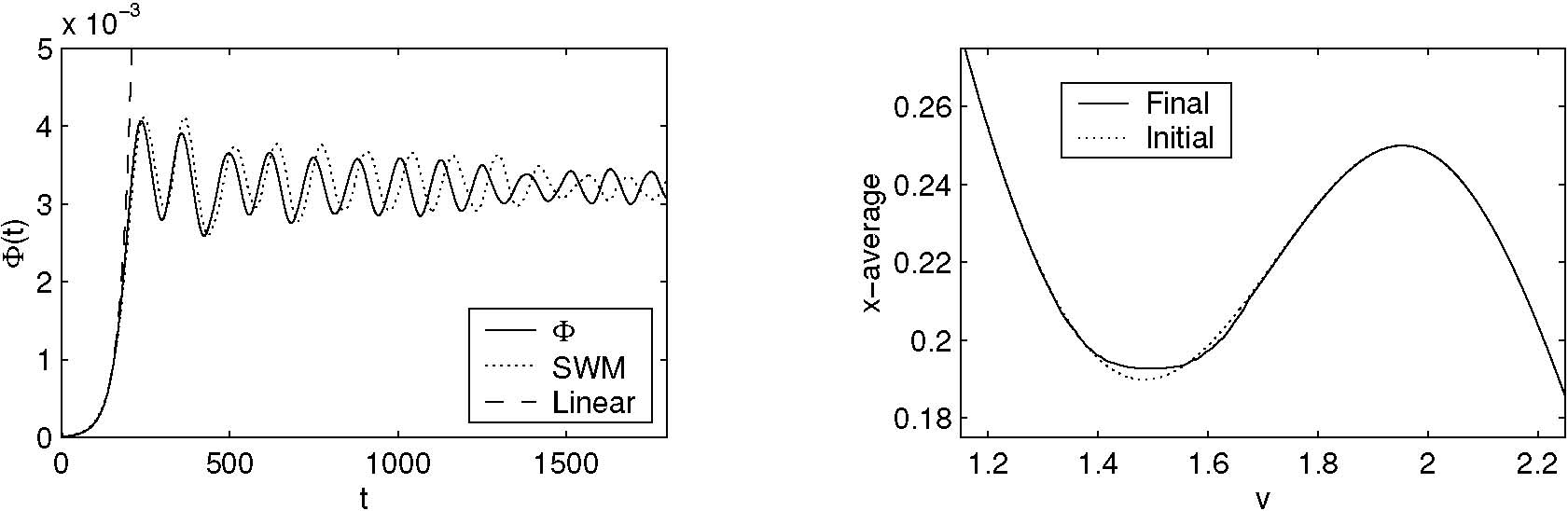}
\end{center}
\caption{
Vlasov simulation. $a=0.23$ and $k=1$ ($L=2\pi$).
The first four pictures (left to right, top to bottom) show snapshots of 
the total distribution function,
$F(v)+f(x,v,t)$ as surfaces above the $(x,v)$-plane
at the times $t=140, 200, 400,$ and $1800$.
The penultimate picture shows the amplitude of
$\Phi(t)  :=  (\int_0^L\!\dint x \, \varphi^2  / L)^{1/2}$;
the dashed line shows the expected exponential growth of
the unstable mode and
the dotted line indicates the 
prediction of the single-wave model. The final panel
shows the spatial average of $F+f$ at the beginning and end of the
computation.
} 
\label{fig:vlasov}
\end{figure}

\begin{figure}[t]
\begin{minipage}[c]{0.5\textwidth}
\centering
\includegraphics[width=7 cm]{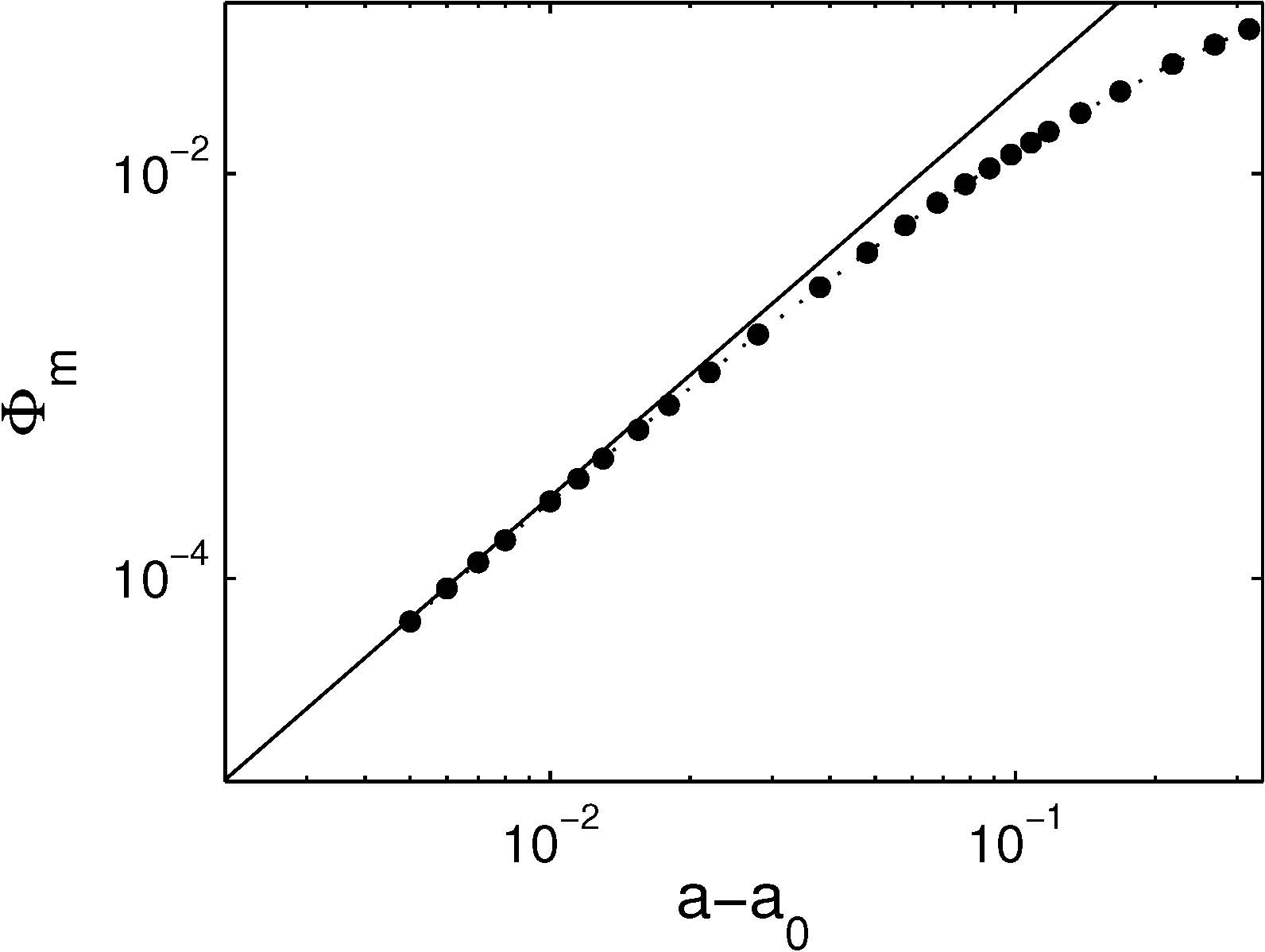}
\end{minipage}%
\begin{minipage}[c]{0.4\textwidth}
\caption{Numerical data for
the saturation measure $\Phi_m$ (the first maximum in 
the amplitude of the potential, $\Phi(t)$), plotted
against the distance to the stability boundary, $a-a_0$.
The prediction of the single-wave
model  (based on trapping scaling) is shown by the solid line.
}
\label{sca}
\end{minipage}%
\end{figure}

Because neither approach was completely convincing, the
effort to distinguish between them fell to numerical
simulation and laboratory experiments, which suggested that trapping
scaling dictates the saturation level (e.g.\ \citealp{denavit81a}).
Numerical computions of the bump-on-tail instability are
presented in Figs.~\ref{fig:vlasov} and \ref{sca}. Figure \ref{fig:vlasov}
gives an illustration of the nonlinear dynamics occuring during
the saturation of an unstable mode. Figure \ref{sca} collects together results 
from a suite of computations that measure the scaling of the saturation
level, and confirms trapping scaling.

Relatively recently, two articles have more convincingly demonstrated that
the correct scaling is trapping \cite{crawford95a,delcastillo98a,delcastillo98b}.
Crawford's approach follows the center-manifold route and attempts to
derive an amplitude equation for the unstable mode. The
approach begins with the Hopf scaling and runs into the
technical complications associated with integrals that diverge at the
wave-particle resonance as one approaches the point of
neutral stability. Rather than
treating the singularities with some {\it ad hoc} rule, Crawford
rescaled the mode amplitude with sufficient powers of $\epsilon$ in
order to counter the divergent integrals.  This procedure rescales the
mode amplitude back to the trapping scaling and avoids singularity,
which formally establishes that trapping is the correct scaling.  
Unfortunately, Crawford's amplitude equation
contains an infinite number of terms of comparable size and cannot be
used to explore how the instability saturates.

Del Castillo-Negrete opts for a different approach by transferring results 
from 
critical-layer theory of fluid mechanics \cite{churilov87a} that exploit 
matched asymptotic expansions to heal the singularity.  The scaling required 
in the matched asymptotics is equivalent to trapping scaling, and 
del Castillo-Negrete thereby derived the single-wave model systematically.   
This is also the tack we take below, but first we highlight the
development of critical layer theory in fluid mechanics.

\subsection{Critical layer theory}

The 1960s also saw much development of weakly nonlinear theories in fluid
mechanics following advances in describing the onset of fluid convection
(e.g.\  \citealp{malkus58a}). The techniques were quickly adapted
to viscous shear flow problem by \textcite{stuart60a,watson60a},
but they could not be adapted for
the corresponding inviscid problem because of the recurring critical-level 
singularities. Some progress was made by assuming that
significant viscosity wiped out the singularity in a slender region
surrounding the critical level \cite{schade64a,huerre87a,%
churilov87b}, an idea dating back to \textcite{lin45a}.
These insights gave some physical support to the idea
that the poles in the asymptotic expansion could be healed
by viscosity. However, viscous effects also inexorably
diffuse away from the critical levels, widening the critical
layer as time proceeds \cite{brown78a}. Overall, this so-called
viscous critical-layer theory glosses over the critical-layer 
dynamics and is not a suitable closer to the inviscid limit, 
the situation of relevance to this review.

The appreciation that the dynamics could be richer led several
authors to develop more complicated ``critical-layer theories''.
These theories all proceed by resolving the dynamics of the critical
level in detail whilst simultaneously 
including various physical effects to relieve 
any divergences. For example, \textcite{benney69a} resolved 
the critical-level singularity by adding nonlinear effects and
thereby built steady nonlinear waves (some criticism
of their detailed construction was given later by \citealp{haberman73a} 
and \citealp{brown78a}).  Other theories were  those of  
\textcite{stewartson78a} and \textcite{warn78a}
for forced critical levels that incorporate
unsteadiness and nonlinearity. In both examples, a key detail is that
nonlinearity first becomes important where the solution is nearly
singular, which is over the critical layer,
the narrow region surrounding the critical level.

\begin{figure}[h]
\begin{center}
\leavevmode
\includegraphics[height=10 cm]{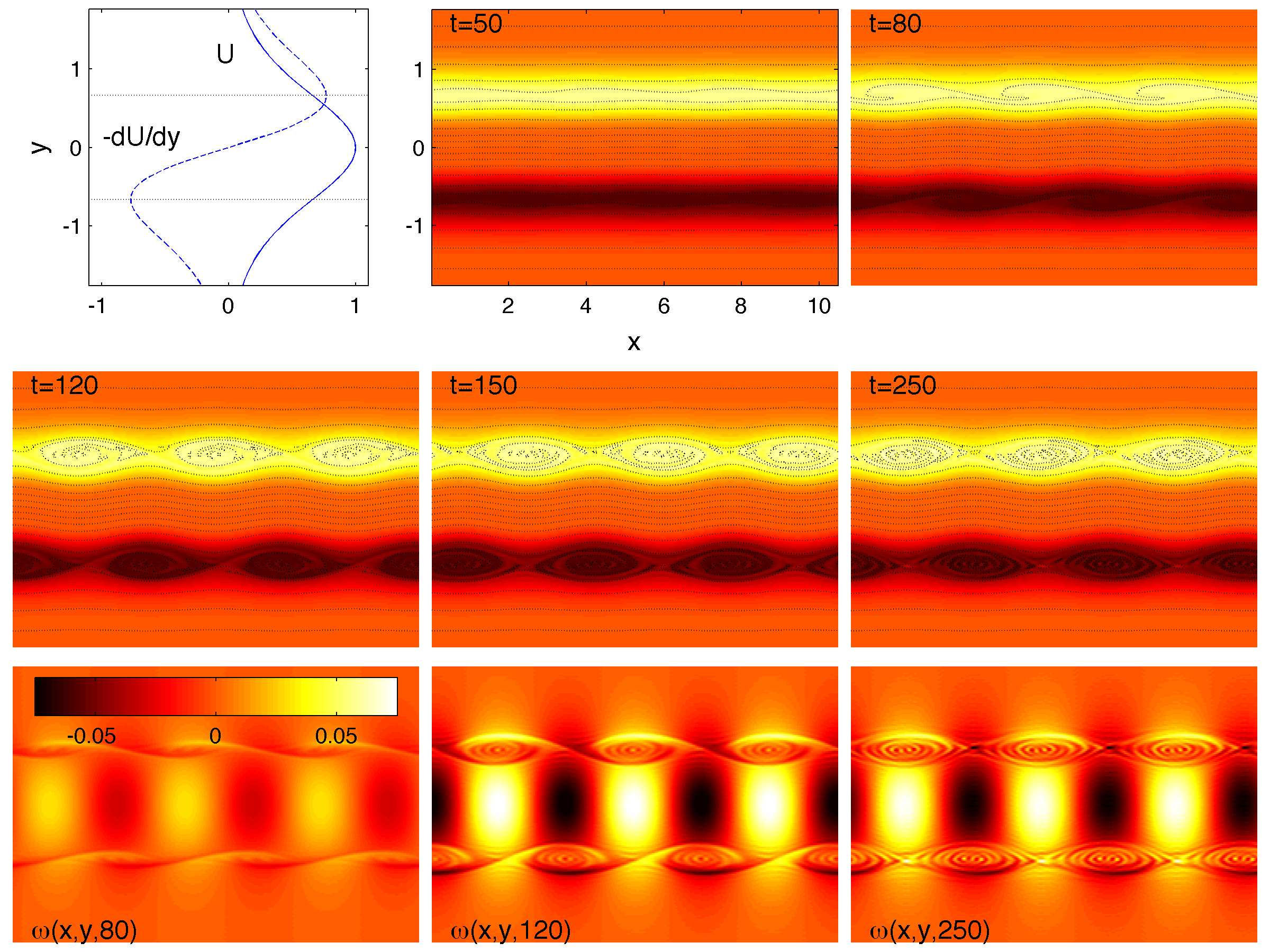}
\end{center}
\caption{
Computed cat's eye patterns formed from the growth of an unstable mode
in a simulation of the Bickley jet with $k=0.6$.  Computation is 
initializing with the third Fourier mode in $x$, and lower modes
excited by round-off errors are explicitly removed. (Computational domain in $y$
is much larger than indicated.  Both  $x$ and $y$ are periodic and a pseudo-spectral scheme with
an explicit viscosity ($\nu=3.75\times10^{-7}$) is used. )
Upper left is  a plot of the background flow
profile, $U(y)$, and the vorticity, $-dU/dy$.  Following (left to right, top to bottom) are five
snapshots of the vorticity field at the times indicated.
Comparison of the vorticity field snapshots with the $-dU/dy$ plot indicates
the shading scheme and contour levels.
The bottom row of pictures shows the perturbation vorticity field, $\omega$, 
for three of the snapshots. 
} 
\label{fig:jet}
\end{figure}

To motivate the critical-layer theory we develop below,
we again exploit numerical computations. Figure \ref{fig:jet}
shows a simulation of instabilities growing on a
shear flow in the form of a jet (the background flow
is the so-called Bickley jet \cite{bickley}, with $U(y) = {\rm sech}^2 y$, 
which is popularly used to model the fluid wake behind an obstacle).
Both this fluid mechanical example and the Vlasov simulation
shown earlier show how the unstable mode grows initially,
but then saturates as a result of pronounced nonlinear effects
inside a narrow region surrounding the critical level.
The principal effect in both examples is the twist up of the
background distribution into vortices over the critical layer.
Note how the perturbation vorticity distribution is clearly composed
of two parts: a global, relatively smooth modal structure,
and finely scaled cat's eye patterns emerging around the critical
levels that coincide with the inflection  points of $U(y)$.

\section{The single-wave model}
\label{sec:derivation}

The stage is now set for a reduction of our general model equations 
to the single-wave model.  From  Sec.~\ref{sec:linear} we  learn that our linear 
theory has an eigenspectrum that is composed of  only the continuum plus a single, distinguished,
embedded neutral mode.  Examples of such embedded modes include
the special plasma oscillations lying on the stability boundary  described 
detailed in Fig.~\ref{stab}.  On adjusting a control parameter, and thereby 
perturbing the equilibrium profile slightly, that mode becomes weakly unstable.  
The single-wave model describes  the ensuing nonlinear development, 
and an essential ingredient is the trapping scaling discussed in  Sec.~\ref{sec:nonlinear}.   

\subsection{Derivation}
\label{ssec:derivation}

At onset, a distinguished neutral mode is embedded within the continuum
and remains regular because its critical level lines up with an
extremal point of the equilibrium profile (where $R'(\pp_*)=0$).
At first sight, this might appear to allow us to proceed with
a conventional weakly nonlinear theory as the critical-level singularity
then disappears from the leading-order solution (the distinguished mode).
However, the continuous spectrum manifests itself at higher
order in the asymptotic expansion where the singularity reappears and
becomes progressively more severe
as one proceeds in the expansion (the core of the difficulty
in early expansions for electrostatic instabilities exploiting
the Hopf scaling of Sec.~\ref{ssec:traphopf}). 
The cure for the critical-level singularity
is to abandon the conventional weakly nonlinear solution over a
slender region surrounding the critical level. Within this critical layer,
a different solution is needed
that varies on a much finer spatial scale.
The recipe is therefore a matched asymptotic expansion in which
one finds an inner solution for the critical layer,
and then matches to the usual weakly nonlinear solution that  remains
valid in the `outer region' further afield.
We sketch out this expansion for the general model; 
additional details can be found in the articles by 
\textcite{churilov87a,goldstein88b,delcastillo98a,delcastillo98b,balmforth2001b}.

To ease the discussion, we will take $\Ome(\pp)$ to be monotonic,
implying that there is only a single critical level, and $\Ome'\geq0$.
We also assume that our periodic domain encompasses
exactly one wavelength of the distinguished neutral mode and 
the system has been suitably scaled so that the wavenumber is $k=1$.
In principle, the expansion requires multiple time scales in order 
to deal with the relatively fast propagation of the neutral mode 
as well as its much slower growth. However, if we first transform into
the frame of the distinguished mode ($\qq \rightarrow \qq - \ome t$), 
where $\ome$ denotes the frequency of the frame  shift,  it then suffices 
to simply rescale and use the slow time, 
$T=\varepsilon^{-1}t$, where $\varepsilon$ is the small parameter
that we use to organize the asymptotics.
In other words, we begin by introducing the transformation,
$\pdt \rightarrow -\ome\,\pdq + \varepsilon\,\pdT$,
into the governing equations  (\ref{eq:eom}) and (\ref{Phi}), whereupon they become, respectively, 
\begin{align}
	&\epsilon \rho_T+ (\Ome-\ome) \rho_\qq - R' \Phi_\theta
	+ [\rho,\Phi]
	= 0 \, ,
	\label{eq:eoma}
\\
&\Phi=\mathcal{K}\rho \ldef \int_\cD\!\!\dint\pp' \int_0^{2\pi}\!\!\!\dint\qq'  
\,  K(\qq-\qq',\pp,\pp') \,\rho(\qq',\pp',t)\, ;
\nonumber
\end{align}
these equations constitute the starting point for our expansion. 
 
\subsubsection{Outer regular expansion}

Outside the critical layer we assume the asymptotic sequences,
\begin{align}
       R(\pp)&= R_0(\pp) + \varepsilon R_1(\pp) + \ldots\\
	\rho &= 
		\varepsilon^2\rho_2(\qq,\pp,T) + \varepsilon^3\rho_3(\qq,\pp,T)
		+ \ldots
		\label{eq:expansionrho}\\
\Phi={\mathcal{K}}\rho &=  
		 \varepsilon^2\psi_2(\qq,\pp,T)
		+ \varepsilon^3\psi_3(\qq,\pp,T) + \ldots,
		\label{eq:expansionK}
\end{align}
where $\psi_m = \mathcal{K} \rho_m$ for~$m=2,3$.
The expansion in (\ref{eq:expansionK}) opens at order $\varepsilon^2$,
which corresponds to the ``trapping scaling'' of  \textcite{oneil71a} discussed in 
Sec.~\ref{ssec:traphopf}, and the sequence for $R(\pp)$ incorporates
the distortion of the equilibrium profile needed to pass from the
marginally stable state to an unstable one (such as by setting 
$a=a_0+\varepsilon$ in the parameterized family of equilibria
of the plasma problem of Sec \ref{ssec:plasmalandau}, as depicted 
in Fig.~\ref{stab}).

Upon introducing the sequences into the governing equations and grouping together
terms of the same order in $\varepsilon$, we obtain to leading order $\varepsilon^2$,  
\begin{equation}
	(\Ome-\ome) \,\pdq\rho_2
		- R_0'\pdq\psi_2 = 0\,,
\end{equation}
which can be integrated with respect to~$\qq$ to obtain
\begin{equation}
	{\mathcal{L}}\rho_2 := 
\left[ 1 - \frac{R_0'}{\Ome-\ome}\,{\mathcal{K}} \right] \rho_2
= 0.
	\label{eq:regmode}
\end{equation}
The solution to (\ref{eq:regmode}) is the embedded neutral mode with
$\Ome(\pp_*)=\ome$ and $R_0'(\pp_*)=0$:
\begin{equation}
	\rho_2(\qq,\pp,T) = \frac{R_0'\psi_2}{\Ome-\ome}\,,
\quad
	\psi_2(\qq,\pp,T) =
		\ampl(T)\,\eta(\pp) \exe^{\ii\qq}  + \compconj ,
	\label{eq:fdisttwosoln}
\end{equation}
where $\eta(\pp)$ is now the relevant ($k=1$) eigensolution to (\ref{njb.1}),
and the complex amplitude, $\ampl(T)$, is not yet determined.
Note that $\exe^{-\imi\qq}\eta(\pp)$ 
is also the null vector of ${\cal L}^\dagger$,
the operator adjoint to $\cal L$.
When evaluated at the critical level~$\pp=\pp_*$, the solution becomes
\begin{equation}
	\rho_2(\qq,\pp_*,T) =
		\frac{R''_{0*}}{\Ome'_*}\,\phimode(\qq,T),
\quad
	\phimode(\qq,T) \ldef \ampl(T)\,\exe^{\ii\qq} + \compconj,
	\label{eq:phidef}
\end{equation}
where the subscript `$*$' notation refers to evaluation at $\pp=\pp_*$ and, for later convenience, 
we have normalized so that $\eta(\pp_*)=1$.

Proceeding to  order~$\varepsilon^3$ we obtain
\begin{equation}
	  (\Ome-\ome)\,\pdq\rho_3
		- R_0'\pdq\psi_3 = R_1'\pdq\psi_2 - \pdT{\rho_2}\, , 
	\label{eq:order3}
\end{equation}
which when integrating with respect to~$\qq$ and dividing through
by~$(\ome-\Ome)$ yields 
\bqy
	{\mathcal{L}}\rho_3 &=& \rho_3 - \frac{R_0'\psi_3}{\Ome-\ome} 
	\nonumber\\
	&=&
	\frac{R_1'\psi_2}{\Ome-\ome} + 
	\frac{R_0'}{(\Ome-\ome)^2}
		\l(\imi  \ampl_T\, \eta\,   \exe^{\ii\qq} + \compconj\r)\,.
	\label{eq:orderthree}
\eqy
In general, although $R'(\pp)$ vanishes at the critical level, neither  $R''(\pp)$ nor $R_1'(\pp)$ 
will vanish. Thus, the right-hand side of (\ref{eq:orderthree}) is singular, which
impedes us from multiplying by the adjoint $\exe^{-\ii\qq}\eta(\pp)$
and integrating to eliminate $\rho_3$ to  find the desired
equation for $\ampl(T)$. To avoid this stumbling block, we resolve the
critical-layer dynamics by using an inner solution.
The scaling of the inner region is determined by noting
that the asymptotic sequence,
$\varepsilon^2\rho_2 + \varepsilon^3\rho_3$ becomes disordered as $\pp$ tends
to $\pp_*$.  In fact, because $\rho_2 \sim O(1)$ 
and $\rho_3 \sim O(\pp-\pp_*)^{-1}$ as $\pp\to\pp_*$,
the asymptotic solution breaks down within a slender layer of thickness
$\varepsilon$ surrounding the critical level.  This is where we must search
for the inner, critical-layer solution.

\subsubsection{Inner critical layer solution}

For the critical layer, we
define a stretched coordinate $Y\ldef(\pp-\pp_*)/\varepsilon$ 
and a new variable, $\zeta(\qq,Y,t)$, 
(not to be confused with the general $\zeta(q,p,t)$ of Sec.~\ref{ssec:badass}),
such that
\begin{equation}
\rho = \varepsilon^2 
	\zeta(\qq,Y,T)+ \varepsilon^2  \rho_2^{\mathrm{outer}}(\qq,\pp_*,T)\,.
	\label{eq:fdistinner}
\end{equation}
Because $\Phi=\mathcal{K}\rho$ is an integral transform of $\rho$
which smoothes that variable,
there is no critical-layer structure in $\Phi$ at leading order
(cf.\  Sec.~\ref{sssec:solvability}), and therefore 
$\Phi\sim\varepsilon^2\psi_2(\qq,\pp_*,T)= \varepsilon^2\phi(\qq,T)$
over this region.
Given these variables, the relation~\hbox{$\pdp=\varepsilon^{-1}\pdy$},  and
Eq.~\eqref{eq:eom}, the equation for~$\zeta$ becomes, to leading order,
\begin{equation}
	\pdT\left(\zeta + \frac{{R_0''}_*}{\Ome_*'} \, \phimode\right)
	+ \Ome_*'Y \, \zeta_{\qq}
	-  \bigl( {R_1'}_* +  \zeta_{Y}\bigr)
	\,\phimode_{\qq} = 0,
	\label{eq:innervorteqn}
\end{equation}
where $\phimode(\qq,T)$ is given by \eqref{eq:phidef}.

Equation (\ref{eq:innervorteqn}) is a fully nonlinear evolution equation
for the critical-layer dynamics and offers no further simplification.
The problem cannot be solved in closed form and must be approached
numerically in general.
Thus, the problem still boils down to the solution of a partial
differential equation, and the single-wave model remains of infinite
dimension.

\subsubsection{Matching}

Expressed in terms of $Y$,
the inner limit of the outer solution can be written in the form,
\begin{equation}
\rho \sim \epsilon^2 {R_{0*}''\phi \over \Ome_*'} 
+ \epsilon^2 {R_{1*}'\phi\over Y \Ome_*'} 
+ \epsilon^2 {R_{0*}'' \phi_{T\qq}\over Y \Ome_*^{\prime 2}}
 + O(\epsilon^3)\,,
\end{equation}
provided $\psi_3={\cal K}\rho_3$ does not diverge faster than
$(\pp-\pp_*)^{-1}$ as $\pp\rightarrow\pp_*$,
which is the case for the $\cal K$-operators of interest here.
On the other hand, the inner solution has an outer limit given
by
\begin{equation}
\rho \sim   \epsilon^2 \zeta + \epsilon^2 {R_{0*}''\phi\over \Ome_*'} 
\,,
\end{equation}
with the second term being the outer contribution to the inner solution.  
The two leading-order solutions can therefore be successfully matched
provided $\zeta\to0$ for $Y\to\pm\infty$, which requires 
consideration the initial condition, $\zeta(\qq,Y,0)$.  In
particular, we take $\zeta(\qq,Y,0)=0$, which corresponds to
kicking the system into action by introducing the discrete mode
at a low amplitude, ${\cal A}(0)={\cal A}_0$
with $|{\cal A}_0|\ll1$, without further rearranging
the equilibrium distribution within or outside the critical layer.
The large $|Y|$ limit of  (\ref{eq:innervorteqn}) then implies that
\begin{equation}
\zeta \sim  {R_{1*}' \phi\over Y \Ome_*'}
+ {R_{0*}'' \phi_{T\qq}\over Y \Ome_*^{\prime 2}} + O(Y^{-2}) \, .
\label{innerY}
\end{equation}
Thus, $\zeta\sim O(Y^{-1})$ for $|Y|\gg1$, ensuring the match.

\subsubsection{Solvability}
\label{sssec:solvability}

The final step is to apply a solvability condition on $\rho_3$
to determine  ${\cal A}(T)$. To accomplish this  task, first consider
\begin{equation}
{\cal K}\rho =
\int_0^{2\pi}\!\! \!\dint\qq' \int_\cD \!\! \dint\pp'\; K(\qq-\qq',\pp,\pp')\,  \rho(\qq',\pp',T)\, .
\label{Krho}
\end{equation}
We break the integral over $\mathcal{D}$ into three pieces:
$[\pp_L,\pp_*-\Delta]$,
$(\pp_*-\Delta,\pp_*+\Delta)$ and
$[\pp_*+\Delta,\pp_R]$,
where $[\pp_L,\pp_R]$ denotes the full integration over $\mathcal{D}$  and
$\varepsilon\ll\Delta\ll1$. The breakages in the integral therefore extend into the matching region
where both the outer and inner solutions remain valid.
For the integral in $\pp$ over $\mathcal{D}$ of (\ref{Krho}), we then obtain
\bqy
&{\ }&
\varepsilon^2 \int_\cD \dint\pp'\; K(\qq-\qq',\pp,\pp') \, 
\rho_2(\qq',\pp',T)
\nonumber\\
&+&
 \varepsilon^3 \pvint_\cD\! \dint\pp' \; K(\qq-\qq',\pp,\pp')\,  \rho_3(\qq',\pp',T)
\nonumber\\
&+ &\varepsilon^3 K(\qq-\qq',\pp,\pp_*)
\int_{-\Delta/\epsilon}^{\Delta/\epsilon}\!\dint Y\,   \zeta(\qq',Y,T) 
+O(\epsilon^4)\,,
\nonumber
\eqy
where we now use the notation,
\begin{equation}
\pvint_\cD \!\dint\pp \,  f(\pp) := 
\int_{\pp_L}^{\pp_*-\Delta}\!\dint\pp \,   f(\pp) +
\int_{\pp_*+\Delta}^{\pp_R}\!\dint\pp \,   f(\pp)
\,.
\end{equation}
Hence,
\begin{equation}
\psi_2 = 
\varepsilon^2 \int_\cD \!\dint\pp'\; K(\qq-\qq',\pp,\pp') \, 
\rho_2(\qq',\pp',T)
\end{equation}
throughout $\cD$, which is therefore free of critical-layer structure
as remarked earlier, and
\bqy
\psi_3 &=& \int_0^{2\pi}\! \!\!\dint\qq \,\pvint_\cD \! \dint\pp' \, 
K(\qq-\qq',\pp,\pp')\,  \rho_3(\qq',\pp',T)
\label{psi3}\\
&&
+ \int_0^{2\pi} \! \!\!\dint\qq\; K(\qq-\qq',\pp,\pp_*)
\int_{-\Delta/\epsilon}^{\Delta/\epsilon} \! \!\! \dint Y\,  \zeta(\qq',Y,T) \,.
\nonumber
\eqy

Now we multiply Eq.~\eqref{eq:orderthree} by
$\eta(\pp) \exe^{-\imi\qq}$, integrate $\qq$ over $[0,2\pi]$,  and $\pp$
over the ranges $[\pp_L,\pp_*-\Delta]$ and $[\pp_*+\Delta,\pp_R]$.
The left-hand side of \eqref{eq:orderthree} becomes, 
\begin{equation}
\int_0^{2\pi}\!\!\! \dint\qq\,  \pvint_\cD\! \dint\pp
\l( \rho_3 - \frac{R_0'\psi_3}{\Ome-\ome} \r) \exe^{-\imi\qq}\,  \eta
\end{equation}
\begin{equation}
= 
-\int_0^{2\pi}\!\!\! \dint\qq\, 
\int_{-\Delta/\varepsilon}^{\Delta/\varepsilon} \!\!\dint Y\, 
\exe^{-\imi\qq} \, \zeta(\qq',Y)+ O(\Delta)\,,
\end{equation}
on introducing $\psi_3$ from (\ref{psi3}) and using the
following properties of $\eta$:
${\mathcal{L}}^\dagger \eta = 0$,
$\eta\exe^{-\imi\theta} = {\cal K} [R_0'\eta\exe^{-\imi\theta}/(\Ome-\ome)]$
and $\eta_*=1$.
The right-hand side of \eqref{eq:orderthree}, on the other hand, becomes
\begin{equation}
	2\pi \imi\, \ampl_T\, \pvint_\cD\!\!\dint\pp\, 
		\frac{R_0'\eta^2}{(\Ome-\ome)^2}
	+ 2\pi\ampl \pvint_\cD\!\!\dint\pp\, 
		\frac{R_1'\eta^2}{\Ome-\ome}\,.
\end{equation}
Thence,
\bqy
	&&\imi\, \ampl_T\,\pvint_\cD\!\!\dint\pp\, 
		\frac{R_0'\eta^2}{(\Ome-\ome)^2}
	+\ampl\int_\cD\!\!\dint\pp\, 
		\frac{R_1'\eta^2}{\Ome-\ome}
		\label{eq:solvcond}\\
	&=&- \frac1{2\pi}\int_0^{2\pi}\!\!\!\dint \qq\!
		\int_{-\Delta/\epsilon}^{\Delta/\epsilon}\!\dint Y\, 
		\exe^{-\imi\qq}\,\zeta(\qq,Y,t) + O(\Delta) \,.
\nonumber
\eqy
Finally, we take the limits $\Delta\rightarrow0$
and $\Delta/\varepsilon\rightarrow\infty$. This
turns the integrals over $\pp$ in (\ref{eq:solvcond})
into standard principal-value integrals,
and the integral over $Y$ into another principal-value integral,
this time at its infinite limits.

\subsection{Normal and degenerate forms}
\label{ssec:forms}

Gathering together the results of Sec.~\ref{ssec:derivation}, we rewrite 
Eqs.~\eqref{eq:phidef},  \eqref{eq:innervorteqn}, and \eqref{eq:solvcond}  compactly  as 
\begin{align}
	&\phimode(\qq,T) =  \ampl(T)\,\exe^{\ii\qq} + \ampl(T)^*\,\exe^{-\ii\qq}
	\,,
	\label{phiT}\\
&\calq_T + [\calq,\mathfrak{E}]= 0\,,
\label{varQT}\\
& \imi\,  \varpi \,  \ampl_T
	+ \nu\,  \ampl 
	=  -\l\langle \zeta\, \exe^{-\imi\qq}\r \rangle \,,
	\label{varAT}
\end{align}
where
\bqy
\calq(\qq,Y,T)&\ldef& \zeta + \frac{{R_0''}_*}{\Ome_*'}\, \phimode +\frac{R_{0*}'' }{2} \, Y^2 + R_{1*}'\,  Y\,,
\label{calq}
\\
\mathfrak{E}(\qq,Y,T)&\ldef&\frac{\Ome_*'}{2}  Y^2 + \phimode(\qq,T)\,,
\label{frake}
\eqy
\bq
 \varpi \ldef  \pvint_{\!\!\!\R}\!\dint\pp \, \frac{R_0'\eta^2}{(\Ome-\ome)^2} \,,
 \qquad{\rm and}\qquad
  \nu\ldef\int_{\R}\! \!\dint\pp\,  \frac{R_1'\eta^2}{\Ome-\ome}\,.
  \label{varpi}
\eq
These equations can be simplified further by scaling, which we do 
in the next section to obtain the normal form. Before setting about this
task, we briefly recall the origin of some of the quantities 
appearing in the system: the variable $\calq$ corresponds to the limit
of the total distribution, $\zep=R(I)+\rho$,
the Hamiltonian system of  Sec.~\ref{ssec:badass} 
within the critical layer. The equilibrium distribution, $R(I)$,
contributes the coefficients, $R_{0*}''$ and $R_{1*}'$
(the leading-order curvature and the first-order slope), to the
definition of $\calq$. Featuring in (\ref{frake}) is the
leading-order `shear' at the critical level, ${\Ome_*'}\equiv h''(I_*)$, 
determined from the one-particle piece of the Hamiltonian.
The integral quantities, $\varpi$ and $\nu$, depend on the shape
of the eigenfunction of the embedded mode.

\subsubsection{Normal}
\label{sssec:normal}

To place the system into a simpler form, we introduce
a characteristic timescale $\tau$, a scale for action $\mu$,
and rescaling of the distribution, $b$. We 
then define the new variables,
\bqy
\tswm &\ldef& \frac{T}{\tau},
\quad
\xswm = \theta + \mu\,\tswm,
\quad
 \yswm \ldef \Ome_*'\tau Y + \mu,
         \quad
         \label{cantrans}\\
\phisw &\ldef& -\Ome_*'\tau^2\phimode,
\quad
         \amplsw \ldef - \Ome_*'\tau^2\ampl \,\exe^{-\imi\mu\,t},
\label{ampscale}\\
\check\zeta&\ldef& b \zeta,
\quad
Q = b \calq - \half \mu^2 \kappa + \mu \gamma ,
\label{zetascale}
\eqy
where
\begin{equation}
         \kappa \ldef -\frac{b{R_0''}_*}{(\tau \Ome_*')^2},\qquad
         \gamma \ldef \mu\kappa+\frac{b {R_1'}_*}{\tau\Ome_*'} .
\end{equation}
If $\varpi\ne0$, we may then make the selections,
\begin{equation}
b = \frac{\tau^2}{\varpi}, \quad
         \mu \ldef\tau \frac{\nu}{\varpi}\,,
         \label{muscale}
\end{equation}
where $\varpi$ and $\nu$ are  given in  (\ref{varpi});
the ``degenerate'' case, with $\varpi=0$ will be dealt with separately
in Sec. \ref{sssec:degenerate}.
Substituting these forms into Eqs.~(\ref{phiT}), (\ref{varQT}), and 
(\ref{varAT}), and then dropping the `$\, \check{\ }\, $'  on $\zeta$  
leads to
\begin{align}
&\phisw(x,t) = \amplsw \exe^{\imi \xswm} + \amplsw^* \exe^{-\imi \xswm} 
 \,,
 \label{varphit}\\
&Q_t + [Q,\mathcal{E}]
         = 0 
         \label{eq:swm1}\\
&\imi \amplsw_\tswm =
         \l\langle\zeta\, \exe^{-\imi \xswm}\r\rangle,\,
         \label{eq:swm2}
\end{align}
where
\bqy
Q(x,y,t)&\ldef& \zeta +\kappa \varphi - \frac{\kappa}{2} y^2 +\gamma y
\label{Q}
\\
\mathcal{E}(x,y,t)&\ldef& \frac{y^2}{2}-\varphi(x,t)\, .
\label{energ}
\eqy
An alternative form of Eq.~(\ref{eq:swm1}) is 
\bq
\zeta_t + y\zeta_x +\varphi_x\zeta_y=-\kappa \varphi_t -\gamma\varphi_x\, .
\label{normal}
\eq

The initial and boundary conditions are
\bqy
\amplsw(0)&=&\amplsw_0, \qquad
\zeta(\xswm,\yswm,0)=0 , \nonumber\\
 \zeta &\sim& \yswm^{-1}(\kappa \phisw_{\xswm\tswm} - \gamma\phisw) 
\quad {\rm for } \ |\yswm| \gg 1,
\label{swmicbc}
\eqy
where $\amplsw_0$ is the initial mode amplitude.
Note that the relatively weak decay of $\zeta$ with $y$ does not
interfere with the construction of 
$\langle \zeta\, \exe^{-\imi \xswm}\rangle$ because the
integral over $y$ is interpreted using the principle value
at its limits ({\it cf.} (\ref{eq:solvcond})).
Similarly, one can eliminate the variable $\zeta$ entirely
in favor of $Q$ by observing that
\bqy
\l\langle Q \exe^{-\imi x}-\kappa A\r\rangle
&=&\l\langle \l( Q-\kappa \varphi\r) \exe^{-\imi x}\r\rangle
\nonumber\\
&=&
\l\langle \l( \zeta - \frac{\kappa}{2} y^2 +\gamma y\r) \exe^{-\imi x}\r\rangle
\nonumber\\
&=&\l\langle  \zeta  \exe^{-\imi x}\r\rangle\, ,
\label{dotA}
\eqy
provided the potentially
divergent integral involving $\half \kappa y^2\exe^{-\imi x}$
is dealt with by first performing the $x$-average.

Equations~\eqref{varphit}--\eqref{energ}
are what we call the normal form of the single-wave model.  
Crucially, the absence of dissipation in the problem has the unappealing
consequence that no dimensional reduction occurs as part of the asymptotic
analysis, and the single-wave model retains a partial differential form.
Of course, no dimensional reduction can be expected here, since at the
onset of instability, the distinguished neutral mode is arbitrarily
close to an infinite number of singular continuum modes with comparable
wavespeeds; the two equations composing the single-wave model describe the
interaction between the distinguished mode and a superposition
of locally resonant continuum modes within the critical layer. 

Note that one further simplification is possible: 
since $\tau$ has not yet been specified, we may exploit this timescale
to set $\gamma=\pm1$. The parameter $\gamma$ is built from the correction
to the equilibrium profile, $R_1$, and,
as shown below, acts as the trigger for instability. After scaling,
the two choices, $\gamma=\pm1$, refer to whether the system is displaced 
from marginality into the unstable ($\gamma=-1$) or stable ($\gamma=+1$) 
regimes.  Aside from the initial amplitude $\amplsw_0$,
this leaves a single parameter in the problem, namely $\kappa$, which
is related to $R_{0*}''$, the curvature of the leading-order equilibrium
profile at the critical level. This parameter encodes 
information regarding the position of the embedded
mode along the stability boundary: at the minimum of the stability
boundary where all wavenumbers are stable, $R_{0*}''=0$ 
(see Sec. \ref{ssec:plasmalandau}) and so $\kappa=0$; elsewhere
on the stability boundary, $\kappa$ does not vanish (and the danger
of additional unstable modes is removed by suitably choosing the
length of the periodic domain).

The normal form as given by (\ref{varphit}), (\ref{eq:swm2}), and (\ref{normal}) was derived for 
various types of fluid shear flows by \textcite{churilov87a,shukhman89a,goldstein88b,
balmforth99b,balmforth2001a,balmforth2001b}.
The system is  similar, though not identical,  to the original single-wave
model proposed for unstable electrostatic waves
\cite{oneil71a,tennyson94a} and nonlinear Landau damping \cite{imamura69a}.
In the original single-wave model, the right-hand side
of (\ref{eq:swm1}) did not appear (i.e.,\ $\kappa\equiv\gamma\equiv 0$)
and instability was introduced into the system
via a suitable initial condition, $\zeta(\xswm,\yswm,0)\ne0$. 
\textcite{oneil71a} used a beam-like distribution (a delta-function
at a fixed velocity); \textcite{onishchenko71a} and \textcite{imamura69a}
used a linear profile, equivalent to the $\kappa\equiv 0$ case.
\textcite{delcastillo98a,delcastillo98b} derived the single-wave model above,
but then elected to drop the parameters $\gamma$ and $\kappa$
and follow the route of \textcite{oneil71a}  to instability by selecting a suitable initial condition.
However, when viewed as the weakly nonlinear description of the distinguished
mode on the stability boundary, there is no freedom in the choice
of initial condition; only the position on the stability boundary
(and therefore $\kappa$) is a free parameter.

\subsubsection{Degenerate}
\label{sssec:degenerate}

It may transpire that  $\varpi=0$. When this happens,  the scaling leading to the normal form of Sec.~\ref{sssec:normal} is precluded because we cannot 
make the selections in (\ref{muscale}).   Upon examination of  (\ref{varpi}),   it is evident that $\varpi=0$ is unlikely unless there is a simplicity or symmetry to the problem, which turns out to be the case for some physical examples.  A case in point is that of the XY model  introduced in Sec.~\ref{sssec:xyinteraction}.  
For this case, $\Omega =I$ and
the interaction has a spatial kernel ($-\pi^{-1}\cos(\theta-\theta')$ is 
independent of  action $I$) so $\eta$ is constant
(see Sec.~\ref{ssec:generaltheory}).
Moreover, for equilibrium profiles
like the Gaussian of Sec. \ref{ssec:xylinear},
$R_0'(I)$ is odd in $I$, the embedded mode has $\ome=0$, and 
$\varpi=0$ follows immediately.
We will treat this example in detail in 
Sec.~\ref{sec:degeneratepatterns}, while here we work out the general degenerate form.  From  Eq.~(\ref{varAT}) it is clear that the mode amplitude for the degenerate form  will no longer be governed by a differential equation, but by an integral constraint.

Instead of (\ref{muscale}), we now make the selections
\bq
b = \frac{\tau}{\nu}, \quad
\mu = - \frac{R_{1*}'}{\nu\Omega_*'}
\label{degzetascale}
\eq
With these choices, $\gamma=0$ and
Eqs.~(\ref{phiT}), (\ref{varQT}), and (\ref{varAT}) now become 
\begin{align}
&\phisw(x,t) = \amplsw \exe^{\imi \xswm} + \amplsw^* \exe^{-\imi \xswm} 
 \,,
 \label{dvarphit}\\
&Q_t + [Q,\mathcal{E}]
         = 0 
         \label{eq:dswm1}\\
& A=
         \l\langle\zeta\, \exe^{-\imi \xswm}\r\rangle,\,
         \label{eq:dswm2}
\end{align}
where  
\bqy
Q(x,y,t)&\ldef& \zeta + \kappa \varphi - \frac{\kappa}{2} y^2 \,. 
\label{degQ}
\\
\mathcal{E}(x,y,t)&\ldef& \frac{y^2}{2}-\varphi(x,t)\, .
\label{degenerg}
\eqy

Finally, we are again free to choose $\tau$, and thereby scale out
a parameter. The only parameter remaining explicitly is
$\kappa$, and so we choose 
\bq
\tau=\frac{1}{(\Omega_*')^2} \left|\frac{R_{0*}''}{\nu}\right|
\eq
so that $\kappa\equiv -$sgn$(R_{0*}''/\nu)=\pm1$.
With this selection, as demonstrated in
Sec.~\ref{sec:linearsinglewave}, we find that the system is
unstable when $\kappa=-1$ (and stable if $\kappa=1$).
Note that, because the mode amplitude is no longer a dependent
variable, but given by the integral constraint (\ref{eq:dswm2}),
the initial amplitude of the mode does not appear as an independent
parameter, and the initial state of the system remains to be specified.
One option is to initialize using the linear normal mode of
Sec. \ref{sec:linearsinglewave}. Alternatively,
a more convenient method in practice is to kick the system
into action by adding a transient forcing term to  (\ref{eq:dswm2})
that smoothly switches on the dynamics. 

\subsection{Hamiltonian structure and conservation laws}
\label{ssec:hamlaws}

\subsubsection{Normal Hamiltonian structure}
\label{ssec:normalHam}

Both the normal and degenerate forms of the single-wave model of 
Sec.~\ref{ssec:forms} inherit the structure of their 
Hamiltonian parent models, discussed in Sec.~\ref{sec:theories}.    
For the normal form of the model, the Hamiltonian 
is given by  
\bq
H_n=\int_{\R}\!dy\int_0^{2\pi}\! \!\!dx \,\left(\mathcal{E} {Q} + \kappa \left|A\right|^2\right)\, .
\label{hnormal}
\eq
The two terms in (\ref{hnormal}) are reminiscent of the particle kinetic 
energy and the electrostatic energy
for the Vlasov-Poisson system (see Sec.~\ref{sec:theories}).
This Hamiltonian together with the Poisson bracket of (\ref{swpb}) gives 
the single-wave model.  To see this we compute the  derivatives,
\bq
\frac{\delta H_n}{\delta Q}= \mathcal{E}\qquad {\rm and}
\qquad \frac{\p H_n}{\p A^*}= -2\pi \l\langle Q e^{ix} -\kappa A\r\rangle\,,
\eq
where $Q$, $A$, and $A^*$ are independent variables, and insert them into (\ref{swpb}) to obtain
\bqy
Q_t&=&\{Q,H_n\}=- [Q,\mathcal{E}], \nonumber
\\
A_t&=& \{A,H_n\}= -i\l\langle Q e^{-ix}-\kappa A\r\rangle
\label{pbkts}
\eqy
and similarly $A^*_t=(A_t)^*$.  This Hamiltonian structure is the continuum 
version of  that derived in \textcite{tennyson94a} (see also \citealp{smith, mynick}).  

Showing that (\ref{hnormal}) exists requires some care, of the type commented 
upon in Sec.~\ref{sssec:normal},  since the second term of $H_{\rm n}$ is 
divergent.  
To show that Hamiltonian (energy) $H_{\rm n}$  is in fact finite we
 return to (\ref{normal}) and analyze more carefully the large $|y|$ behavior 
of $\zeta$, obtaining 
\bq
\zeta \sim
\frac{1}{y} (\kappa \varphi_{xt} - \gamma \varphi)
+ \frac{1}{y^2} (\kappa |A|^2 - \kappa \varphi_{tt} - \gamma \varphi_{xt})
+ ...\, .
\label{largey}
\eq
The  $x$-independent, $O(y^{-2})$ term in (\ref{largey})
follows upon  averaging (\ref{normal}) over $x$, and is sufficient to
cancel the divergent second piece of $H_{\rm n}$, leaving
a convergent double integral.

Given the Hamiltonian structure, it is natural to look for invariants.  Since 
both the single wave and the medium can carry momentum we expect a 
total momentum of the following form to be conserved (cf.\ \textcite{tennyson94a}):
\bq
P=2\pi\, |A|^2-\int_{\R}\!dy\int_0^{2\pi}\! \!\!dx \,y\, \zeta\,, 
\label{mominv}
\eq
where the first term of (\ref{mominv}) will shortly be identified with the wave action,  while the second has the familiar form of  the (relative) $x$-component of the momentum in vortex dynamics (e.g.\ \citealp{flierl11}).  It can be shown that $P$ commutes with $H_{\rm n}$, i.e.\ $\{P,H_{\rm n}\}=0$, or directly that this is equivalent to 
\[
P_t=\l(\langle \yswm\zeta\rangle-|\amplsw|^2\r)_\tswm = 0\,. 
\]

An immediate consequence of the Poisson bracket of (\ref{swpb}) is the existence of 
the following infinite family  of Casimir invariants (see e.g.\  \citealp{morrison98}):
\bq
C=\int_{\R}\!dy\int_0^{2\pi}\! \!\!dx\,  \mathcal{C}(Q)\,,
\eq
where $\mathcal{C}$ is an arbitrary function of $Q$. An important Casimir is the `enstrophy'
\bq
C_2:=\int_{\R}\!dy\int_0^{2\pi}\! \!\!dx\, Q^2\,.
\eq
With some manipulation this can be shown to be equivalent to 
\bq
C_2=\int_{\R}\!dy\int_0^{2\pi}\! \!\!dx\,  \zeta^2 - 2\kappa H_{\rm n} + 2\gamma\big(2\pi\, |A|^2- P\big)\,.
\eq
Since $H_{\rm n}$ and $P$ are constants of motion, this also implies the
further invariant,
\bq
I= \int_{\R}\!dy\int_0^{2\pi}\! \!\!dx\,  \zeta^2 + 4\pi \gamma \, |A|^2\,.
\label{iinv}
\eq
From (\ref{iinv}) we obtain a stability condition, viz., for $\gamma=+1$ in the vicinity of the equilibrium state $A=\zeta =0$, the invariant $I$ provides a norm for bounding solutions to an infinitesimal neighborhood of this equilibrium.  Also, with the initial conditions $\zeta(x,y,0)=0$  and $A(0)=A_0$ we have 
\begin{equation}
\gamma |\amplsw|^2 + {1\over2} \langle \zeta^2 \rangle
= \gamma |\amplsw_0|^2\,.
\end{equation}
Thus, for  $\gamma=+1$, we have the finite amplitude bounds $|\amplsw|^2\leq |\amplsw_0|^2$ and $\langle \zeta^2 \rangle\leq 2|\amplsw_0|^2$.   This amounts to energy-Casmir stability for this example (see e.g.\  \citealp{morrison98}). 

To show that $\mathcal{J}=2\pi\, |A|^2$ is equivalent to
the  wave action, we introduce the  coordinate change,
\bq
A=\sqrt{\frac{\mathcal{J}}{2\pi}} \, e^{i\Psi} \qquad {\rm and }\qquad A^*=\sqrt{\frac{\mathcal{J}}{2\pi}} \, e^{-i\Psi}\,.
\eq
The bracket of (\ref{swpb}) becomes
\bqy
\{F,G\}&=&  \left(
\frac{\p F}{\p \Psi} \frac{\p G}{\p \mathcal{J}}-\frac{\p G}{\p \Psi}\frac{\p F}{\p \mathcal{J}} 
\right)
\nonumber\\
&{\ }& \hspace{.5 cm} + \int_{\mathcal{\R}}\!  dy \int_0^{2\pi}\!\!\! dx \, 
Q\left[\frac{\delta F}{\delta Q},\frac{\delta G}{\delta Q}\right] \,,
\eqy
where now $\Psi$ and $\mathcal{J}$ are a genuine canonically conjugate pair 
of wave variables for the wave.  In terms of these variables the potential $\varphi$ is given by
\[
\varphi=\sqrt{\frac{2\mathcal{J}}{\pi}} \cos(x+\Psi)
\]
and Eq.~(\ref{eq:swm2}) is replaced by the following two real-valued equations:
\bqy
\dot\Psi&=&  \frac{\p H_n}{\p \mathcal{J}}= -\l\langle \sqrt{\frac{2\pi}{\mathcal{J}}}\,  Q \, \cos(x+ \Psi) - \kappa \r\rangle
\nonumber\\
 \dot{\mathcal{J}}&=&-\frac{\p H_n}{\p \Psi}= - 2 \l\langle \sqrt{2\pi\mathcal{J}}\,  Q\,  \sin(x+ \Psi)\r\rangle\,. 
\eqy

\subsubsection{Degenerate Hamiltonian structure}
\label{ssec:degHam}

It is evident from  Sec.~\ref{ssec:normalHam} that divergences associated with system size must be treated with some delicacy;  e.g., observing that  $\langle\varphi\rangle$ vanishes requires integration over $x$, giving zero because of periodicity,  before integration over $y$, which would give a divergent result,  as discussed  in  obtaining Eq.~(\ref{dotA}).  Such system size divergences are common for two-dimensional theories such as vortex dynamics, quasigeostropy, etc., which possess infinite kinetic energy for unbounded domains.   For this reason, relative energies and momenta are often used (see, e.g., \ \citealp{flierl11} for an example of numerically tracking such quantities), but for our presentation of the Hamiltonian form for the degenerate case,  we find it convenient to introduce a system size so that $\langle 1 \rangle= \cald$.   The need for this artifice is  compounded by the fact that  for the  degenerate form, the theory is written entirely in terms of the variable  $Q$ - the amplitude being  determined by Eq.~(\ref{eq:dswm2}), which is   a constraint equation.     Tools for handling such constraints are now well-developed   \cite{TaChaMo09,ChaMoTa12,Cha13},  but we will present the   Hamiltonian structure  for this degenerate case in a manner  akin to that for the  Vlasov-Poisson system \cite{morrison80}, which  has a bracket of the form of (\ref{eq:eom0}).

Using (\ref{dvarphit}) and (\ref{eq:dswm2}) it is evident that  the constraint amounts to $\varphi$ being functionally dependent on $\ze$, i.e.,
\bq 
\varphi(x)=2\langle \ze \cos(x-x')\rangle=:  \underline{\varphi}[\ze](x)
\eq
where the underline will be used to denote $\varphi$ as an operator.  Upon introducing  the linear operator 
\[
\call_{\kappa}:= \cali + \kappa   \underline{\varphi}\,,
\]
 where $\cali$ is the identity operator, the equation of motion (\ref{eq:dswm1}) becomes
\bq
\call_{\kappa}[\ze_t]=- y \, \ze_x -   \underline{\varphi}[\ze]_x\,  \ze_y 
\eq
and thus it is necessary to assume $\call_{\kappa}[\ze_t]^{-1}$ exists, which we write as
\bq
\call_{\kappa}^{-1}=\frac{1}{\cali +\kappa \underline{\varphi}}= \cali - \kappa  \underline{\varphi} + \kappa^2 \underline{\varphi}^2- \dots\,.
\label{Linverse}
\eq

Various properties of the operators $ \underline{\varphi}$ and  $\call_{\kappa}$ will be needed, and so we present them now.  Observe, both operators are formally self-adjoint, e.g, 
\[
\langle g\, \underline{\varphi}[h]\rangle = \langle h \, \underline{\varphi}[g]\rangle
\]
 for  functions $g,h(x,y)$, and we have the  following:
 \bqy
 \mathcal{E}&=&{y^2}/{2} -  \underline{\varphi}[\ze]
 \\
 Q&=& \call_{\kappa}[\ze] -\kappa {y^2}/{2}
 \\
 \ze&=&\call_{\kappa}^{-1}[Q +\kappa {y^2}/{2}]\,.
 \eqy
 Also, by periodicity, $\underline{\varphi}[1]=\underline{\varphi}[g(y)]=0$, where $g$ is a function of $y$ alone.  A short   calculation gives
 \bq
\underline{\varphi}\left[\underline{\varphi}[\ze]\right]=\cald \underline{\varphi}[\ze]\ 
\Rightarrow\ 
\underline{\varphi}^n[\ze] =\cald^{n-1}\underline{\varphi}[\ze]\,.
\eq
Thus, upon making use of (\ref{Linverse}) we obtain
\bqy
\call_{\kappa}^{-1}[y^2/2]&=&y^2/2 \,,
\nonumber\\
 \call_{\kappa}^{-1}\left[[ \underline{\varphi}[\ze]\right]&=& \frac1{1 + \kappa \cald}\,   \underline{\varphi}[\ze]\,,
\label{props}
 \eqy
and a functional chain rule calculation (e.g.\   \citealp{morrison13}) yields
 \bq
\frac{\de F}{\de \ze}= \call_{\kappa}\left[\frac{\de F}{\de Q}\right]
\quad {\rm or}\quad 
\frac{\de F}{\de Q}= \call_{\kappa}^{-1}\left[\frac{\de F}{\de \ze}\right]\,,
\label{chnrl}
\eq
which we will find convenient.  
 
Now, the Hamiltonian  of (\ref{hnormal}) for the normal form suggests a Hamiltonian for the degenerate case with terms as follows:
\bqy
H_d[\ze]&=&\frac{1}{2} \int_{\cald}\!dy\int_0^{2\pi}\! \!\!dx \,\, y^2\, {\ze} 
\nonumber\\
&-& \frac{(1+\kappa\cald) }{2\pi} 
\int_{\cald}\!dy\!\int_0^{2\pi}\!\!\!\!\!dx \int_{\cald}\!dy'\!\!\int_0^{2\pi}\!\!\!\!\!dx' \,
\nonumber\\
&&\times \ \ze(x,y)\ze(x',y') \, \cos(x-x')\,,
 \label{hdegnew}
\eqy
which has a form reminiscent of the Hamiltonian for the Vlasov-Poisson system.  Evidently,
\bqy
\frac{\de H_d}{\de \ze}&=&  \frac{y^2}{2} - \frac{(1+\kappa\cald) }{\pi}  \int_{\cald}\!dy'\!\!\int_0^{2\pi}\!\!\!\!\!dx' \,
\ze(x',y') \, \cos(x-x')
\nonumber\\
&=& \frac{y^2}{2} -(1+\kappa \cald)\underline{\varphi}[\ze]
\label{zederiv}
\eqy
Making use of (\ref{zederiv}),   (\ref{chnrl}), and (\ref{zederiv}) gives 
\bq
\frac{\delta H_d}{\delta Q}=  \frac{\,y^2}{2} - \underline{\varphi}=\mathcal{E}\,;
\label{degder}
\eq
whence, we obtain
\bq
Q_t=\{Q,H_d\}= -[Q,\cale]\,,
\eq
the equation of motion of (\ref{eq:dswm1}).   This Hamiltonian structure is similar to  that given in  \textcite{chandre09} for the  XY model.

\subsection{Dissipative effects within the critical layer}
\label{ssec:dissipation}

Depending on the physics under consideration, the Hamiltonian forms described 
in Sec.~\ref{ssec:hamlaws} can be amended  by adding terms that describe 
dissipative effects.  In particular, dissipation terms can be included
on the right-hand side of the critical-layer equation (\ref{normal}),
corresponding to various kinetic theory transport processes:
\bq
\zeta_t + \yswm\,\zeta_\xswm + \phisw_\xswm \zeta_\yswm
         +\kappa \phisw_\tswm + \gamma\phisw_\xswm
         =  
(\lambda \zeta_\yswm)_\yswm \,. 
\label{eq:vswm1}
\eq 
The dissipative term 
corresponds to viscous dissipation in the fluids context, 
or a Fokker-Planck operator in the kinetic theory context.
For the latter, $\lambda$ may depend on the coordinate $y$.

The inclusion of such dissipative terms within the ordering scheme of 
Sec.~\ref{ssec:derivation} requires justification.
In each case, there needs to be a distinguished 
scaling that ensures the term enters at the
same order as the other physical effects inside the critical layer,
whilst simultaneously not affecting the outer solution.  We do not 
pursue this here in generality, but 
for the fluid problem the kinematic viscosity $\nu$ of the parent 
model, which provides a term $\nu(\omega_{xx}+\omega_{yy})$ on the right-hand 
side of Eq. (\ref{f.1}), must be scaled as  $\nu=\varepsilon^3\nu_3$,
and $\lambda$ is then a suitably scaled version of $\nu_3$  
(cf.\ \citealp{churilov87a,goldstein88a}).  Note that if $\lambda\gg1$, 
one can perform some additional
reductions of the model to furnish simpler amplitude equations 
(see \citealp{churilov87a}), but here we 
focus on the less viscous cases with $\lambda$ order unity or smaller.

\section{Linear  single-wave dynamics}
\label{sec:linearsinglewave}

We now  explore the linear theory of the single-wave model, 
focussing mostly on the normal form of Sec. \ref{sssec:normal}.
On discarding the nonlinear term, $\varphi_x\zeta_y$, 
the system can be written as
\begin{align}
&\zeta_\tswm + \yswm\,\zeta_\xswm 
         +\kappa \phisw_\tswm +\gamma\phisw_\xswm
         = 
(\lambda \zeta_y)_y
         \,,
\label{eq:lswm1}\\
&\imi \amplsw_\tswm =
         \l\langle\zeta\, \exe^{-\imi \xswm}\r\rangle,\,\qquad
\phisw = \amplsw \exe^{\imi \xswm} + \compconj
\,,
 \label{eq:lswm2}
\end{align}
where we have included the  dissipative terms of Sec.~\ref{ssec:dissipation}. 
These equations can be solved analytically
for both normal modes and the initial-value problem.
As a result of the latter, we arrive at one of the simplest examples
of a system exhibiting Landau damping.

\subsection{Normal modes}

We consider first the ideal case where the dissipative terms of 
Sec.~\ref{ssec:dissipation} are set to zero (i.e.
$\lambda=0$), and then we investigate what happens when we 
include them.   

\subsubsection{Ideal}

Assuming that  nondissipative normal modes have the dependencies, 
$\zeta \propto e^{\imi(x-c_1t)}$  and $A\propto e^{-\imi c_1t}$ (recall 
our single-wave has wavenumber $k=1$),  where $c_{1}$ 
is the (complex) wavespeed, Eqs.~(\ref{eq:lswm1}) and (\ref{eq:lswm2})  
reduce to
\begin{equation}
\zeta = \frac{\kappa c_1 - \gamma} {\yswm-c_1}
\, \amplsw \exe^{\imi \xswm} + c.c.
\eq
and
\bq
c_1 \amplsw  =  (\kappa c_1 - \gamma) A
\int_{\R}\!  dy \, 
\frac1{\yswm-c_1}
=\ii \pi (\kappa c_1 - \gamma) \amplsw \; {\rm sgn}(c_{1i})\,.
\label{idii}
\end{equation}
The second equality of (\ref{idii}) is valid for all $c_{1i}\neq 0$; it is easily obtained by  
closing the contour with a semicircle in the upper or lower half plane, depending on the sign of   
$c_{1i}$, and  using Cauchy's theorem.
Solving  (\ref{idii}) for $c_1$ gives  
\begin{equation}
c_1 = {\pi\gamma  \over 1+\pi^2 \kappa^2} 
\big[\pi\kappa-\ii\;{\rm sgn}(c_{1i})\big],
\label{idi}
\end{equation}
Thus, $c_{1i} = - \pi\gamma \sgn(c_{1i})/(1+\pi^2\kappa^2)$, 
 which is inconsistent unless $\gamma<0$.  Since we have normalized 
 $\gamma=\pm1$, this indicates  that the system  is unstable when $\gamma=-1$.  
 However,   if $\gamma=+1$, there is no consistent solution for $c_{1i}$, and so 
 there are no normal modes and the system is stable.

The factor of ${\rm sgn}(c_{1i})$ in (\ref{idii})  arises because the integral  has a 
branch cut  along the real axis of the complex $c_1$-plane,  which as noted in 
Sec.~\ref{sec:linear}  reflects the continuous spectrum lying along the contour of integration. 
If we start from the consistent unstable solution with $c_{1i}>0$ ($\gamma=-1$) 
we can follow Landau and  analytically continue $c_1$ into the lower have plane 
by deforming the contour of integration so as to capture the pole.    This gives
$$
\int_{\!\mathbf{\hspace{-3 pt}_{{\ }^{-}\!\hspace{-1 pt}
\mathbf{\cup}^{\hspace{-1.5 pt}-}}}} \!\! dz\,  \frac1{z-c_1} = i \pi
$$
where the symbol on the integral is selected to show the contour is deformed so 
as to enclose  the descending pole.    This procedure of Landau  amounts to 
analytically continuing  through the branch cut and onto a different Riemann sheet with $c_{1i}<0$.  
Thus, the analytic continuation of the dispersion
relation in (\ref{idi}), which would
appear in the long time limit of the Laplace transform solution of the initial-value
problem (as in Landau's classical analysis), has no sgn$(c_{1i})$.  The  damped solutions  
one obtains in this case for $\gamma>0$ correspond to the quasi-modes discussed in Sec.~\ref{ssec:plasmalandau}. 

Following suite for the degenerate normal form of Sec. \ref{sssec:degenerate},
we find that the normal-mode solutions satisfy
\begin{equation}
\zeta = \frac{\kappa c_1} {\yswm-c_1}
\, \amplsw \exe^{\imi \xswm} + c.c.
\eq
and
\bq
\amplsw  =  \kappa c_1 A
\int_{\R}\!  dy \, 
\frac1{\yswm-c_1}
= \ii \pi \kappa c_1 \amplsw \; {\rm sgn}(c_{1i})\,.
\end{equation}
That is,
\bq
c=\ii c_i, \qquad |c_i| = - \frac{1}{\pi\kappa}.
\eq
Hence, there are normal modes and the system is unstable if $\kappa=-1$.

\subsubsection{Dissipative}

Assuming that $\lambda$ is constant, the
dissipative modes  satisfy the following equation: 
\begin{equation}
\lambda \zeta_{\yswm\yswm} 
- \ii(\yswm-c_1)\zeta = \ii (\gamma-\kappa c_1) \varphi\,.
\label{lindis}
\end{equation}
Equation (\ref{lindis}) can be solved by using Airy functions or Fourier transforms 
(generalizing the case with only the viscosity term as in  \citealp{benney69a} and \citealp{balmforth98a}):
\bq
\zeta = \imi
(\kappa c_1-\gamma) \varphi \; \calj(\yswm,c_1;\lambda_0,\lambda_1,\lambda_2),
\label{j.5}
\eq
where
\[
\calj\ldef\int_0^\infty \!\!dq\, \exe^{- \lambda q^3/3 + 
-  \imi q(y-c_1 
)} \,.
\]
Upon inserting (\ref{j.5}) into (\ref{eq:lswm2}), and then  interchanging 
the order of integration, we obtain $\delta(q)$ for the integral over $\yswm$.
Using this to evaluate  the integral over $q$ gives the eigenvalue,
\begin{equation}
c_1 =  {\pi\gamma  \over 1+\pi^2 \kappa^2} (\pi\kappa-\ii)
\,.
\label{j.6}
\end{equation}
Equation (\ref{j.6})
is identical to the inviscid result, save that
${\rm sgn}(c_{1i})$ no longer appears.
Thus, if $\gamma=-1$, the system is unstable, and the 
inviscid and viscous growth rates coincide. But when $\gamma=+1$,
there is a decaying viscous normal mode with no inviscid counterpart.
However, it does correspond to
the fake eigenvalue (quasi-mode) obtained by analytical continuation
of the inviscid dispersion relation.
 
\begin{figure}
\begin{center}
\leavevmode
\includegraphics[height=8 cm]{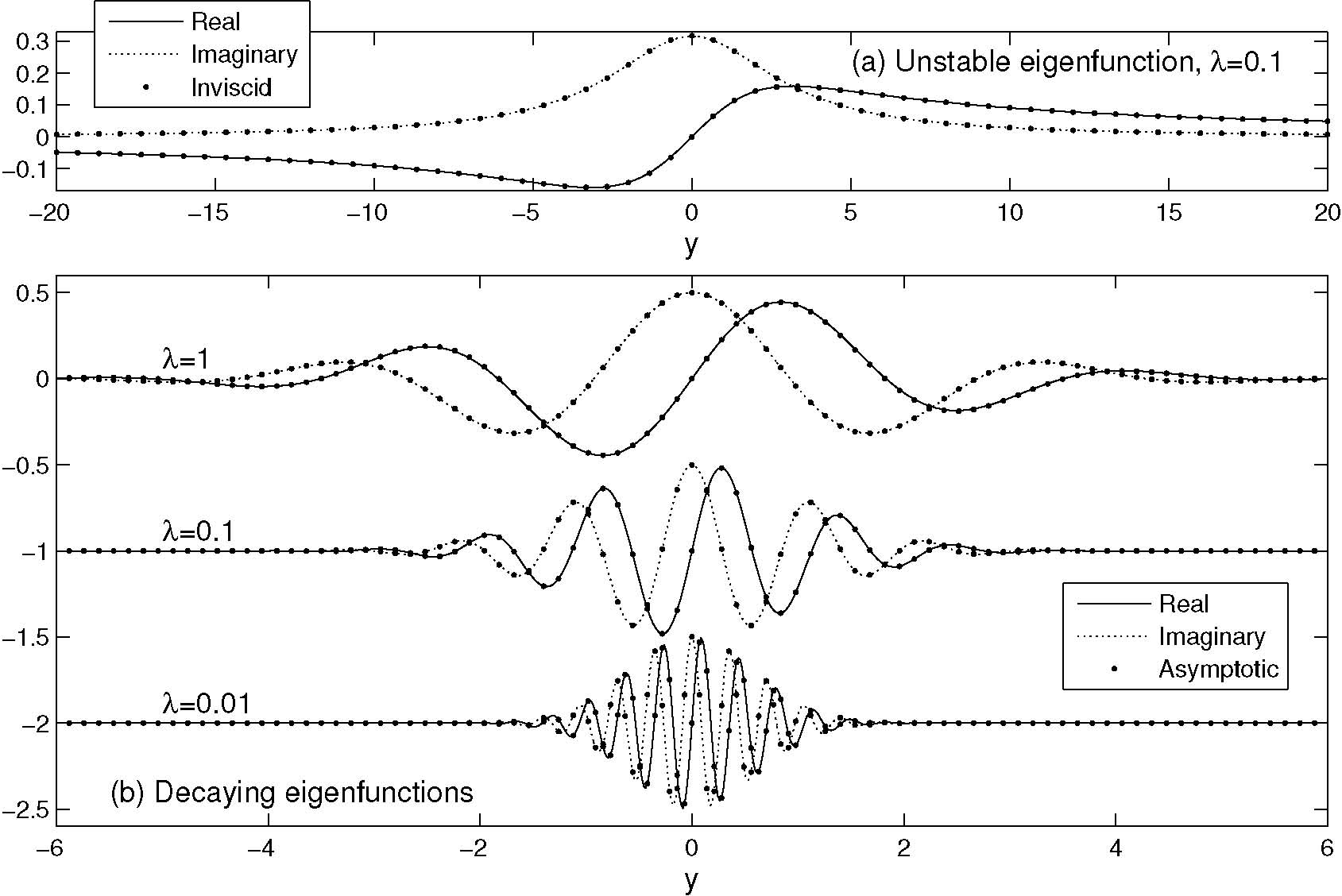}
\end{center}
\caption{
Eigenfunctions for $\kappa=0$.
Panel (a) shows the unstable mode for $\lambda=0.1$ and 0,
and panel (b) the decaying modes for
$\lambda=1$, $0.1$ and $0.01$. Also shown in  (b)
is the asymptotic result of Eq.~(\ref{j.7}).
}
\label{eigs}
\end{figure}

The eigenfunctions associated with the unstable and decaying eigenvalues
are illustrated in Fig.~\ref{eigs}.
The unstable viscous mode has a regular shape and smoothly limits to
the inviscid eigenfunction. However, the decaying viscous modes
oscillate spatially with a wavelength that becomes ever shorter as
the viscous parameter $\lambda$ is decreased. 
For $\lambda\ll1$, the approximate shape
of the decaying eigenfunctions can be found using the method of
stationary phase on the integral in (\ref{j.5}):
\begin{equation}
\zeta \sim
{\sqrt\pi (\kappa c_1-\gamma) \varphi\over \lambda^{1/4} (\yswm-c_1
)^{1/4}}
\exp \left[ {5\imi\pi\over8} - {2\over3} \exe^{\imi\pi/4}
\sqrt{(\yswm-c_1
)^{3}\over\lambda} \right]
.
\label{j.7}
\end{equation}
Clearly, this viscous mode has no regular inviscid limit.

\subsection{The initial-value problem and Landau damping}

The solution of the initial-value problem for $\zeta$ 
can be written in the
form,
\begin{equation}
\zeta = - \exe^{\imi \xswm} \int_0^\tswm [\kappa \amplsw_s(s) 
+ \imi \gamma \amplsw(s)]
\exe^{-\imi \yswm(\tswm-s) - \lambda (\tswm-s)^3/3} \dint s
+ \compconj
\label{j.8}
\end{equation}
Again, the integral over $\yswm$ generates a delta-function, allowing the
mode amplitude to be found, 
\begin{equation}
\amplsw = \amplsw_0\exe^{\Gamma \tswm} , \qquad
\Gamma={\pi\gamma \over (\ii\pi\kappa-1)}\,, 
\label{j.9}
\end{equation}
where $\Gamma$ is the viscous normal-mode eigenvalue.
The complete solution is then given by
\begin{equation}
\zeta =  { \amplsw_0 \exe^{\Gamma \tswm + \imi \xswm} \over \imi\pi\kappa-1}
\int_0^\tswm \exe^{-\Gamma s - \imi \yswm s - \lambda s^3/3} \dint s 
+ \compconj\,. 
\label{j.10}
\end{equation}

For $\lambda\to0$, Eq.~(\ref{j.10}) reduces to
\begin{equation}
\zeta =  {\ii\gamma A_0 \exe^{\ii \xswm} ( \exe^{\Gamma \tswm} 
- \exe^{-\ii \yswm\tswm} )
\over \pi\gamma-\pi\kappa \yswm - \ii \yswm}  + \compconj,
\label{j.11}
\end{equation}
Alternatively, for $t\gg1$, we may replace the upper limit
of the integral in (\ref{j.10}) by infinity to leading order, recovering
the viscous eigenfunction in (\ref{j.5}).
Thus, if the system is unstable, the solution
of the initial-value problem converges to the
inviscid or viscous normal mode. When the system is stable, 
on the other hand, the initial-value solution
converges to an exponentially decaying disturbance which is
equivalent to the decaying viscous mode. In the inviscid problem,
the decaying disturbance corresponds to a Landau pole or quasi-mode.
It is clear from (\ref{j.11}) that the quasi-mode has a form which
is not separable in space and time (the combination $i(\xswm-\yswm\tswm)$ 
appears), and consequently it cannot be a true normal mode.
Moreover, the quasi-mode is evidently the inviscid limit of the
decaying mode, as found in some related problems
\cite{stewartson81a,balmforth98a}.

Mathematically, Landau damping results from phase mixing in
integral superpositions of modes of the continuous spectrum.
The solution in (\ref{j.9}) and (\ref{j.11}) is, in fact, 
one such superposition. In (\ref{j.11}), we observe the
non-decaying, non-separable factor, $\exp \imi(x-yt)$, reflective of
the basic advection and the continuous spectrum
(see Sec. \ref{ssec:generaltheory}). This non-separable
structure of the superposition is also directly
responsible for Landau damping: the long-time
limit of integrals of the form,
\[
\int_{\R}\!\dint y \,\hat F(y) e^{\imi(x-yt)}
\, ,
\]
vanishes according to the Riemann-Lebesgue lemma \cite{stein},
provided $|F(y)|$ is integrable.  Moreover, in the case where 
$F(y)$ is analytic in a strip containing the real $y$ axis
(as is the case for (\ref{j.11})), 
the decay is exponential, and this the mathematical essence of 
Landau damping.  

\begin{figure}
\begin{center}
\leavevmode
\includegraphics[width=12 cm]{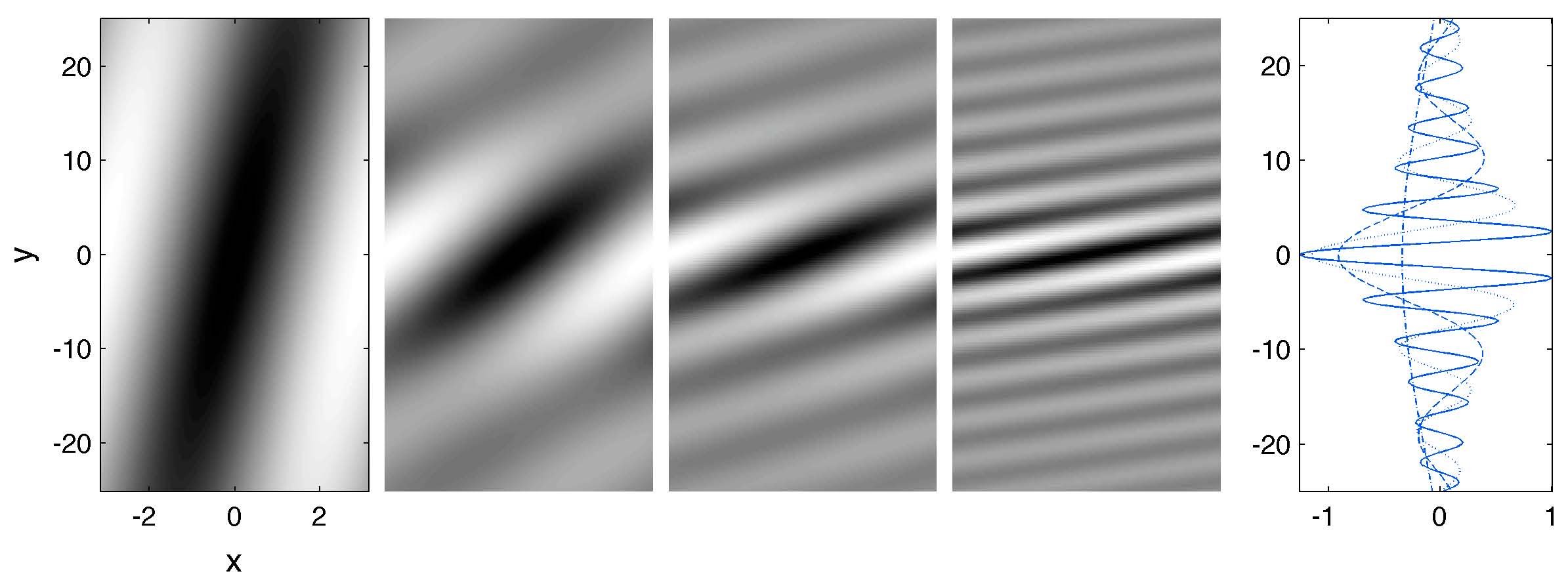}
\end{center}
\caption{
Snapshots of the solution to the linear version of the
single-wave model at the times, $t=0.1$, $0.4$, $0.75$, 
and $1.5$, showing the tilting of a
initially periodic disturbance in a stable flow
($\gamma=1$ and $\kappa=0$). The right-hand picture shows
the increasingly crenellated (phase mixed) distribution along $x=0$.
}
\label{pic2}
\end{figure}

Physically, the non-separable superposition of
continuum modes describes how a perturbation in the plasma or fluid tilts
over as a result of the action of particles orbiting at different
speeds or from the shearing flow of fluid elements.
A tilting, initially periodic, disturbance then evolves into
an increasingly crenellated (phase mixed) structure 
as shown in Fig.~\ref{pic2}.
There is no decay in the tilting process:
the plasma distribution function or fluid vorticity
is simply re-arranged. However, integral averages of
these quantities, such as the electric field or streamfunction,
do decay as a result of increasing cancellations. 
This does not imply that energy dissipation takes places;
merely that energy is transfered between the various components
of  the energy as shown in \textcite{MP92, balmforth2002a}. 

Note that, although exponential decay dominates the decay of the streamfunction
in the single-wave model, shear tilting normally leads to
algrabraic decay in fluid shear flows (e.g.\ \citealp{brown80a}).
A full solution of the linear initial-value problem for shear flow
does indeed uncover
both algebraic and exponential behavior \cite{briggs70a}. The
former disappears
in the single-wave expansion because it lies at higher order
in the asymptotic scheme. Thus, small disturbances
to stable flows would decay exponentially for a period, as
the quasi-modes appear and subside, but ultimately a slower, more
protracted algebraic decay sets in. This behavior is seen both
in solutions of the linear initial-value problem for
differentially rotating vortices, and in electron plasma
experiments designed to imitate them \cite{schecter2000a}.

\section{Normal form pattern formation}
\label{sec:normalpatterns}

We now consider the nonlinear patterns described by the single-wave model;
we first consider the unstable case 
($\gamma=-1$), and then discuss the stable case ($\gamma=1$).  

\subsection{Pattern formation via unstable modes}

\begin{figure}[t]
\centering
\includegraphics[height=9 cm]{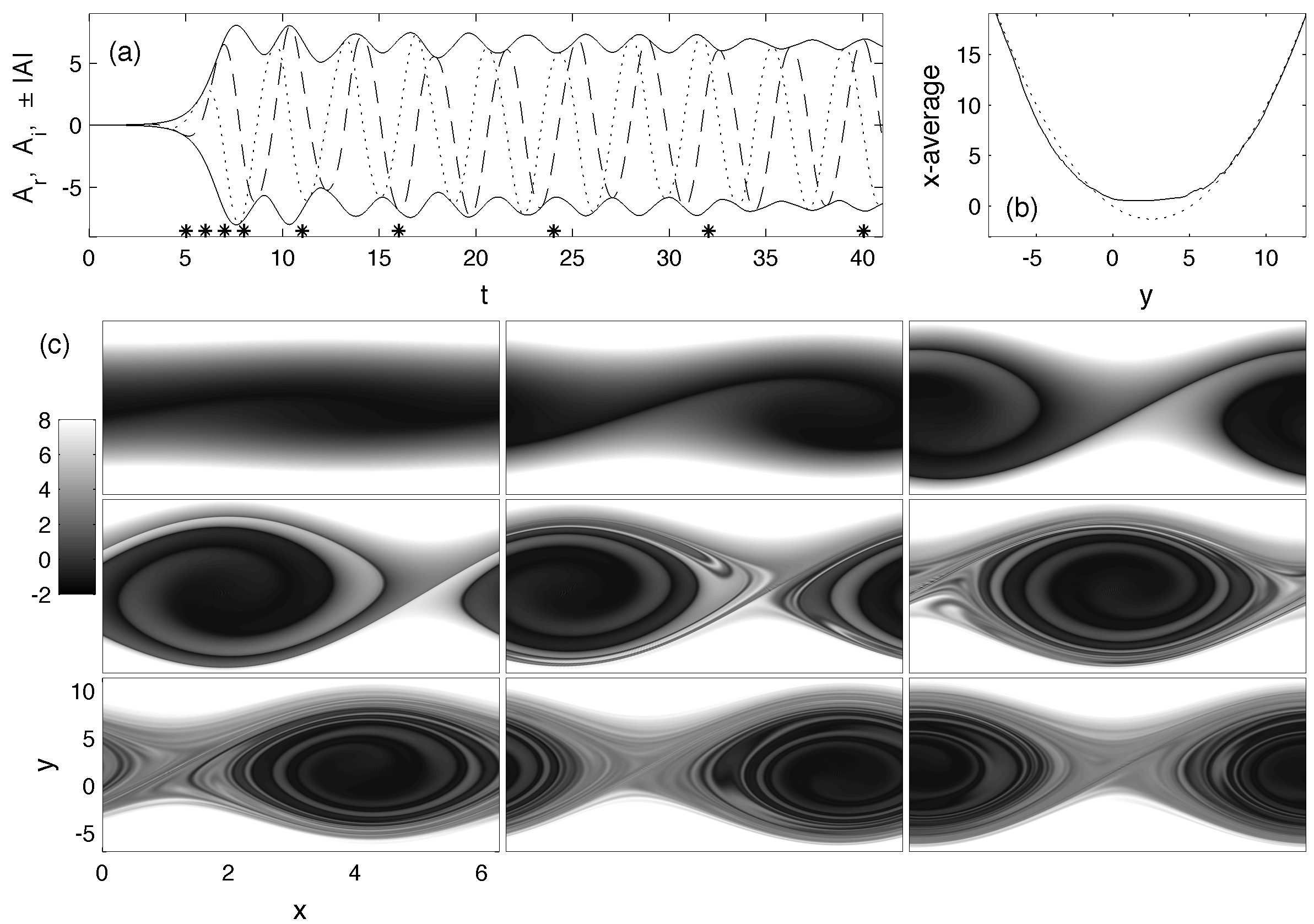}
\caption {
Computation of an unstable mode in the single-wave model
for $\gamma=-1$ and $\kappa=-0.4$.
(a) shows a time evolution of $A_r$ (dashed), $A_i$ (dotted)
and $\pm |A|$ (solid),
and (b) shows the initial and final $x$-averages of 
the total critical-layer distribution,
$Q=-\kappa \yswm^2/2+\gamma \yswm + \kappa\varphi + \zeta$.
The stars in (a) denote the times at which snapshots of $q$ are
shown in (c) as densities on the $(\xswm,\yswm)$-plane,
}
\label{swmex}
\end{figure}

A typical example of the nonlinear evolution of an unstable mode is shown in
Fig.~\ref{swmex}, computed numerically
(for a description of the scheme, see
 \citealp{balmforth2001a} and \citealp{balmforth2001b}).
Initially, $\amplsw(\tswm)$ grows exponentially as
predicted by linear theory. However, when the mode amplitude
reaches order one values, that growth is interrupted
and the linear instability saturates. The saturation occurs 
when the mode twists up the total critical-layer distribution $Q$ of (\ref{Q}) 
into a localized pattern. Beyond saturation, the mode continues to wind 
up the distribution, creating a well-defined cats-eye structure in $Q$ that    
corresponds to vortices in the fluid and electron holes in the plasma. 
The final $x$-averaged $Q$ in Fig.~\ref{swmex}
shows a suppression of the original equilibrium gradient at the center
of the critical region.
Thus, saturation occurs when the mode removes the
unstable mean gradients surrounding its critical level. Importantly,
the flattened equilibrium distribution is contained well within the
critical layer, and therefore occupies a much narrower region that the
area spanned by the unstable gradients of the equilibrium (e.g.
the bump in the plasma problem).
A complete suppression of those gradients 
requires much more extensive re-arrangements, which could only be
achieved if the mode amplitude saturated at a level
dictated by the Hopf scaling.

Overall, the dynamics culminating in the formation of the localized 
structures matches that seen  with  full
simulations.  The electron hole formation displayed in Fig.~\ref{fig:vlasov}  is  ubiquitous in  
Vlasov-Poisson simulations of unstable equilibria (see also, e.g., \citealp{cheng76a,Heath} 
and references therein).  Similarly, the cat's eye structures   seen in Fig.~\ref{fig:jet} are a 
common pattern that forms in simulations of unstable shear flow.  A benefit of the single-wave 
model is the attainment of  finer resolution with equivalent computational cost.

Saturation also does not produce a steady mode amplitude.
In fact, aperiodic ``trapping'' oscillations ensue (as they are
often called by plasma physicists, since they reflect oscillations
in the population of particles that are resonantly  trapped in the
potential created by the wave). The initial trapping
oscillations are connected to the first overturns of the 
level sets of $Q$. These oscillations subside as the localized structure (the cat's eye) becomes
well developed. The persistent trapping oscillations that occur
much later appear to correspond to the
sporadic formation of secondary structures within the localized structure that 
 are sometimes called ``macro-particles''
\cite{tennyson94a,delcastillo2002a}.
Unfortunately, numerical computations with finite resolution
are unable to determine the ultimate fate of the system
and its trapping oscillations. Phase mixing ideas (cf.\ 
\citealp{oneil65a,killworth85a,goldstein88b}) 
suggest that the secondary structures eventually become sheared out
and disappear to leave a smooth, steady cat's eye. 
However, numerical simulations do not show much sign
of the expected algebraic decay and numerical diffusion
obscures the true dynamics \cite{balmforth2001b}.

In fact, explicit dissipation like the viscous term in (\ref{eq:vswm1})
has an especially destructive effect on the cat's pattern:
when one adds viscosity, the rearrangements of the
distribution within the critical region begin to diffuse outwards.
Consequently, the cat's eyes spread and the critical layer
thickens continually with time. This allows the mode to access
fresh, unstable equilibrium gradients, prompting a secular
growth in $\amplsw(\tswm)$; asymptotic estimates and numerical 
computations indicate that $\amplsw\sim \tswm^{2/3}$ over long times 
\cite{churilov87a,goldstein88a,balmforth2001b}.
Secular growth continues until the system passes out of the
the single-wave regime. In this situation, the single-wave model only captures
an initial transient and the trapping scaling is eventually broken.

\subsection{Pattern formation in stable systems; nonlinear quasi-modes}
\label{ssec:stablepatterns}

In dissipative systems, the occurrence
of instability and the resulting dynamics is usually the
most interesting problem; stable systems relax back to
equilibrium exponentially quickly and the only complication
that can occur is if there are multiple possible end-states
that ``frustrate'' the evolution of the system.
Nondissipative systems like the pendulum, on the other hand, have 
an infinite number of possible periodic orbits adjacent to
the stable equilibrium position, a general property of finite-dimensional 
Hamiltonian systems  (see, e.g.,  \citealp{moser76}),  and the Darwinian 
choice amongst them is determined by the energy of the initial condition.
Stable ideal plasmas and inviscid shears also support an infinite 
number of adjacent wave-like nonlinear states. For Vlasov plasmas, the 
explicit construction of such states (the so-called BGK modes) dates back to
 \textcite{bernstein58a}; in shear flows, Kelvin-Stuart cat's eye patterns \cite{drazin81a}
are other examples. Nevertheless, 
despite this wealth of equilibria and the absence of dissipation,
linear theory predicts that stable ideal plasma and inviscid shear flow
relax through Landau damping. 
However, using the single-wave model, we now illustrate how sufficiently
nonlinear perturbations do not continually decay.

\begin{figure}
\begin{center}
\leavevmode
\includegraphics[height=9.6 cm]{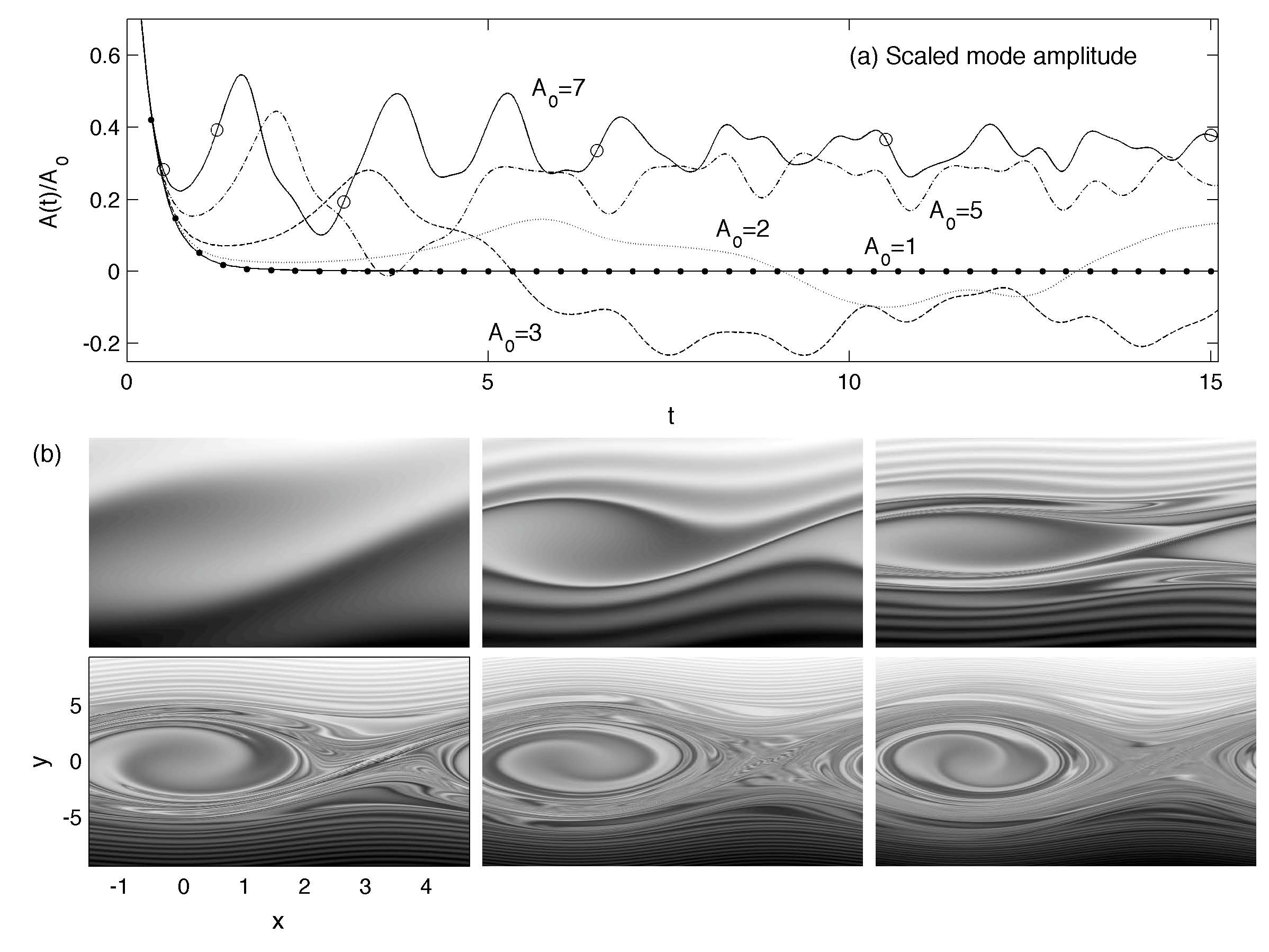}
\end{center}
\caption{
Computation of a stable mode in the single-wave model
for  $\gamma=+1$ and $\kappa=0$. Panel (a) shows
scaled mode amplitudes, $\amplsw(t)/\amplsw_0$,
for perturbations with different initial amplitude, $\amplsw_0$.
The dots show the trend of Landau damping. Panel (b) shows 
the evolving cat's eye pattern for $A(0)=7$ at the times indicated
by the circles in (a). In the particular case
$\kappa=0$, the single-wave model possesses solutions 
with a real mode amplitude, $\amplsw(\tswm)$, and the reflection symmetry,
$\yswm\to-\yswm$, $\xswm\to-\xswm$,  and $\zeta\to-\zeta$.
}
\label{quasimodo}
\end{figure}

Numerical solutions of the nonlinear initial-value problem
with $\gamma=+1$ and $\kappa=0$ and differing initial amplitudes
are shown in Fig.~\ref{quasimodo}. 
When the initial amplitude is sufficiently low, the numerical
solutions do indeed display increasing crenellation of the
distribution and the mode amplitude decays as
predicted by linear theory. However,
for larger initial amplitudes,
the damping becomes interrupted, and the mode amplitude returns to
larger values where it fluctuates unsteadily in a
sequence of aperiodic trapping oscillations.

The interruption of Landau damping corresponds to the creation of another
cat's eye pattern. Thus,
sufficiently strong nonlinear perturbations realize Kelvin's
objections (see Sec.~\ref{ssec:raleigh}) to inviscid shear flow dynamics.
In the plasma context, the interruption of Landau damping
has been observed in many computations with the Vlasov equation and
in experiments (e.g., 
\citealp{franklin72a,sugihara72a,feix94a,manfredi97a,lancelotti,brunetti00,danielson04a}),
and the nonlinear structures are
assumed to converge to steady BGK modes.
However, as for the unstable problem, the trapping oscillations persist
throughout the computations in Fig.~\ref{quasimodo}, 
and the ultimate fate of the system is again unclear.

The critical threshold in $\amplsw_0$
below which Landau damping continues unabated
occurs for initial amplitudes of about unity.  However,
this threshold does not provide a clean criterion for the
formation of a cat's eye because for slightly larger initial amplitudes
(e.g.\ $\amplsw_0=2$ and 3 in Fig.~\ref{quasimodo}),
the mode amplitude repeatedly switches sign. This implies that the mode twists up
the distribution in one direction for a period, but then
the sense of overturning switches and the mode unwinds the twisted
background; it is not clear whether a coherent cat's eye ever forms.
Only for somewhat larger initial amplitudes ($\amplsw_0=5$ and 7 in 
Fig.~\ref{quasimodo}) does the mode amplitude remain of one sign and
the critical-layer distribution winds  up continually into a cat's eye.

\begin{figure}
\begin{center}
(a) 
\leavevmode
\includegraphics[width=8 cm]{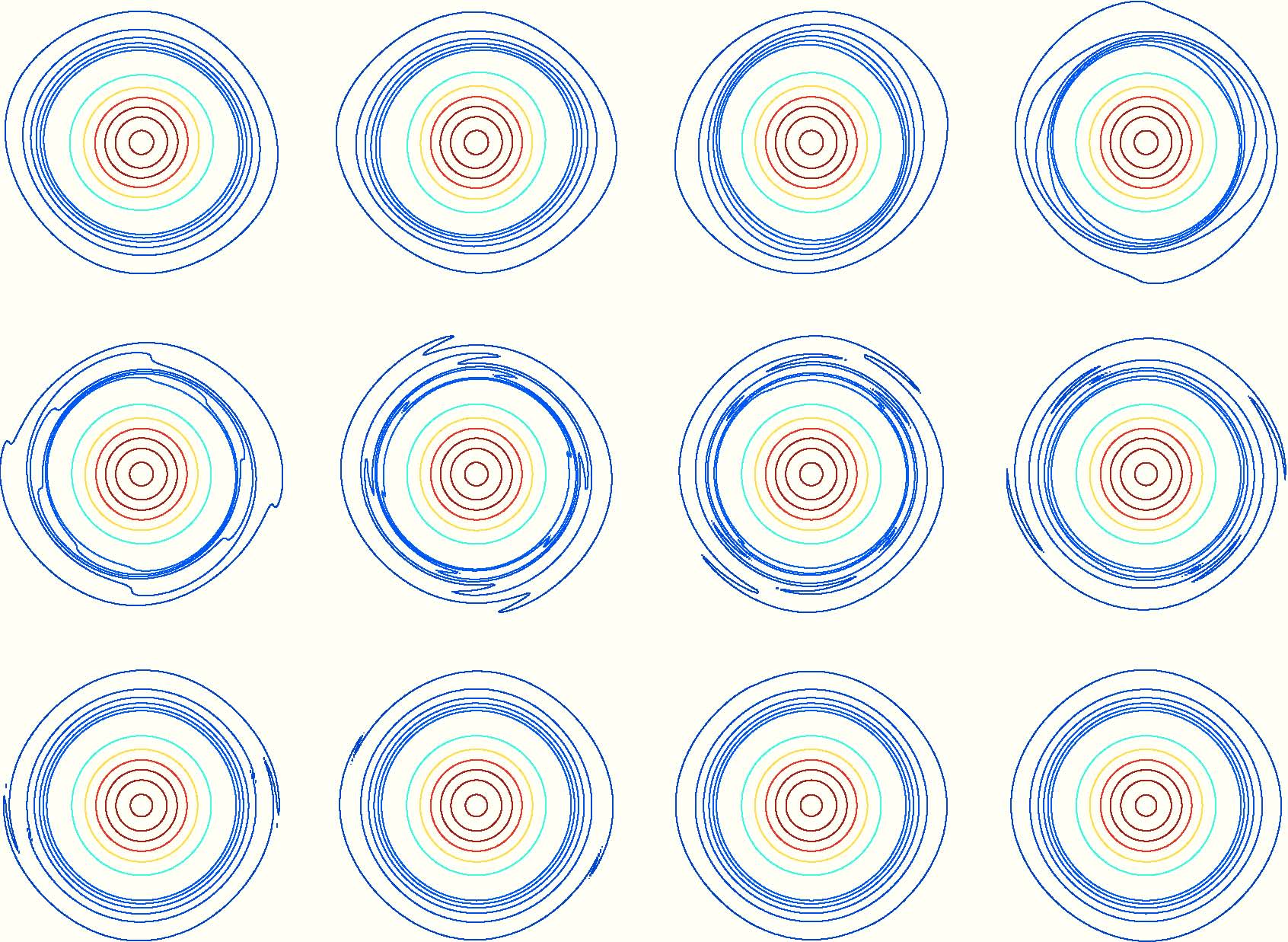}
\end{center}
\begin{center}
(b) 
\includegraphics[width=8 cm]{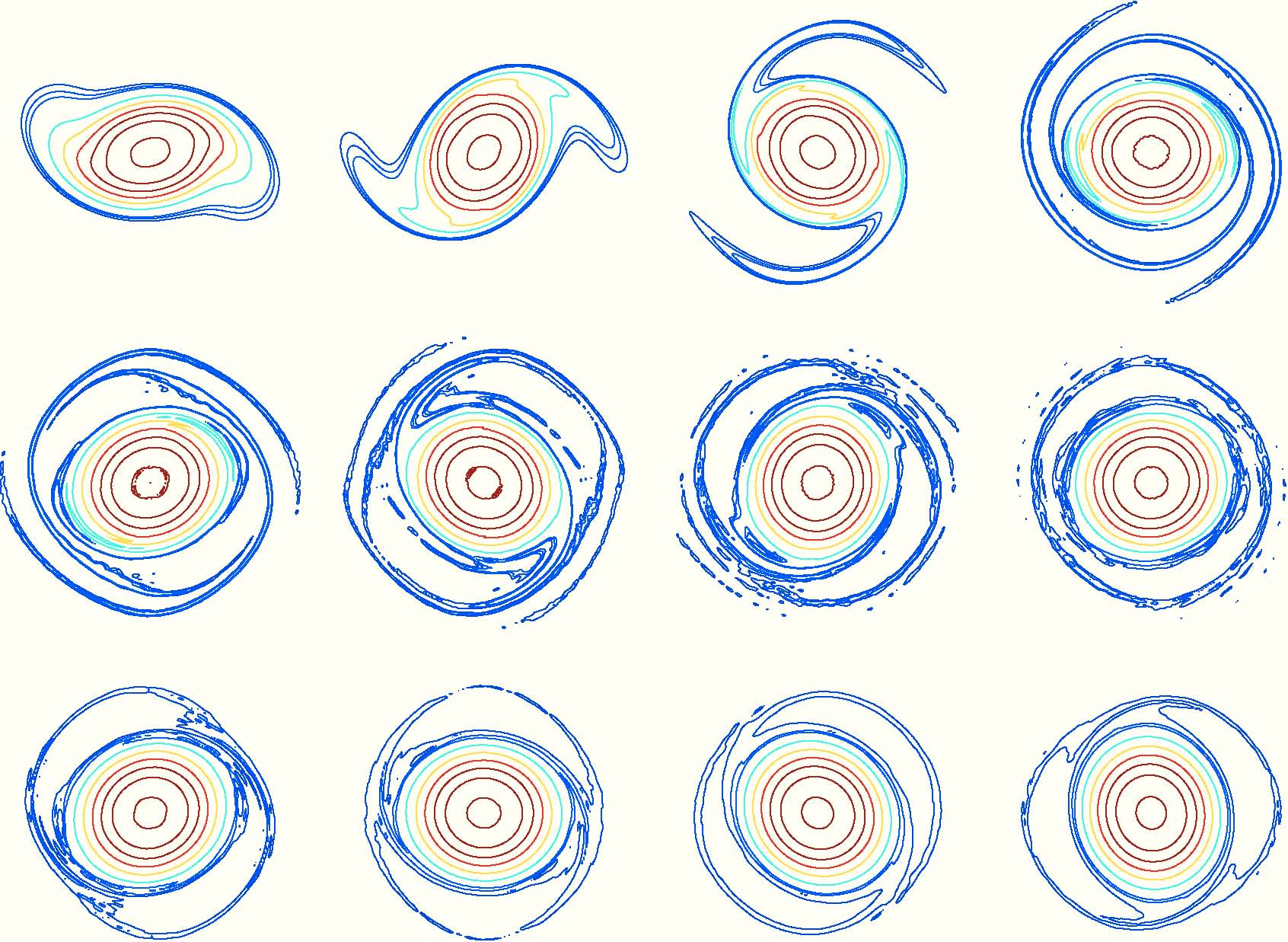}
\end{center}
\caption{Simulation results of elliptical perturbations to a stable (Gaussian) vortex.
Shown are contours of constant vorticity. In panel (a) the perturbation is small and 
the vortex  axisymmetrizes, while in panel (b) the perturbation is above a 
threshold for the formation of a multipole state. 
}
\label{vux}
\end{figure}

With a simple reinterpretation of variables, the single-wave model can be applied 
to problems in polar coordinates (cf.\ Sec.~\ref{ssec:shear_flow}).  Thus, thresholds 
observed in the single-wave model apply to the formation of nonlinear structures 
in desymmetrized vortices, both in numerical computations \cite{rossi97a,montgomery99} 
and in non-neutral plasma experiments \cite{driscoll90a}.  The interest in these thresholds  
was sparked off by the idea that Landau damping
(or, more generally, ``continuum'' damping, which includes
algebraically decaying perturbations) could act to
cause disturbed vortices to return to axisymmetry, 
rationalizing the preponderance of almost circular vortices in
two-dimensional turbulence and atmospheric weather systems. 
However, if the vortex is too distorted,
then it does not return to axisymmetry,  a 
cat's eye develops and a multipolar vortex forms. Two
numerical simulations of the two-dimensional Euler  equations,  
showing the evolution of an elliptically distorted,
Gaussian vortex are shown in Fig.~\ref{vux}. 
The more weakly distorted vortex displays sub-threshold axisymmetrization,
whereas the larger distortion creates a multipole.
By modeling the disturbance as the response of a weakly nonlinear
quasi-mode and then deriving an appropriate version of the  single-wave model to describe the
dynamics,  \textcite{balmforth2001a} predicted that the
critical amplitude threshold translates to an initial aspect ratio
of about $1.1$ (a ten percent departure from
circularity); the two simulations in Fig.~\ref{vux}
do indeed straddle this value.  
\textcite{turner07,turner08} present a much more thorough
analysis of the threshold for desymmetrized vortices.

\section{XY model; patterns of degenerate form}
\label{sec:degeneratepatterns}

As noted in Sec.~\ref{sssec:degenerate} the XY model 
(also called the {\it Hamiltonian mean field model} by \textcite{campa} 
and others) provides an example of the 
degenerate form of the single-wave model; in this section
we explore the pattern formation described by both the XY model
and its degenerate single-wave form near onset.
For the task, we consider the Gaussian family of equilibrium profiles,
\beq 
R=R_0+\varepsilon R_1 =
\frac{1+\varepsilon a_1}{\sqrt{2\pi}} \exp\left(-\half I^2\right).
\eeq
For this profile,
$I_*=0$, $R'_{1*}=0\Rightarrow \hat \gamma=0$.
Moreover, $\Omega_*'=1$ and 
(because this example has a spatial kernel) $\eta=1$,
indicating
\bq
\nu = \pvint_{\R}\!\!\dint I\,   \frac{R_1'}{I} = a_1 .
\eq
The degenerate form of the single-wave model for this model is therefore
\beq
\zeta_t + y \zeta_x + \varphi_x \zeta_y = - \kappa \varphi_t\,,
\qquad
A = \langle e^{-\ii x}  \zeta \rangle\,,
\label{hmf.9}
\eeq
where $\kappa=-{\rm sgn}(\nu /R_{0*}'')$. 

In Sec.~\ref{sec:linearsinglewave}, we established that 
the system is unstable when $\kappa=-1$,
and the growth rate is $\pi^{-1}$, or $- \Omega_*' \nu / (\pi R_{0*}'')
\longrightarrow (a-1)\sqrt{2/\pi}$, in terms of the original variables.
When $\kappa=+1$, the system is linearly stable and small perturbations
decay through Landau damping; again, the exponent
is $(a-1)\sqrt{2/\pi}$ for the original variables.
Note that, except for the two choices, $\kappa=\pm1$, 
determining the stability of the system, the degenerate form of the 
single-wave model of (\ref{hmf.9}) has no free parameters.

\subsection{Unstable modes and degenerate phase-space patterns}

Consider now  patterns that occur in the XY model with the Gaussian
equilibrium profile by  increasing  the amplitude parameter $a$ through unity to trigger instability.
We first solve the  XY model numerically by using the operator splitting 
scheme of \textcite{cheng76a}.  We begin the simulation by  kicking the 
equilibrium state by adding the forcing,  $10^{-4} t e^{-10t^2} \cos\theta$,
to the potential $\Phi$ in (\ref{hmf.2}). This procedure prepares an initial 
condition for the dynamics after the forcing term decays to zero.  
Such prepared initial conditions,
were  called {\it dynamically accessible} in  \textcite{mp89, mp90,MP92}, 
where they were advocated because they  preserve phase space constraints. 

\begin{figure}[t]
\centering
\includegraphics[height=7 cm]{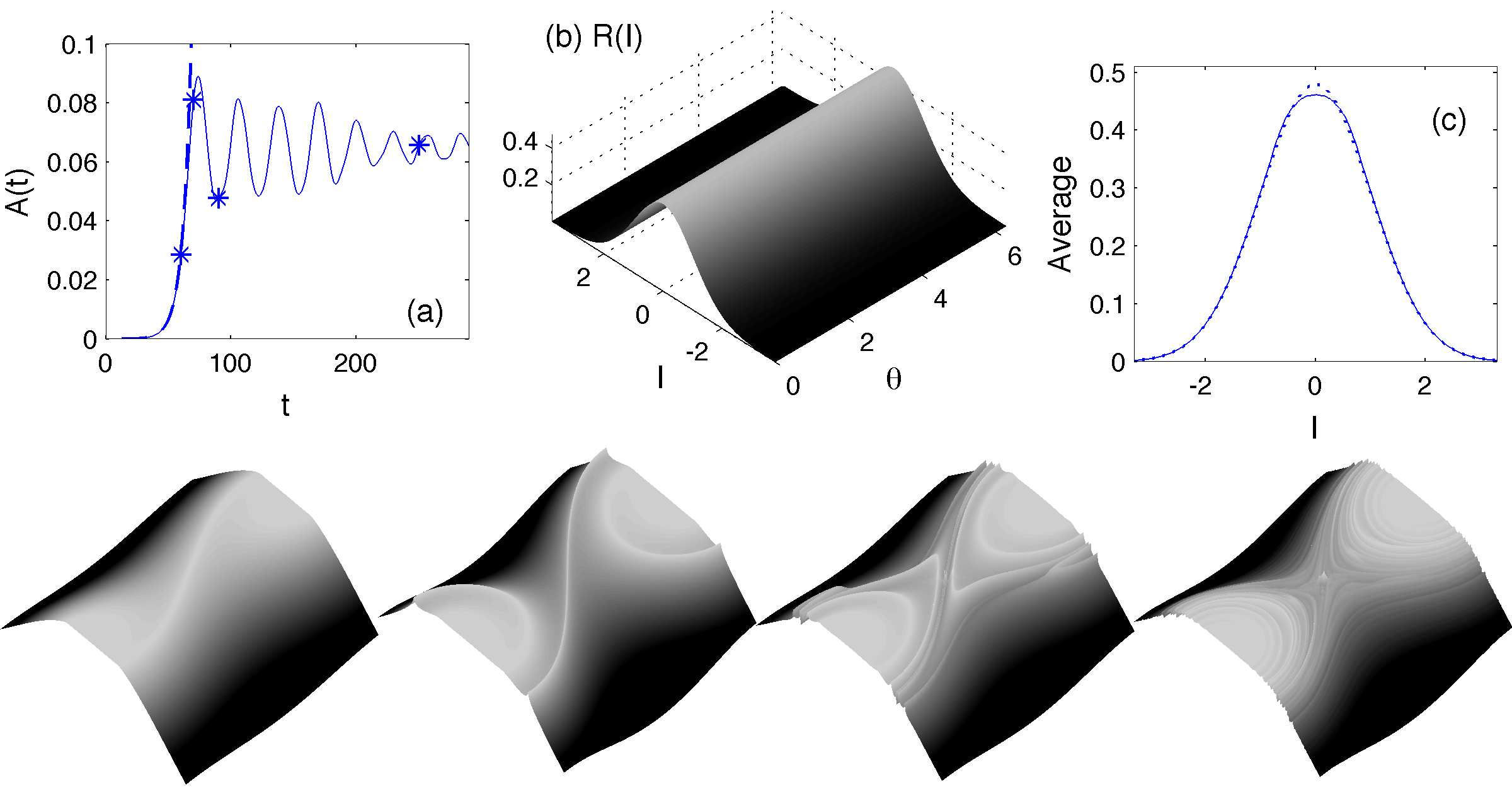}
\caption{
Kicked (dynamically accessible) solution of the XY model 
of (\ref{hmf.9}) for the Gaussian 
equilibrium distribution of  (\ref{hmf.4}) with  $a=1.2$.
Panel (a) shows the mode amplitude, $\ampl(t)$, along with the expected linear
growth (dashed line). Panel (b) shows the equilibrium distribution as 
a surface over the $(\theta,I)$-plane.
Panel (c) shows the initial and final 
$\theta$-averages of the total density. 
The stars in (a) indicate the times of the
snapshots of the  total density $R(I)+\rho(\theta,I,t)$ 
shown in the lower row as surfaces over 
the $(\theta,I)$-plane, focussing on the central maximum. 
}
\label{hmfpic}
\end{figure}

Figure \ref{hmfpic} shows results for such a kick-started simulation with 
$a=1.2$.  Specifically, Fig.~\ref{hmfpic}(a) displays  the temporal behavior 
of the amplitude, $\ampl(t)$, of the potential, defined in
\beq
\Phi = \ampl(t) e^{\ii\theta} + c.c.,
\label{hmf.5}
\eeq
along with the 
behavior of linear growth (dashed line), for the Gaussian equilibrium 
depicted in Fig.~\ref{hmfpic}(b).  Figure \ref{hmfpic}(a)  shows the 
solution initially following the linear growth, followed by  oscillations 
in $\ampl(t)$ that are symptomatic of the  twist-up of 
$\rho(\theta,I,t)$ into a cat's eye pattern around the peak of the 
distribution.  
In Fig.~\ref{hmfpic}(c) we see that the corresponding average in $\theta$ 
of the total distribution  $R(I)+\rho(\theta,I,t)$ is thereby flattened, 
suggesting how the instability is able to saturate itself nonlinearly.   
Not surprisingly, beyond saturation, trapping oscillations continue  
in $\ampl(t)$.  Figure \ref{hmfpic}(d) depicts snapshots of the total 
distribution at the times indicated by the stars in Fig.~\ref{hmfpic}(a),
and shows how cat's eyes develop during saturation. 

\begin{figure}[t]
\centering
\includegraphics[height=5cm]{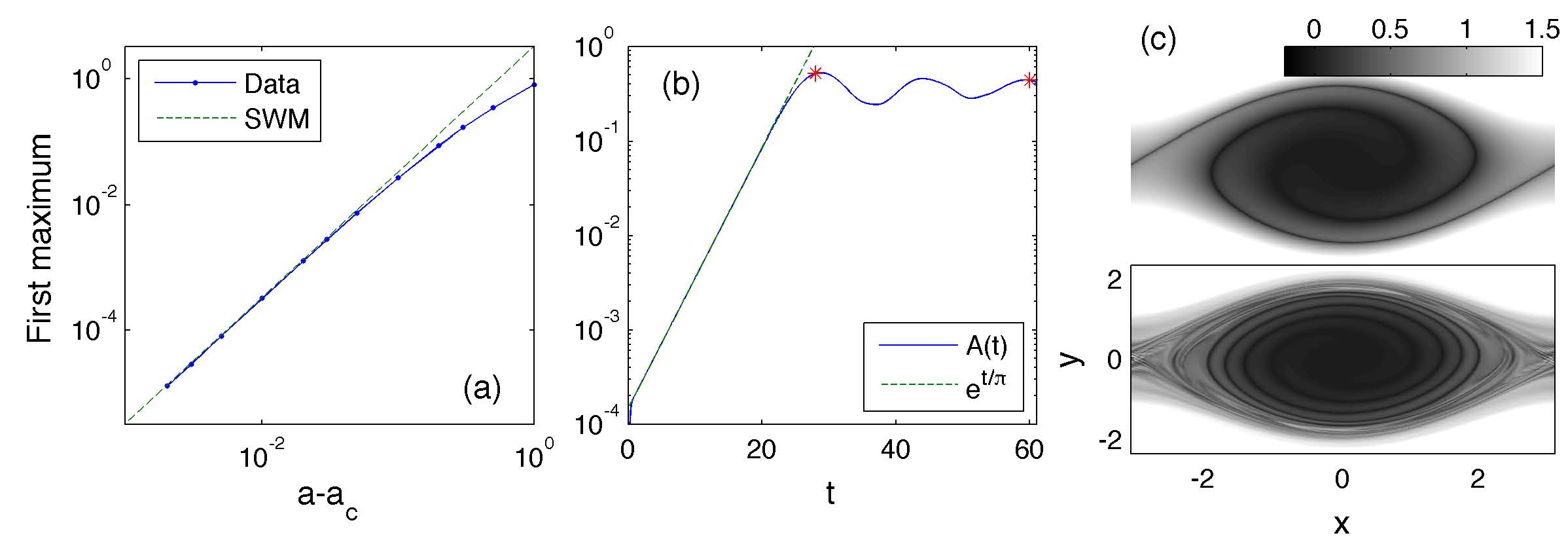}
\caption{
Numerical results for the XY model with Gaussian equilibrium and kick 
start (cf.\ Fig.~\ref{hmfpic}).  Panel (a) shows scaling data for the 
XY  model, with the height of the first maximum in $\ampl(t)$,  the amplitude
of the potential, plotted against the control parameter 
$a-1$. Also shown is the trapping scaling prediction of the single-wave 
model (SWM, dashed line).  Panels (b) and (c) show the numerical
solution of the degenerate single-wave model:
panel (b) shows the mode amplitude, 
$A(t)$, vs.\ time, $t$, along with the expected linear growth (dashed line); 
the stars of this panel mark times for the snapshots of the 
solution, $Q=\half y^2 + \zeta - \varphi$, 
shown as densities on the $(\theta,I)$-plane in panel (c).
}
\label{maxi}
\end{figure}

Figure \ref{maxi}(a) shows scaling data obtained from a suite of computations 
of the type depicted in Fig.~\ref{hmfpic};
the first maximum of $\ampl(t)$ is plotted 
(cf.  Fig.~\ref{hmfpic}(a)) against the control parameter $a$.  
This figure verifies the trapping scaling of the saturation level.

The remaining two panels of Figure \ref{maxi} show the numerical solution
of the degenerate form of the single-wave model in
(\ref{dvarphit})--(\ref{degenerg}). The system is again kicked into
action by suitably forcing the potential $\varphi$.
The unstable mode grows and saturates in a similar manner to
the dynamics seen for the full XY model in Fig.~\ref{hmfpic}.
The mode amplitude, $A(t)$, peaks
first for $A_m\approx0.52$, which leads to a prediction for
the first maximum of $\ampl(t)$ in terms of the original variables:
$\ampl_{max}\approx2\pi A_m (a-1)^2$. 
This prediction is included in Fig.~\ref{maxi}(a);
the scaling data for the amplitude 
maxima from the full XY model computations
converge satisfyingly to the prediction.
Note that trapping scaling 
means that the rearrangment of $\rho(\theta,I,t)$
is unable to suppress the distribution to 
the point where the $\theta$-average shown in Fig.~\ref{hmfpic}(c)
corresponds to a linearly stable equilibrium.

\subsection{Patterns in stable systems; nonlinear Landau damping in the XY model}

The parallels between the XY model and the Vlasov equation, together with the 
fact that both can be reduced to the single-wave model in the vicinity of 
onset, suggests that the XY dynamics should also display pattern formation
even when $a<1$, provided the system is forced with sufficient
strength. That is, the XY model should display a nonlinear arrest of 
Landau damping akin to that discussed in Sec.~\ref{ssec:stablepatterns}. 
To confirm this expectation we once more
solve the initial-value problem for the full XY model, again 
initiating the computations by applying a suitable kick to the potential.

\begin{figure}[t]
\centering
\includegraphics[width=14cm]{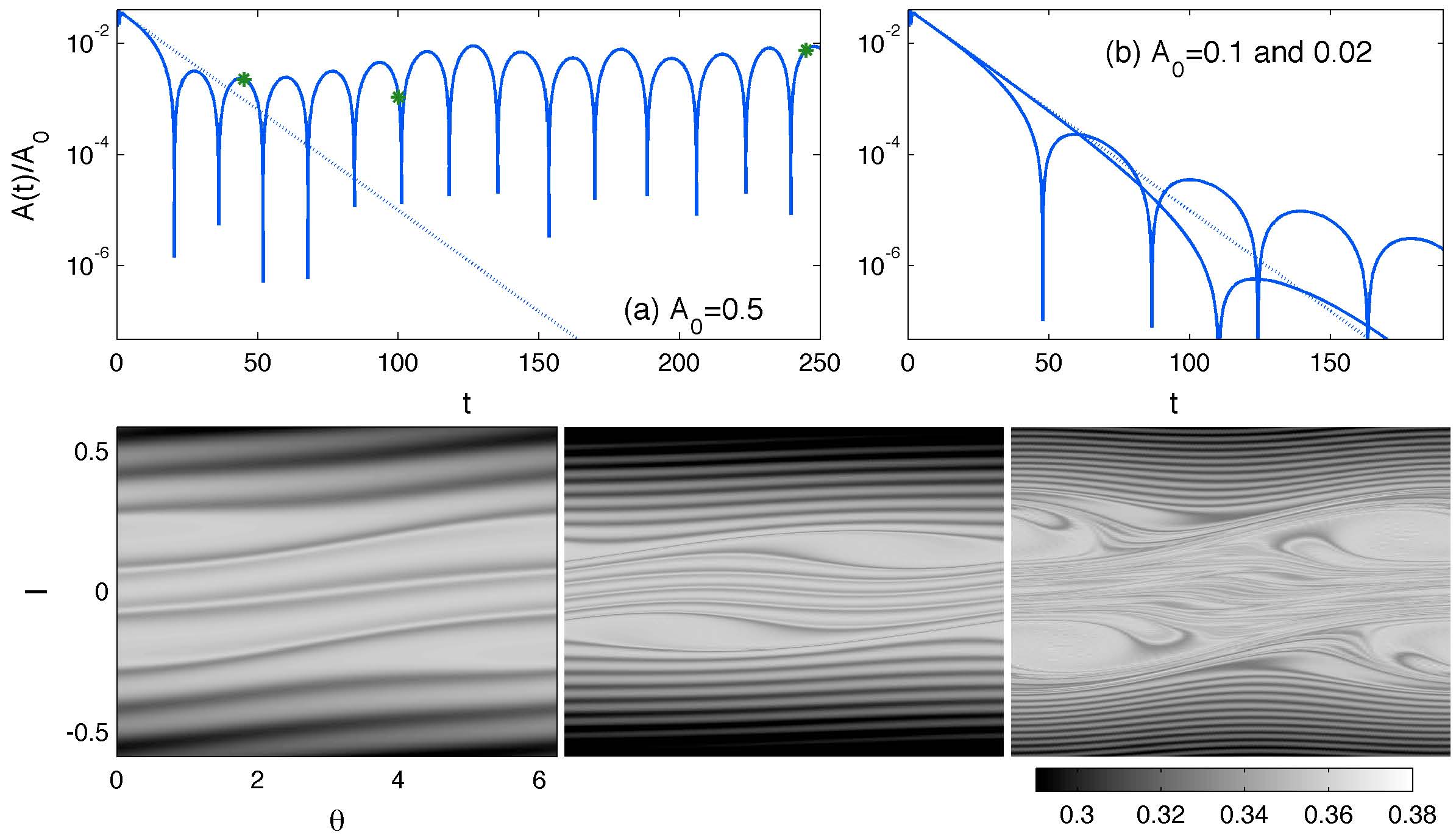}
\caption{
Solutions to the nonlinear XY  model for $a=0.9$, initiated by
adding the external forcing (\ref{xydrive})
to the potential $\Phi(\theta,t)$. 
In panel (a) $\ampl_0=0.5$, and the stars indicates the
times at which snapshots of $\rho(\theta,I,t)$ are shown below in panel (c) 
as densities on the $(\theta,I)$-plane.  In  panel (b),  the smaller values 
of $\ampl_0=0.1$ and $0.02$ are used, and longer intervals of Landau
damping are observed.
The dotted lines in (a) and (b) indicate the  solution to the linear problem.
}
\label{hmfs}
\end{figure}

Figure \ref{hmfs} shows a computation
in which a stable Gaussian equilibrium with $a=0.9$
is forced by adding to $\Phi(\theta,t)$ the external potential,
\bq
\Phi_e=2 \ampl_0 t e^{-10t^2} \cos\theta \, .
\label{xydrive}
\eq
In Fig.~\ref{hmfs}(a) a plot of the logarithm of $\ampl$ vs $t$  is shown for 
a drive amplitude $\ampl_0=0.5$. This plot shows how the initial decay becomes 
interrupted for a sufficiently large initial perturbation, as in
Vlasov-Poisson simulations (e.g., \citealp{cheng76a,Heath}).
Fig.~\ref{hmfs}(c) displays plots of the total density 
at the times indicated by the stars in Fig.~\ref{hmfs}(a);
the interruption of Landau damping is caused by trapping and the emergence
of the nonlinear quasi-modes visible in these images. These 
cat's eye structures are symmetrically
displaced about the maximum of the original equilibrium profile 
at $I=0$. Reducing the amplitude of the initial kick as in 
Fig.~\ref{hmfs}(b) prolongs the
transient in which the system decays through Landau damping, but we have
not attempted to compute any threshold below which that damping
continues unabated. Further explorations of nonlinear Landau damping 
in the XY model are presented by \textcite{barre09}.

\subsection{Smoothing the top-hat profile}

The XY model is convenient for exploring the differences  
between the dynamics underlying
trapping and Hopf scaling.  In this section we explore these differences
in detail by considering
the instabilities of a family of equilibrium profiles in
which we smooth out
the discontinuities of the top-hat profile of (\ref{hmf.1}),
As discussed in Sec.~\ref{ssec:xylinear},
the top-hat profile (analogous to a broken-line shear flow
or ``water-bag'' Vlasov equilibria) 
has a bifurcation to instability at $a=1$ with $c=0$, much
like the Gaussian profile. However, we argued earlier
(Sec. \ref{ssec:weakanalysis}), the instabilities of
the top-hat profile are expected to be affected by nonlinearity at 
strengths characterized by the Hopf scaling and be {\sl subcritical}.
In other words, the transition to instability for the top-hat profile
is quite different from that for a smooth equilibrium profile like
the Gaussian.

The family of equilibrium profiles we adopt are given by
\beq
R = \frac{a}{4}\left[
\tanh\left(\frac{I+1}{\Upsilon}\right)
-\tanh\left(\frac{I-1}{\Upsilon}\right)\right]\,,
\label{tophat}
\eeq
where $\Upsilon$ is a parameter controlling the narrowness of the
hyperbolic tangent functions. When $\Upsilon$ is decreased from
order one values, the steps in the profile at $I=\pm1$ sharpen,
and for $\Upsilon\to0$ limit to the discontinuities of the top-hat.
Using these profiles, we again conduct suites of initial-value problems;
as for the Gaussian examples above, we begin these
computations  with $\rho(\theta,I,0)=0$ and kick the system into action 
by adding the forcing (\ref{xydrive}) to the potential, $\Phi(\theta,t)$,
with $\ampl_0$ taken to be small ($10^{-4}$), at least initially. 

\begin{figure}[t]
\centering
\includegraphics[height=6 cm]{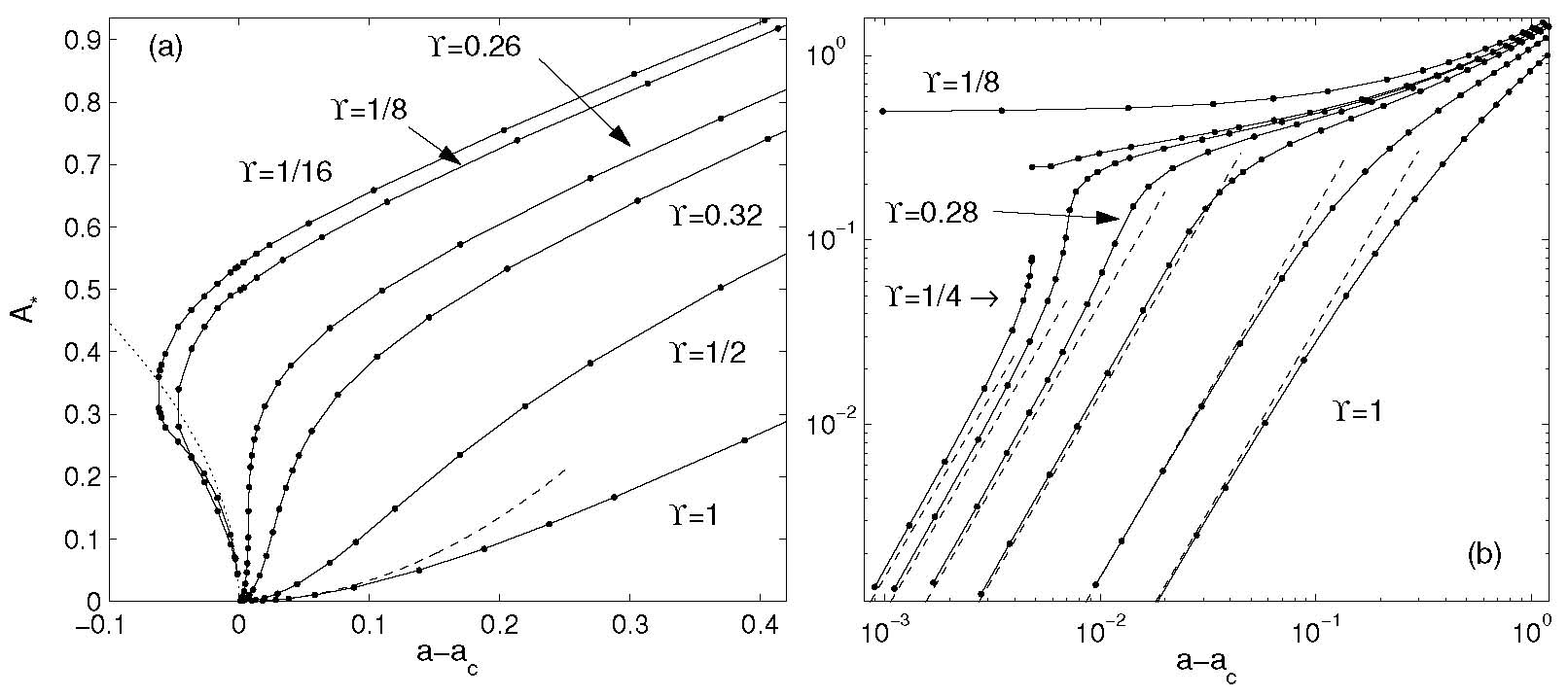}
\caption{
Scaling data for the first maximum, $\ampl_*$, plotted against distance to the
stability boundary, $a-a_c$, for the smoothed top-hat profiles
of  (\ref{tophat}) for various values of the
smoothing parameter, $\Upsilon$.  Panel (a) shows data for
$\Upsilon=1/16$, $1/8$, $0.26$, $.32$, $1/2$ and 1.  Panel 
(b) plots the same data logarithmically (omitting $\Upsilon=1/16$), 
and includes additional
data for $\Upsilon=0.25$ and $0.28$. The dashed lines indicate the
prediction of the single-wave model, and the dotted curve in (a) is the
subcritical weakly nonlinear solution expected for the top-hat.
}
\label{scallywag}
\end{figure}

Figure \ref{scallywag} displays
scaling data for the smoothed top-hat profiles of (\ref{tophat}).
For $\Upsilon>1/4$, the first maximum, $\ampl_*$,
follows the trapping scaling predicted by the single-wave model
sufficiently close to the stability boundary (see panel (b)).
However, as the profile becomes sharper, the trapping prediction for 
$\ampl_*$ increases exponentially quickly (with $\ampl_*$ proportional
to $(R_{0*}'')^{-2}\propto \Upsilon^2\cosh^4(1/\Upsilon)$), and the
system is unable to remain in the trapping regime much beyond
onset. Instead, the saturation level flattens off quickly beyond the 
trapping regime and $\ampl_*$
approaches a different scaling that we interpret to be  
more like that for a genuine top-hat profile (e.g. $\Upsilon=0.28$). 
Indeed, for the lowest values of $\Upsilon$ (e.g.\ $\Upsilon=1/8$),
the trapping regime is no longer observed and $\ampl_*$ remains
finite as $a\to a_c$, which is
consistent with the subcritical bifurcation of the top hat profile.

The detection of a finite-amplitude branch 
of solutions for $a\geq a_c$ with low $\Upsilon$ suggests that
an unstable subcritical branch is born in the bifurcation at $a=a_c$,
and that this branch subsequently turns around at smaller values of $a$, 
That is, there must be a second, saddle-node bifurcation for $a<a_c$,
which generates the larger-amplitude stable solutions that return
to higher $a$ and provide the end-states of the initial-value
computations. Thus, for the top-hat profile, or its relatively sharp
smooth relatives of  (\ref{tophat}) with $\Upsilon\ll1$,
the instability has an abrupt onset.

For intermediate values of $\Upsilon$ ({\it e.g.} $\Upsilon=1/4$), the 
system first follows the trapping scaling  and, therefore, initially shows a
relatively smooth transition. However, the branch's amplitude subsequently 
increases quickly and becomes vertical. The branch then disappears
(no doubt by turning back on itself), leaving the system to grow to
higher amplitudes and achieve a first maximum closer to that of the
corresponding top-hat solution. The
amplitude data in Fig.~\ref{scallywag}(b) is thereby rendered discontinuous.
In other words, although the profile with $\Upsilon=1/4$ initially
suffers a second-order phase transition following the trapping scaling,
the transition can appear to become first-order if the system is 
pushed too far beyond the single-wave regime.

\begin{figure}[t]
\centering
\includegraphics[height=5 cm]{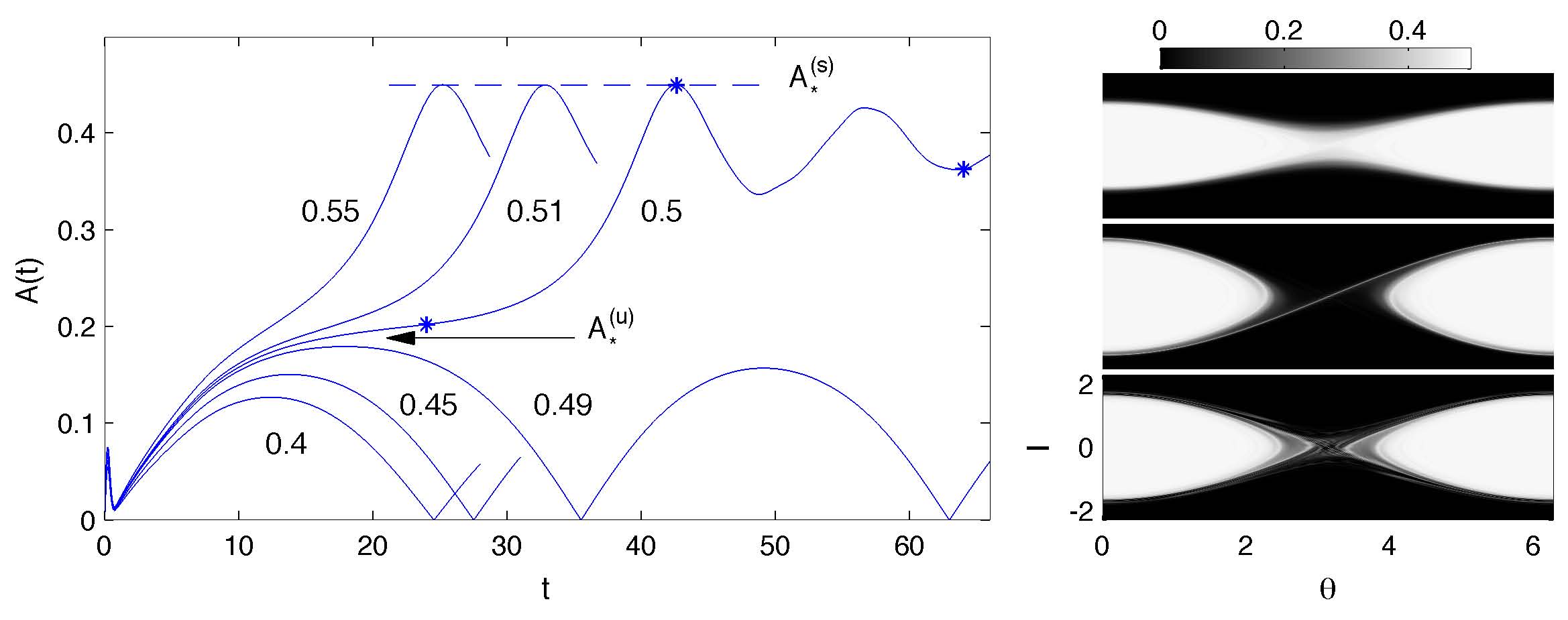}
\caption{
Panel (a) depicts $\ampl(t)$, for $a=0.96<a_c\approx0.986$ and $\Upsilon=1/8$,
for computations kicked with different amplitudes $\ampl_0$ for the forcing
(\ref{xydrive}) added to $\Phi(\theta,t)$.  
The  values of $\ampl_0$  are as shown and the ``stable'' and ``unstable'' 
first maxima, $\ampl_*^{(s)}$ and $\ampl_*^{(u)}$, are also indicated.  
For $\ampl_0=0.5$, the stars indicate
the times of the three snapshots of $\rho(\theta,T,t)$ shown in panel (b)
 to the right.
}
\label{subo}
\end{figure}

Because we begin the initial-value computations
from a weakly perturbed equilibrium profile 
parameterized by $a$, the solutions do not
converge to any finite amplitude solutions for $a<a_c$,
but Landau damp back to the unpatterned equilibrium.
Nevertheless, it is possible to initialize computations
with stronger kicks in this parameter regime (i.e. 
perturb the system with larger values for $\ampl_0$). The resulting
initial-value problems then display  a clear
threshold behavior, as illustrated in Fig.~\ref{subo}: for lower values of
$\ampl_0$, the system executes regular, weakly decaying oscillations
with an amplitude set by $\ampl_0$, 
as seen for  the cases with $\ampl_0=0.4$, $0.45$
and $0.49$ in Fig.~\ref{subo}. 
Simultaneously, $\rho(\theta,I,t)$ develops no
persistent phase-space pattern. Larger forcings, however,
trigger a growth of the mode amplitude upto a level, $\ampl_*^{(s)}$,
that is otherwise independent of $\ampl_0$, as seen for  
$\ampl_0=0.5$ and higher in the figure. The system subsequently
executes trapping oscillations and $\rho(\theta,I,t)$
develops a cat's eye pattern much like the end states generated
in the linearly unstable regime.

The threshold forcing amplitude $\ampl_0$ corresponds to a
first maximum, $\ampl_*^{(u)}$ that we assume is characteristic of the unstable
subcritical nonlinear state. On the other hand, 
the nonlinear states identified by $\ampl_*^{(s)}$, are plausibly 
the stable solutions on the upper branch.
By recording the two first maxima, $\ampl_*^{(u)}$ and
$\ampl_*^{(s)}$, we  are therefore able to continue the scaling data
of  Fig.~\ref{scallywag} to $a<a_c$ for the cases with
sharper equilibrium profiles ($\Upsilon=1/16$ and $1/8$).
As expected, the data for $\ampl_*^{(s)}$
continue the original results for the
first maxima from $a>a_c$ into the linearly stable regime.
Moreover, this stable branch eventually terminates
at lower $a$ by colliding
with the unstable branch characterized by $\ampl_*^{(u)}$. 
As illustrated in  Fig.~\ref{scallywag}(a), the latter
converge towards the weakly nonlinear solution branch predicted
for a true top-hat profile (as given by the weakly nonlinear theory
of  Sec.~\ref{sssec:tophat}  which, after a translation of $|A|=\sqrt{-2a_2}$
back into the original unscaled variables,  furnishes
$\mathcal{A}_*^{(u)}=\sqrt{2(1-a)}$).

In summary, top-hat (or broken-line and water-bag) equilibrium profiles 
display a very different nonlinear dynamics at onset to that of smooth profiles.
The differences between the two in linear theory is well appreciated: 
the linear eigenspectra of smooth, stable profiles are characterized purely by
the  continuous spectra, which leads to Landau damping (or, more generally,
to ``continuum damping'' for shear flows, for which algebraic decay also
appears), whereas piecewise linear profiles support persistent discrete waves
riding on the interfaces of the equilibrium. That the weakly nonlinear
states are also very different is not, however, 
so well acknowledged.

\section{Variants and deviants}
\label{sec:deviants}

So far we have seen several examples of  single-wave model dynamics, viz.,  
the patterns that form when  equilibrium gradients inside the 
critical layer are wound up to saturate
an unstable mode or arrest Landau damping. We next mention
a number of variations on this single-wave theme.

\subsection{Long-wave theories and subharmonic instability}

An important ingredient in our derivation of the single-wave model
is the imposition of periodicity in $x$, which leads to
a wave-like global mode with a critical layer. 
Different boundary conditions would imply a different global mode structure
in $x$ and furnish a ``single-mode'' theory near the onset of instability.
Alternatively, in spatially extended systems
entire bands of wavenumber become unstable,
rather than a distinguished neutral mode. 
In asymptotic analysis, one accounts for a finite bandwidth by introducing
multiple lengthscales, with the long length scale describing the
spatial modulation of unstable patterns. 

For systems suffering a long-wave instability the situation is more 
straightforward, since then one retains only a single, long length scale.
When the unstable system also suffers critical-level singularities,
the global mode amplitude depends on both the slow time and the long 
length scale, and the resulting weakly nonlinear theory couples
familiar nonlinear wave equations to the critical-layer dynamics.
Following this route in shear flows, \textcite{stewartson81a,balmforth97b}   
coupled critical layers to the
nonlinear Schr\"{o}dinger equation and Boussinesq equations, respectively.
More recent developments are provided by \textcite{shagalov09}.

When the unstable wavenumber is finite, an alternative to solving the full
multiple scale theory is to reconstitute the two spatial variables
back into a single one. A single-wave-like model is then recovered,
but contains additional spatial derivatives capturing larger-scale
spatial modulations of the wave patterns.
This route has been followed by \textcite{shagalov10}.

\begin{figure}
\begin{center}
\leavevmode
\includegraphics[width=12 cm]{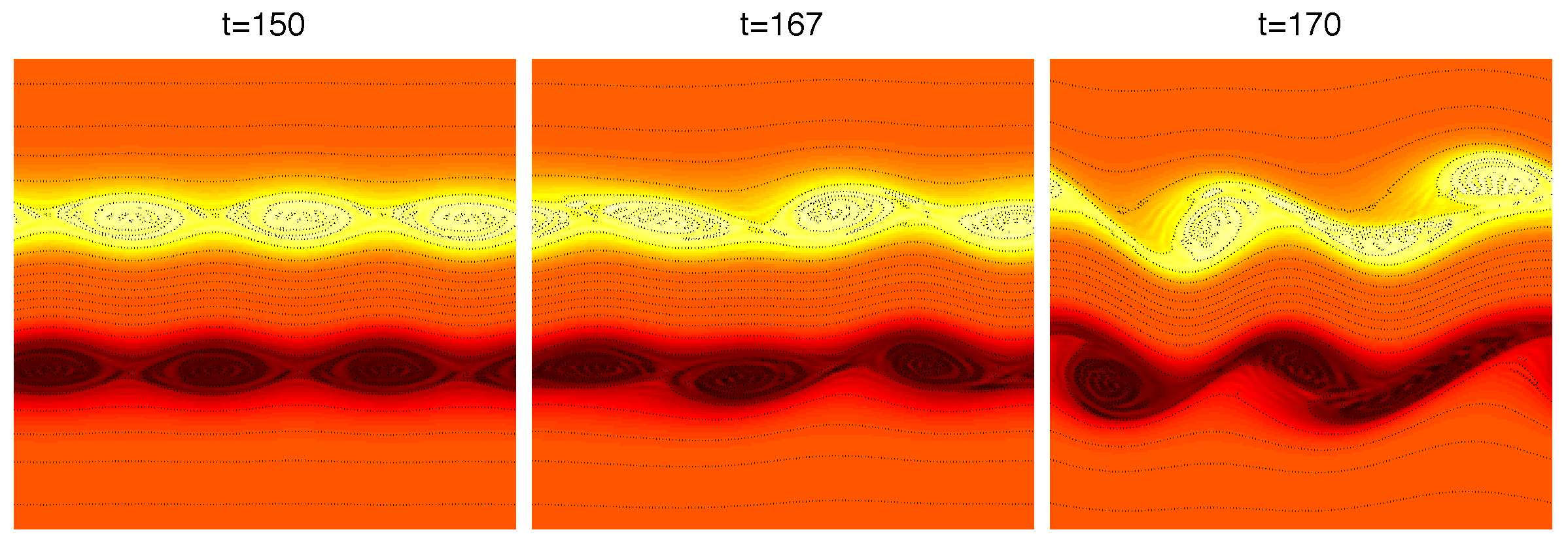}
\end{center}
\caption{
Subharmonic instability in an unstable jet.
The panels show the late evolution of the jet
after an unstable mode with $x$-wavenumber three
grows to create a symmetrically placed pair of cat's eye patterns.
In contrast to Fig.~\ref{fig:jet}, the lower overtones
were allowed to be excited by round-off error during the
earlier parts of the computation, resulting in a subharmonic,
vortex pairing instability.
} 
\label{fig:jetsub}
\end{figure}

Models with spatial modulation incorporate a key physical effect:
the subharmonic secondary instability of a chain of cat's eyes
that can lead to pairings of the vortices or plasma holes
(e.g.\  \citealp{flierl87a} and \citealp{feix94a}, respectively).
This instability is easily illustrated for  the simulation of a jet pictured earlier in
Fig.~\ref{fig:jet}.  In that simulation, the third Fourier mode in $x$
is unstable and grows to create the cat's eye pattern displayed 
in the figure; the lower overtones initially had zero amplitude, and
any round-off errors that might otherwise excite these modes
were deliberately suppresed. Once those lower overtones are
allowed to develop, on the other hand, we arrive at the evolution 
shown in Fig.~\ref{fig:jetsub}, in which
a pairing instability of the cat's eyes emerges to destroy the pattern.
This dynamics is not captured by the single-wave description,
and is one of its main limitations.

\subsection{Modified single-wave models and forced critical layers}

The instability in the single-wave model results
from the equilibrium gradients within the critical layer; once these
gradients are twisted up by the global mode, the instability saturates
(unless it can be reactivated by viscous diffusion).
In other situations, the global mode can be rendered unstable
by a different mechanism, and the instability competes with
Landau damping in a critical layer  
(\citealp{balmforth2002b}; see also \citealp{wg}).
The weakly nonlinear theory describing this situation is much
like the single-wave model (with $\gamma=+1$), except that the amplitude
equation for $A(t)$ contains a linear growth term. However,
the unstable mode is not then able to saturate: as the mode grows,
it twists up the critical-layer distribution into cat's eyes,
removing the stabilizing equilibrium gradients that  damp the mode. 
Consequently, the mode amplifies even 
faster, twisting up an ever wider critical layer until the trapping scaling is
broken.

The opposite physical situation arises when the mode
is driven by unstable equilibrium gradients within the critical
layer, but damped globally (such as in the two-species plasma model
with disparate ion masses explored by \textcite{balmforth2003a}).
Then, the amplitude equation contains a linear decay term, and saturation 
again arises when the destabilizing equilibrium gradients
in the critical layer are twisted up into cat's eyes.
However, the global damping is unaffected by those
critical layer rearrangements and continues
to damp the mode beyond saturation, forcing a subsequent decline in
mode amplitude. The ultimate decay of the
mode appears to be halted by the migration of the 
cat's eye across the critical region, which allows the
mode to access fresh equilibrium gradients \cite{balmforth2003a}.

Another natural variation on the theme
is to add more  waves to the problem (multi-wave models).  
In fluid mechanics this has been done for coupled Rossby waves
\cite{vanneste}, wind-excited water waves \cite{alexakis2003a,reutov},
over-reflectional instabilities in shallow-water and compressible
shear flows \cite{balmforth98a}, and even-odd mode interactions
in jets \cite{leib89a}. In plasma physics, additional waves have been added
in relativistic beam-plasma systems \cite{evstatiev1,evstatiev2}  
and in models of plasma turbulence \cite{cary, caryII}. 
Other problems that could be attacked in this way (but have not yet been)
include beat-wave resonance in plasma theory \cite{crawford86},
vortices \cite{mitchell94} and astrophysical disks \cite{mattor96},
ion acceleration in the central plasma sheet of the Earth's 
geotail \cite{padhye}, and the instability
of inviscid flow over compliant walls \cite{landahl62a,benjamin63a}.  
All these cases involve at least one wave with
a critical layer, together with wave-wave interactions captured within global
amplitude equations.

The critical layer in the single-wave model is rearranged 
due to the growth of a global mode. In other problems,
there is no such mode and the critical layer is directly forced
by an external perturbation. Models describing these situations
contain a critical-layer equation in which the potential
or streamfunction, $\varphi$, is determined not by an evolution
equation but directly from the
integral critical-layer distribution and the global forcing
(as in the degenerate form of the single-wave model;
Sec. \ref{sssec:degenerate}). 
The classic example of such a  problem in fluid mechanics
is the forced Rossby wave critical layer where  waves excited 
by a uneven boundary enter a shear flow and
impinge on the critical level \cite{stewartson78a,warn78a,redekopp}.
However, the theory has also been applied to a differentially rotating
accretion disk perturbed by a forming planet
\cite{balmforth2001c}, which is key in modeling angular momentum exchange 
and the migration of extra-solar planets. 
A novelty of forced critical layers is that the global potential
or streamfunction contains a richer wavenumber content than the
single-wave model. As a result, the filamentation of the
critical layer distribution can
suffer secondary instability \cite{killworth85a}, 
with the filaments rolling up into
smaller vortices and generating more filamentation. 
A cascade ensues, leading to what \textcite{haynes85a}
has coined ``critical-layer turbulence''.

All of the critical-layer models described above have the common feature
that inside the critical region, the differential advection in $\yswm$
is locally linear. Theories that consider a critical layer
located at the tip of a fluid jet where the flow is locally parabolic
\cite{brunet90a,brunet95a,balmforth2001e}
lead to a shearing term with quadratic form $\yswm^2\zeta_\xswm$.
Disturbances in the
linear initial-value problem are then found to decay algebraically
as well as exponentially. Thus Landau damping is supplemented
by a weaker form of continuum damping near the shearless point
of the velocity profile. Moreover, the phase-space patterns 
that are generated can show a richer topology (see \citealp{delcastillo93a}), taking the form of
either stacked cat's eyes or other kinds of patterns
\cite{balmforth2001e}.

\subsection{Singular neutral modes}

Another key feature of the single-wave model is that the distinguished
neutral mode that we perturb off the stability boundary is smooth.
However, the distinguished mode need not be regular. For example,
although symmetrical jets like that illustrated in Fig.~\ref{fig:jet}
have smooth neutral modes, when the profile is made asymmetrical
it generally becomes impossible to satisfy the condition,
$U''(y_*)=U(y_*)-c_r=0$, at both critical levels simultaneously.
The neutral modes on the stability boundary then become singular,
and the inner expansion for the critical layer is
needed in the leading order of the asymptotics.
Other examples include instabilities of Rossby waves
\cite{hickernell84a}, stratified and compressible flow 
\cite{goldstein89a,churilov99a}, and two-species plasmas
\cite{berk99,balmforth2003a}.

Because one has to re-order the asymptotic expansion for singular
distinguished modes, the trapping scaling is no longer the
relevant measure of the mode amplitude. It turns out that
nonlinearity first becomes important at the level $\epsilon^{5/2}$
(a conclusion also reached by \citealp{crawford96a,crawford99a},
using center-manifold methods), and the analysis, which must be taken to
much higher order than in the single-wave case,  eventually
predicts a delay-differential equation for the mode amplitude
\cite{hickernell84a,goldstein89a,churilov99a,berk99,balmforth2003a}.
Unfortunately, that equation predicts
that nonlinearity is destabilizing: beyond the initial phase of
linear growth, the nonlinear terms promote an explosive growth
of the mode amplitude (unless there is a significant viscosity as in 
\textcite{goldstein89a,churilov99a,berk99}).
As a result, the mode rapidly leaves the ``singular scaling''
and only a transient is described by the model.

\section{Concluding remarks}
\label{conclusion}

For a pattern theorist, the single-wave model could be regarded as
unsatisfactory because it offers no reduction in the
dimension of the problem. Yet despite the lack of dimensional reduction,
the single-wave model has several virtues over the original
equations. First, mathematically, the model equations offer
some simplifications over the full problem, but nevertheless
captures  the important dynamical behavior, such as
the creation of cat's eyes. Second, the model appears in a wide range
of different problems, and therefore provides
a universal description of pattern formation in a variety systems with continuous spectra.

Given this universality, it is worthwhile emphasizing
the key properties of the original problem that are required
for the single-wave model to emerge as the weakly nonlinear
description in the vicinity of onset. As we have outlined,
the single-wave model requires
periodic boundary conditions in the direction of the
equilibrium streaming or flow, and a regular neutral
mode lying on the stability boundary to open the expansion of
asymptotic theory. However, by themselves
these properties are not sufficient to lead to the single-wave model:
the so-called ``Kuramoto model'' describes
a population of coupled oscillators with fixed amplitude
and varying phase and natural frequency that is
coupled through a mean field \cite{kuramoto75a,strogatz}. 
The continuum limit of this model furnishes
a partial differential system that possesses equilibrium
solutions corresponding to populations
with randomly distributed phases. The associated linear problem
has a continuous spectrum, stable populations exhibit Landau
damping, and instabilities arise when a distinguished neutral
mode detaches from the continuum \cite{strogatz92a,crawford99b}.
However, the nonlinear dynamics of the instability,
and the phase-synchronized population patterns that form, 
are quite different from the cat's eyes of our
fluid and plasma problems \cite{balmforth2000a}. The similarity
in the linear problem is therefore misleading, and simply the
presence of a distinguished mode bifurcating from the continuum
is not sufficient to anticipate the single-wave model. 
 
In fact, a key difference between the Kuramoto model and
our plasma and fluid problems is that the former does not
possess a Hamiltonian structure.
An underlying Hamiltonian structure
undoubtedly impacts any weakly nonlinear description.
Nevertheless, the complete classification of models 
for which the single-wave model is an appropriate normal form 
has not been performed, and an underlying
Hamiltonian structure may be neither necessary nor sufficient.  There is,
however,   a Hamiltonian  interpretation of the bifurcations to instability of the single-wave model.  For example, the normal from case can be viewed a a continuum version of the Hamiltonian-Hopf bifurcation, which generalizes the discrete Kre\u{i}n \cite{krein}  transition  that is known to occur in a variety of fluid and plasma systems (e.g., \citealp{sturrock,cairns,mackay,morrison90,kueny,casti,tassi1,tassi2}).  This will be the subject of a future publication \cite{HagMor}.

Some other open questions are as follows.
First, the single-wave model describes the saturation of an unstable
mode that twists up the destabilizing equilibrium gradients in its
critical layer. However, in the single-wave scaling, the
critical layer is much thinner than the range of unstable gradients,
and saturation occurs when the background equilibrium has been flattened
over only a small fraction of that range.
Why does the mode not continue to grow and feed off the
unstable gradients outside the narrow critical region?
The answer surely lies in the fact that the system 
is constrained by the need to conserve the 
infinite number of Casimir invariants of Sec.~\ref{ssec:badass}.  
Also, when dissipation is added to the
system, the mode begins to grow secularly beyond saturation.
Thus, the saturated single-wave system looks very much like
a negative energy state of a finite-dimensional Hamiltonian
problem with dissipation-induced instability \cite{morrison90,morrison98,krech}. 

Second, as mentioned earlier, it is not clear
what is the ultimate fate of the unstable mode. Beyond saturation,
aperiodic trapping oscillations occur that attenuate
somewhat with time. However, one cannot decide whether the
oscillations persist, or if the system relaxes to a steady
BGK mode. One often observes the
intermittent creation and subsidence of secondary
structures, or macroparticles, which contribute to the aperiodic
motion \cite{tennyson94a},  albeit possibly with a dominant frequency. If the trapping oscillations continue indefinitely,
then the critical region is in a constant state of
agitation, and prolonged chaotic mixing occurs across
the cat's eyes \cite{delcastillo00a}.

\section*{Acknowledgments}

We would like to acknowledge the hospitality of the Geophysical Fluid Dynamics Program held each summer at the Woods Hole Oceanographic Institution,
which is supported by the National Science Foundation
and the Office of Naval Research.  The program has allowed us to work together on this project over many summers of the past ten years.  PJM was supported by U.S.~Dept.\ of Energy Contract \# DE-FG02-04ER54742 and he would like to acknowledge useful conversations with D. del-Castillo-Negrete, E. Knobloch, E. Tassi, and W. R. Young.

\bibliographystyle{apsrmp}

\bibliography{130219singlewave}

\begin{thebibliography}{183}
\expandafter\ifx\csname natexlab\endcsname\relax\def\natexlab#1{#1}\fi
\expandafter\ifx\csname bibnamefont\endcsname\relax
  \def\bibnamefont#1{#1}\fi
\expandafter\ifx\csname bibfnamefont\endcsname\relax
  \def\bibfnamefont#1{#1}\fi
\expandafter\ifx\csname citenamefont\endcsname\relax
  \def\citenamefont#1{#1}\fi
\expandafter\ifx\csname url\endcsname\relax
  \def\url#1{\texttt{#1}}\fi
\expandafter\ifx\csname urlprefix\endcsname\relax\def\urlprefix{URL }\fi
\providecommand{\bibinfo}[2]{#2}
\providecommand{\eprint}[2][]{\url{#2}}

\bibitem[{\citenamefont{Alexakis} \emph{et~al.}(2004)\citenamefont{Alexakis,
  Rosner, and Young}}]{alexakis2003a}
\bibinfo{author}{\bibnamefont{Alexakis}, \bibfnamefont{A.}},
  \bibinfo{author}{\bibfnamefont{R.}~\bibnamefont{Rosner}}, and
  \bibinfo{author}{\bibfnamefont{Y.-N.} \bibnamefont{Young}},
  \bibinfo{year}{2004}, \bibinfo{journal}{J.~Fluid Mech.}
  \textbf{\bibinfo{volume}{503}}, \bibinfo{pages}{171}.

\bibitem[{\citenamefont{Allis}(1959)}]{All59}
\bibinfo{author}{\bibnamefont{Allis}, \bibfnamefont{W.~P.}},
  \bibinfo{year}{1959}, in \emph{\bibinfo{booktitle}{Electrons, Ions, and
  Waves, Selected Works of {W}illiam {P}helps {A}llis}}, edited by
  \bibinfo{editor}{\bibfnamefont{S.~C.} \bibnamefont{Brown}}
  (\bibinfo{publisher}{MIT Press, Cambridge Massachuetts}),
  volume~\bibinfo{volume}{1}, pp. \bibinfo{pages}{269--279}.

\bibitem[{\citenamefont{Allis} \emph{et~al.}(1963)\citenamefont{Allis,
  Buchsbaum, and Bers}}]{All63}
\bibinfo{author}{\bibnamefont{Allis}, \bibfnamefont{W.~P.}},
  \bibinfo{author}{\bibfnamefont{S.~J.} \bibnamefont{Buchsbaum}}, and
  \bibinfo{author}{\bibfnamefont{A.}~\bibnamefont{Bers}}, \bibinfo{year}{1963},
  \emph{\bibinfo{title}{Waves in anisotropic plasmas}} (\bibinfo{publisher}{MIT
  Press}, \bibinfo{address}{Cambridge, Massachuetts}).

\bibitem[{\citenamefont{Antoni and Ruffo}(1995)}]{antoni95}
\bibinfo{author}{\bibnamefont{Antoni}, \bibfnamefont{M.}}, and
  \bibinfo{author}{\bibfnamefont{S.}~\bibnamefont{Ruffo}},
  \bibinfo{year}{1995}, \bibinfo{journal}{Phys. Rev. E}
  \textbf{\bibinfo{volume}{52}}, \bibinfo{pages}{2361}.

\bibitem[{\citenamefont{Antoniazzi}
  \emph{et~al.}(2007)\citenamefont{Antoniazzi, Fanelli, Ruffo, and
  Yamaguchi}}]{antoniazzi07}
\bibinfo{author}{\bibnamefont{Antoniazzi}, \bibfnamefont{A.}},
  \bibinfo{author}{\bibfnamefont{D.}~\bibnamefont{Fanelli}},
  \bibinfo{author}{\bibfnamefont{S.}~\bibnamefont{Ruffo}}, and
  \bibinfo{author}{\bibfnamefont{Y.~Y.} \bibnamefont{Yamaguchi}},
  \bibinfo{year}{2007}, \bibinfo{journal}{Phys. Rev. Lett.}
  \textbf{\bibinfo{volume}{99}}, \bibinfo{pages}{040601}.

\bibitem[{\citenamefont{Bachelard} \emph{et~al.}(2009)\citenamefont{Bachelard,
  Chandre, Ciani, Fanelli, and Yamaguchi}}]{chandre09}
\bibinfo{author}{\bibnamefont{Bachelard}, \bibfnamefont{R.}},
  \bibinfo{author}{\bibfnamefont{C.}~\bibnamefont{Chandre}},
  \bibinfo{author}{\bibfnamefont{A.}~\bibnamefont{Ciani}},
  \bibinfo{author}{\bibfnamefont{D.}~\bibnamefont{Fanelli}}, and
  \bibinfo{author}{\bibfnamefont{Y.~Y.} \bibnamefont{Yamaguchi}},
  \bibinfo{year}{2009}, \bibinfo{journal}{Phys. Lett. A}
  \textbf{\bibinfo{volume}{373}}, \bibinfo{pages}{4239}.

\bibitem[{\citenamefont{Backus}(1960)}]{Bac60}
\bibinfo{author}{\bibnamefont{Backus}, \bibfnamefont{G.}},
  \bibinfo{year}{1960}, \bibinfo{journal}{J. Math. Phys.}
  \textbf{\bibinfo{volume}{1}}, \bibinfo{pages}{178}.

\bibitem[{\citenamefont{Balmforth}(1998)}]{balmforth98a}
\bibinfo{author}{\bibnamefont{Balmforth}, \bibfnamefont{N.~J.}},
  \bibinfo{year}{1998}, \bibinfo{journal}{J.~Fluid Mech.}
  \textbf{\bibinfo{volume}{357}}, \bibinfo{pages}{199}.

\bibitem[{\citenamefont{Balmforth}(1999)}]{balmforth99b}
\bibinfo{author}{\bibnamefont{Balmforth}, \bibfnamefont{N.~J.}},
  \bibinfo{year}{1999}, \bibinfo{journal}{J.~Fluid Mech.}
  \textbf{\bibinfo{volume}{387}}, \bibinfo{pages}{97}.

\bibitem[{\citenamefont{Balmforth} \emph{et~al.}(1997)\citenamefont{Balmforth,
  {del-Castillo-Negrete}, and Young}}]{balmforth97a}
\bibinfo{author}{\bibnamefont{Balmforth}, \bibfnamefont{N.~J.}},
  \bibinfo{author}{\bibfnamefont{D.}~\bibnamefont{{del-Castillo-Negrete}}}, and
  \bibinfo{author}{\bibfnamefont{W.~R.} \bibnamefont{Young}},
  \bibinfo{year}{1997}, \bibinfo{journal}{J.~Fluid Mech.}
  \textbf{\bibinfo{volume}{333}}, \bibinfo{pages}{197}.

\bibitem[{\citenamefont{Balmforth and Kerswell}(2002)}]{balmforth2003a}
\bibinfo{author}{\bibnamefont{Balmforth}, \bibfnamefont{N.~J.}}, and
  \bibinfo{author}{\bibfnamefont{R.~R.} \bibnamefont{Kerswell}},
  \bibinfo{year}{2002}, \bibinfo{journal}{J. Plasma Phys.}
  \textbf{\bibinfo{volume}{68}}(\bibinfo{number}{2}), \bibinfo{pages}{87}.

\bibitem[{\citenamefont{Balmforth and Korycansky}(2001)}]{balmforth2001c}
\bibinfo{author}{\bibnamefont{Balmforth}, \bibfnamefont{N.~J.}}, and
  \bibinfo{author}{\bibfnamefont{D.}~\bibnamefont{Korycansky}},
  \bibinfo{year}{2001}, \bibinfo{journal}{Mon. Not. Roy. Astron. Soc.}
  \textbf{\bibinfo{volume}{326}}, \bibinfo{pages}{833}.

\bibitem[{\citenamefont{Balmforth}
  \emph{et~al.}(2001{\natexlab{a}})\citenamefont{Balmforth, {Llewellyn Smith},
  and Young}}]{balmforth2001a}
\bibinfo{author}{\bibnamefont{Balmforth}, \bibfnamefont{N.~J.}},
  \bibinfo{author}{\bibfnamefont{S.~G.} \bibnamefont{{Llewellyn Smith}}}, and
  \bibinfo{author}{\bibfnamefont{W.~R.} \bibnamefont{Young}},
  \bibinfo{year}{2001}{\natexlab{a}}, \bibinfo{journal}{J.~Fluid Mech.}
  \textbf{\bibinfo{volume}{426}}, \bibinfo{pages}{95}.

\bibitem[{\citenamefont{Balmforth and Morrison}(1999)}]{balmforth1999}
\bibinfo{author}{\bibnamefont{Balmforth}, \bibfnamefont{N.~J.}}, and
  \bibinfo{author}{\bibfnamefont{P.~J.} \bibnamefont{Morrison}},
  \bibinfo{year}{1999}, \bibinfo{journal}{Stud. Applied Math.}
  \textbf{\bibinfo{volume}{102}}, \bibinfo{pages}{309}.

\bibitem[{\citenamefont{Balmforth and Morrison}(2001)}]{balmforth2002a}
\bibinfo{author}{\bibnamefont{Balmforth}, \bibfnamefont{N.~J.}}, and
  \bibinfo{author}{\bibfnamefont{P.~J.} \bibnamefont{Morrison}},
  \bibinfo{year}{2001}, in \emph{\bibinfo{booktitle}{Large-Scale
  Atmosphere-Ocean Dynamics 2: Geometric Methods and Models}}, edited by
  \bibinfo{editor}{\bibfnamefont{J.}~\bibnamefont{Norbury}} and
  \bibinfo{editor}{\bibfnamefont{I.}~\bibnamefont{Roulstone}}
  (\bibinfo{publisher}{Cambridge University Press},
  \bibinfo{address}{Cambridge, U.K.}).

\bibitem[{\citenamefont{Balmforth and Piccolo}(2001)}]{balmforth2001b}
\bibinfo{author}{\bibnamefont{Balmforth}, \bibfnamefont{N.~J.}}, and
  \bibinfo{author}{\bibfnamefont{C.}~\bibnamefont{Piccolo}},
  \bibinfo{year}{2001}, \bibinfo{journal}{J.~Fluid Mech.}
  \textbf{\bibinfo{volume}{449}}, \bibinfo{pages}{85}.

\bibitem[{\citenamefont{Balmforth}
  \emph{et~al.}(2001{\natexlab{b}})\citenamefont{Balmforth, Piccolo, and
  Umurhan}}]{balmforth2001e}
\bibinfo{author}{\bibnamefont{Balmforth}, \bibfnamefont{N.~J.}},
  \bibinfo{author}{\bibfnamefont{C.}~\bibnamefont{Piccolo}}, and
  \bibinfo{author}{\bibfnamefont{M.}~\bibnamefont{Umurhan}},
  \bibinfo{year}{2001}{\natexlab{b}}, \bibinfo{journal}{J.~Fluid Mech.}
  \textbf{\bibinfo{volume}{449}}, \bibinfo{pages}{115}.

\bibitem[{\citenamefont{Balmforth} \emph{et~al.}(2012)\citenamefont{Balmforth,
  Roy, and Caulfield}}]{balmforth12}
\bibinfo{author}{\bibnamefont{Balmforth}, \bibfnamefont{N.~J.}},
  \bibinfo{author}{\bibfnamefont{A.}~\bibnamefont{Roy}}, and
  \bibinfo{author}{\bibfnamefont{C.~P.} \bibnamefont{Caulfield}},
  \bibinfo{year}{2012}, \bibinfo{journal}{J.~Fluid Mech.}
  \textbf{\bibinfo{volume}{694}}, \bibinfo{pages}{292}.

\bibitem[{\citenamefont{Balmforth and Sassi}(2000)}]{balmforth2000a}
\bibinfo{author}{\bibnamefont{Balmforth}, \bibfnamefont{N.~J.}}, and
  \bibinfo{author}{\bibfnamefont{R.}~\bibnamefont{Sassi}},
  \bibinfo{year}{2000}, \bibinfo{journal}{Physica D}
  \textbf{\bibinfo{volume}{143}}, \bibinfo{pages}{21}.

\bibitem[{\citenamefont{Balmforth and Young}(1997)}]{balmforth97b}
\bibinfo{author}{\bibnamefont{Balmforth}, \bibfnamefont{N.~J.}}, and
  \bibinfo{author}{\bibfnamefont{W.~R.} \bibnamefont{Young}},
  \bibinfo{year}{1997}, \bibinfo{journal}{Phys. Rev. Lett.}
  \textbf{\bibinfo{volume}{79}}, \bibinfo{pages}{4155}.

\bibitem[{\citenamefont{Balmforth and Young}(2002)}]{balmforth2002b}
\bibinfo{author}{\bibnamefont{Balmforth}, \bibfnamefont{N.~J.}}, and
  \bibinfo{author}{\bibfnamefont{Y.-N.} \bibnamefont{Young}},
  \bibinfo{year}{2002}, \bibinfo{journal}{J.~Fluid Mech.}
  \textbf{\bibinfo{volume}{450}}, \bibinfo{pages}{131}.

\bibitem[{\citenamefont{Barr\'e and Yamaguchi}(2009)}]{barre09}
\bibinfo{author}{\bibnamefont{Barr\'e}, \bibfnamefont{J.}}, and
  \bibinfo{author}{\bibfnamefont{Y.~Y.} \bibnamefont{Yamaguchi}},
  \bibinfo{year}{2009}, \bibinfo{journal}{{Phys. Rev. E}}
  \textbf{\bibinfo{volume}{79}}, \bibinfo{pages}{036208}.

\bibitem[{\citenamefont{Benjamin}(1963)}]{benjamin63a}
\bibinfo{author}{\bibnamefont{Benjamin}, \bibfnamefont{T.~B.}},
  \bibinfo{year}{1963}, \bibinfo{journal}{J.~Fluid Mech.}
  \textbf{\bibinfo{volume}{16}}, \bibinfo{pages}{436}.

\bibitem[{\citenamefont{Benney and Bergeron}(1969)}]{benney69a}
\bibinfo{author}{\bibnamefont{Benney}, \bibfnamefont{D.~J.}}, and
  \bibinfo{author}{\bibfnamefont{R.~F.} \bibnamefont{Bergeron}},
  \bibinfo{year}{1969}, \bibinfo{journal}{Stud. Appl. Math.}
  \textbf{\bibinfo{volume}{48}}, \bibinfo{pages}{181}.

\bibitem[{\citenamefont{Berk} \emph{et~al.}(1999)\citenamefont{Berk, Breizmann,
  Candy, Pekker, and Petviashvili}}]{berk99}
\bibinfo{author}{\bibnamefont{Berk}, \bibfnamefont{H.~L.}},
  \bibinfo{author}{\bibfnamefont{B.~N.} \bibnamefont{Breizmann}},
  \bibinfo{author}{\bibfnamefont{J.}~\bibnamefont{Candy}},
  \bibinfo{author}{\bibfnamefont{M.}~\bibnamefont{Pekker}}, and
  \bibinfo{author}{\bibfnamefont{N.~V.} \bibnamefont{Petviashvili}},
  \bibinfo{year}{1999}, \bibinfo{journal}{Phys. Plasmas}
  \textbf{\bibinfo{volume}{6}}, \bibinfo{pages}{3102}.

\bibitem[{\citenamefont{Berk and Roberts}(1967)}]{berk}
\bibinfo{author}{\bibnamefont{Berk}, \bibfnamefont{H.~L.}}, and
  \bibinfo{author}{\bibfnamefont{K.~V.} \bibnamefont{Roberts}},
  \bibinfo{year}{1967}, \bibinfo{journal}{Phys. Fluids}
  \textbf{\bibinfo{volume}{10}}, \bibinfo{pages}{1595}.

\bibitem[{\citenamefont{Bernstein} \emph{et~al.}(1958)\citenamefont{Bernstein,
  Greene, and Kruskal}}]{bernstein58a}
\bibinfo{author}{\bibnamefont{Bernstein}, \bibfnamefont{I.~B.}},
  \bibinfo{author}{\bibfnamefont{J.~M.} \bibnamefont{Greene}}, and
  \bibinfo{author}{\bibfnamefont{M.~D.} \bibnamefont{Kruskal}},
  \bibinfo{year}{1958}, \bibinfo{journal}{Phys. Rev.}
  \textbf{\bibinfo{volume}{108}}, \bibinfo{pages}{546}.

\bibitem[{\citenamefont{Bickley}(1937)}]{bickley}
\bibinfo{author}{\bibnamefont{Bickley}, \bibfnamefont{W.~C.}},
  \bibinfo{year}{1937}, \bibinfo{journal}{Philos. Mag.}
  \textbf{\bibinfo{volume}{23}}, \bibinfo{pages}{727}.

\bibitem[{\citenamefont{Binney and Tremaine}(2008)}]{binney}
\bibinfo{author}{\bibnamefont{Binney}, \bibfnamefont{J.}}, and
  \bibinfo{author}{\bibfnamefont{S.}~\bibnamefont{Tremaine}},
  \bibinfo{year}{2008}, \emph{\bibinfo{title}{Galactic Dynamics}}
  (\bibinfo{publisher}{Princeton University Press},
  \bibinfo{address}{Princeton, N. J.}), \bibinfo{edition}{second} edition.

\bibitem[{\citenamefont{Bohm and Gross}(1949{\natexlab{a}})}]{Boh49a}
\bibinfo{author}{\bibnamefont{Bohm}, \bibfnamefont{D.}}, and
  \bibinfo{author}{\bibfnamefont{E.~P.} \bibnamefont{Gross}},
  \bibinfo{year}{1949}{\natexlab{a}}, \bibinfo{journal}{Phys. Rev.}
  \textbf{\bibinfo{volume}{75}}, \bibinfo{pages}{1851}.

\bibitem[{\citenamefont{Bohm and Gross}(1949{\natexlab{b}})}]{Boh49b}
\bibinfo{author}{\bibnamefont{Bohm}, \bibfnamefont{D.}}, and
  \bibinfo{author}{\bibfnamefont{E.~P.} \bibnamefont{Gross}},
  \bibinfo{year}{1949}{\natexlab{b}}, \bibinfo{journal}{Phys. Rev.}
  \textbf{\bibinfo{volume}{75}}, \bibinfo{pages}{1864}.

\bibitem[{\citenamefont{Bohm and Gross}(1950)}]{Boh50}
\bibinfo{author}{\bibnamefont{Bohm}, \bibfnamefont{D.}}, and
  \bibinfo{author}{\bibfnamefont{E.~P.} \bibnamefont{Gross}},
  \bibinfo{year}{1950}, \bibinfo{journal}{Phys. Rev.}
  \textbf{\bibinfo{volume}{79}}, \bibinfo{pages}{992}.

\bibitem[{\citenamefont{Briggs} \emph{et~al.}(1970)\citenamefont{Briggs,
  Daugherty, and Levy}}]{briggs70a}
\bibinfo{author}{\bibnamefont{Briggs}, \bibfnamefont{R.~J.}},
  \bibinfo{author}{\bibfnamefont{J.~D.} \bibnamefont{Daugherty}}, and
  \bibinfo{author}{\bibfnamefont{R.~H.} \bibnamefont{Levy}},
  \bibinfo{year}{1970}, \bibinfo{journal}{Phys. Fluids}
  \textbf{\bibinfo{volume}{13}}, \bibinfo{pages}{421}.

\bibitem[{\citenamefont{Brown and Stewartson}(1978)}]{brown78a}
\bibinfo{author}{\bibnamefont{Brown}, \bibfnamefont{S.~N.}}, and
  \bibinfo{author}{\bibfnamefont{K.}~\bibnamefont{Stewartson}},
  \bibinfo{year}{1978}, \bibinfo{journal}{Geophys. Astrophys. Fluid Dyn.}
  \textbf{\bibinfo{volume}{10}}, \bibinfo{pages}{1}.

\bibitem[{\citenamefont{Brown and Stewartson}(1980)}]{brown80a}
\bibinfo{author}{\bibnamefont{Brown}, \bibfnamefont{S.~N.}}, and
  \bibinfo{author}{\bibfnamefont{K.}~\bibnamefont{Stewartson}},
  \bibinfo{year}{1980}, \bibinfo{journal}{J. Fluid Mech.}
  \textbf{\bibinfo{volume}{100}}, \bibinfo{pages}{811}.

\bibitem[{\citenamefont{Brunet and Haynes}(1995)}]{brunet95a}
\bibinfo{author}{\bibnamefont{Brunet}, \bibfnamefont{G.}}, and
  \bibinfo{author}{\bibfnamefont{P.~H.} \bibnamefont{Haynes}},
  \bibinfo{year}{1995}, \bibinfo{journal}{J.~Atmos. Sci.}
  \textbf{\bibinfo{volume}{52}}, \bibinfo{pages}{464}.

\bibitem[{\citenamefont{Brunet and Warn}(1990)}]{brunet90a}
\bibinfo{author}{\bibnamefont{Brunet}, \bibfnamefont{G.}}, and
  \bibinfo{author}{\bibfnamefont{T.}~\bibnamefont{Warn}}, \bibinfo{year}{1990},
  \bibinfo{journal}{J.~Atmos. Sci.} \textbf{\bibinfo{volume}{47}},
  \bibinfo{pages}{1173}.

\bibitem[{\citenamefont{Brunetti} \emph{et~al.}(2000)\citenamefont{Brunetti,
  Califano, and Pegoraro}}]{brunetti00}
\bibinfo{author}{\bibnamefont{Brunetti}, \bibfnamefont{M.}},
  \bibinfo{author}{\bibfnamefont{F.}~\bibnamefont{Califano}}, and
  \bibinfo{author}{\bibfnamefont{F.}~\bibnamefont{Pegoraro}},
  \bibinfo{year}{2000}, \bibinfo{journal}{Physical Rev. E}
  \textbf{\bibinfo{volume}{62}}, \bibinfo{pages}{4109}.

\bibitem[{\citenamefont{Bryuno}(1988)}]{bryuno}
\bibinfo{author}{\bibnamefont{Bryuno}, \bibfnamefont{A.~D.}},
  \bibinfo{year}{1988}, \bibinfo{journal}{Russian Math. Surveys}
  \textbf{\bibinfo{volume}{43}}, \bibinfo{pages}{25}.

\bibitem[{\citenamefont{de~Buyl} \emph{et~al.}(2009)\citenamefont{de~Buyl,
  Fanelli, Bachelard, and Ninno}}]{buyl09}
\bibinfo{author}{\bibnamefont{de~Buyl}, \bibfnamefont{P.}},
  \bibinfo{author}{\bibfnamefont{D.}~\bibnamefont{Fanelli}},
  \bibinfo{author}{\bibfnamefont{R.}~\bibnamefont{Bachelard}}, and
  \bibinfo{author}{\bibfnamefont{G.~D.} \bibnamefont{Ninno}},
  \bibinfo{year}{2009}, \bibinfo{journal}{{Phys. Rev. Special Topics -
  Accelerators and Beams}} \textbf{\bibinfo{volume}{12}},
  \bibinfo{pages}{060704}.

\bibitem[{\citenamefont{Cairns}(1979)}]{cairns}
\bibinfo{author}{\bibnamefont{Cairns}, \bibfnamefont{R.~A.}},
  \bibinfo{year}{1979}, \bibinfo{journal}{J. Fluid Mech.}
  \textbf{\bibinfo{volume}{92}}, \bibinfo{pages}{1}.

\bibitem[{\citenamefont{Campa} \emph{et~al.}(2009)\citenamefont{Campa, Dauxois,
  and Ruffo}}]{campa}
\bibinfo{author}{\bibnamefont{Campa}, \bibfnamefont{A.}},
  \bibinfo{author}{\bibfnamefont{T.}~\bibnamefont{Dauxois}}, and
  \bibinfo{author}{\bibfnamefont{S.}~\bibnamefont{Ruffo}},
  \bibinfo{year}{2009}, \bibinfo{journal}{Phys. Repts.}
  \textbf{\bibinfo{volume}{480}}, \bibinfo{pages}{57}.

\bibitem[{\citenamefont{Cary} \emph{et~al.}(1992)\citenamefont{Cary, Doxas,
  Escande, and Verga}}]{caryII}
\bibinfo{author}{\bibnamefont{Cary}, \bibfnamefont{J.~R.}},
  \bibinfo{author}{\bibfnamefont{I.}~\bibnamefont{Doxas}},
  \bibinfo{author}{\bibfnamefont{D.~F.} \bibnamefont{Escande}}, and
  \bibinfo{author}{\bibfnamefont{A.~D.} \bibnamefont{Verga}},
  \bibinfo{year}{1992}, \bibinfo{journal}{Phys. Fluids B}
  \textbf{\bibinfo{volume}{4}}.

\bibitem[{\citenamefont{Case}(1960)}]{case60a}
\bibinfo{author}{\bibnamefont{Case}, \bibfnamefont{K.~M.}},
  \bibinfo{year}{1960}, \bibinfo{journal}{Phys. Fluids}
  \textbf{\bibinfo{volume}{3}}, \bibinfo{pages}{143}.

\bibitem[{\citenamefont{Casti} \emph{et~al.}(1998)\citenamefont{Casti,
  Morrison, and Spiegel}}]{casti}
\bibinfo{author}{\bibnamefont{Casti}, \bibfnamefont{A.}},
  \bibinfo{author}{\bibfnamefont{P.~J.} \bibnamefont{Morrison}}, and
  \bibinfo{author}{\bibfnamefont{E.~A.} \bibnamefont{Spiegel}},
  \bibinfo{year}{1998}, \bibinfo{journal}{Annals of the New York Academy of
  Sciences} \textbf{\bibinfo{volume}{867}}, \bibinfo{pages}{93}.

\bibitem[{\citenamefont{Chaikin and Lubensky}(1995)}]{chaikin}
\bibinfo{author}{\bibnamefont{Chaikin}, \bibfnamefont{P.~M.}}, and
  \bibinfo{author}{\bibfnamefont{T.~C.} \bibnamefont{Lubensky}},
  \bibinfo{year}{1995}, \emph{\bibinfo{title}{Principles of condensed matter
  physics}} (\bibinfo{publisher}{Cambridge U. Press},
  \bibinfo{address}{Cambridge}).

\bibitem[{\citenamefont{Chandre} \emph{et~al.}(2013)\citenamefont{Chandre,
  de~Guillebon, Back, Tassi, and Morrison}}]{Cha13}
\bibinfo{author}{\bibnamefont{Chandre}, \bibfnamefont{C.}},
  \bibinfo{author}{\bibfnamefont{L.}~\bibnamefont{de~Guillebon}},
  \bibinfo{author}{\bibfnamefont{A.}~\bibnamefont{Back}},
  \bibinfo{author}{\bibfnamefont{E.}~\bibnamefont{Tassi}}, and
  \bibinfo{author}{\bibfnamefont{P.~J.} \bibnamefont{Morrison}},
  \bibinfo{year}{2013}, \bibinfo{journal}{J. Phys. A} .

\bibitem[{\citenamefont{Chandre} \emph{et~al.}(2012)\citenamefont{Chandre,
  Morrison, and Tassi}}]{ChaMoTa12}
\bibinfo{author}{\bibnamefont{Chandre}, \bibfnamefont{C.}},
  \bibinfo{author}{\bibfnamefont{P.~J.} \bibnamefont{Morrison}}, and
  \bibinfo{author}{\bibfnamefont{E.}~\bibnamefont{Tassi}},
  \bibinfo{year}{2012}, \bibinfo{journal}{Phys. Lett. A}
  \textbf{\bibinfo{volume}{376}}, \bibinfo{pages}{737}.

\bibitem[{\citenamefont{Chavanis and Delfini}(2009)}]{chavanis}
\bibinfo{author}{\bibnamefont{Chavanis}, \bibfnamefont{P.~H.}}, and
  \bibinfo{author}{\bibfnamefont{L.}~\bibnamefont{Delfini}},
  \bibinfo{year}{2009}, \bibinfo{journal}{{Euro. Phys. J. B}}
  \textbf{\bibinfo{volume}{69}}, \bibinfo{pages}{389}.

\bibitem[{\citenamefont{Cheng and Knorr}(1976)}]{cheng76a}
\bibinfo{author}{\bibnamefont{Cheng}, \bibfnamefont{C.~Z.}}, and
  \bibinfo{author}{\bibfnamefont{G.}~\bibnamefont{Knorr}},
  \bibinfo{year}{1976}, \bibinfo{journal}{J.~Comp. Phys.}
  \textbf{\bibinfo{volume}{22}}, \bibinfo{pages}{330}.

\bibitem[{\citenamefont{Churilov}(1999)}]{churilov99a}
\bibinfo{author}{\bibnamefont{Churilov}, \bibfnamefont{S.~M.}},
  \bibinfo{year}{1999}, \bibinfo{journal}{J.~Fluid Mech.}
  \textbf{\bibinfo{volume}{392}}, \bibinfo{pages}{233}.

\bibitem[{\citenamefont{Churilov and
  Shukhman}(1987{\natexlab{a}})}]{churilov87a}
\bibinfo{author}{\bibnamefont{Churilov}, \bibfnamefont{S.~M.}}, and
  \bibinfo{author}{\bibfnamefont{I.~G.} \bibnamefont{Shukhman}},
  \bibinfo{year}{1987}{\natexlab{a}}, \bibinfo{journal}{Geophys. Astrophys.
  Fluid Dyn.} \textbf{\bibinfo{volume}{38}}, \bibinfo{pages}{145}.

\bibitem[{\citenamefont{Churilov and
  Shukhman}(1987{\natexlab{b}})}]{churilov87b}
\bibinfo{author}{\bibnamefont{Churilov}, \bibfnamefont{S.~M.}}, and
  \bibinfo{author}{\bibfnamefont{I.~G.} \bibnamefont{Shukhman}},
  \bibinfo{year}{1987}{\natexlab{b}}, \bibinfo{journal}{Proc. Roy. Soc. A}
  \textbf{\bibinfo{volume}{409}}, \bibinfo{pages}{351}.

\bibitem[{\citenamefont{Crawford}(1991)}]{crawford91a}
\bibinfo{author}{\bibnamefont{Crawford}, \bibfnamefont{J.~D.}},
  \bibinfo{year}{1991}, \bibinfo{journal}{Rev. Mod. Phys.}
  \textbf{\bibinfo{volume}{63}}, \bibinfo{pages}{991}.

\bibitem[{\citenamefont{Crawford}(1995)}]{crawford95a}
\bibinfo{author}{\bibnamefont{Crawford}, \bibfnamefont{J.~D.}},
  \bibinfo{year}{1995}, \bibinfo{journal}{Phys. Plasmas}
  \textbf{\bibinfo{volume}{2}}, \bibinfo{pages}{97}.

\bibitem[{\citenamefont{Crawford and Davies}(1999)}]{crawford99b}
\bibinfo{author}{\bibnamefont{Crawford}, \bibfnamefont{J.~D.}}, and
  \bibinfo{author}{\bibfnamefont{K.~T.~R.} \bibnamefont{Davies}},
  \bibinfo{year}{1999}, \bibinfo{journal}{Physica D}
  \textbf{\bibinfo{volume}{125}}, \bibinfo{pages}{1}.

\bibitem[{\citenamefont{Crawford and Jayaraman}(1996)}]{crawford96a}
\bibinfo{author}{\bibnamefont{Crawford}, \bibfnamefont{J.~D.}}, and
  \bibinfo{author}{\bibfnamefont{A.}~\bibnamefont{Jayaraman}},
  \bibinfo{year}{1996}, \bibinfo{journal}{Phys. Rev. Lett}
  \textbf{\bibinfo{volume}{77}}, \bibinfo{pages}{3549}.

\bibitem[{\citenamefont{Crawford and Jayaraman}(1999)}]{crawford99a}
\bibinfo{author}{\bibnamefont{Crawford}, \bibfnamefont{J.~D.}}, and
  \bibinfo{author}{\bibfnamefont{A.}~\bibnamefont{Jayaraman}},
  \bibinfo{year}{1999}, \bibinfo{journal}{Phys. Fluids}
  \textbf{\bibinfo{volume}{6}}, \bibinfo{pages}{666}.

\bibitem[{\citenamefont{Crawford} \emph{et~al.}(1986)\citenamefont{Crawford,
  Kaufman, Oberman, and Johnston}}]{crawford86}
\bibinfo{author}{\bibnamefont{Crawford}, \bibfnamefont{J.~D.}},
  \bibinfo{author}{\bibfnamefont{A.~N.} \bibnamefont{Kaufman}},
  \bibinfo{author}{\bibfnamefont{C.}~\bibnamefont{Oberman}}, and
  \bibinfo{author}{\bibfnamefont{S.}~\bibnamefont{Johnston}},
  \bibinfo{year}{1986}, \bibinfo{journal}{{Phys. Fluids}}
  \textbf{\bibinfo{volume}{29}}, \bibinfo{pages}{3219}.

\bibitem[{\citenamefont{Cross and Hohenberg}(1993)}]{cross}
\bibinfo{author}{\bibnamefont{Cross}, \bibfnamefont{M.~C.}}, and
  \bibinfo{author}{\bibfnamefont{P.~C.} \bibnamefont{Hohenberg}},
  \bibinfo{year}{1993}, \bibinfo{journal}{Rev. Mod. Phys.}
  \textbf{\bibinfo{volume}{65}}, \bibinfo{pages}{851}.

\bibitem[{\citenamefont{Danielson} \emph{et~al.}(2004)\citenamefont{Danielson,
  Anderegg, and Driscoll}}]{danielson04a}
\bibinfo{author}{\bibnamefont{Danielson}, \bibfnamefont{J.~R.}},
  \bibinfo{author}{\bibfnamefont{F.}~\bibnamefont{Anderegg}}, and
  \bibinfo{author}{\bibfnamefont{C.~F.} \bibnamefont{Driscoll}},
  \bibinfo{year}{2004}, \bibinfo{journal}{Phys.~Fluids}
  \textbf{\bibinfo{volume}{92}}, \bibinfo{pages}{245003}.

\bibitem[{\citenamefont{Degond}(1986)}]{degond}
\bibinfo{author}{\bibnamefont{Degond}, \bibfnamefont{P.}},
  \bibinfo{year}{1986}, \bibinfo{journal}{Trans. Am. Math. Soc.}
  \textbf{\bibinfo{volume}{294}}, \bibinfo{pages}{435}.

\bibitem[{\citenamefont{{del-Castillo-Negrete}}(1998{\natexlab{a}})}]{delcastillo98a}
\bibinfo{author}{\bibnamefont{{del-Castillo-Negrete}}, \bibfnamefont{D.}},
  \bibinfo{year}{1998}{\natexlab{a}}, \bibinfo{journal}{Phys. Lett.}
  \textbf{\bibinfo{volume}{241}}, \bibinfo{pages}{99}.

\bibitem[{\citenamefont{{del-Castillo-Negrete}}(1998{\natexlab{b}})}]{delcastillo98b}
\bibinfo{author}{\bibnamefont{{del-Castillo-Negrete}}, \bibfnamefont{D.}},
  \bibinfo{year}{1998}{\natexlab{b}}, \bibinfo{journal}{Phys. Plasmas}
  \textbf{\bibinfo{volume}{5}}, \bibinfo{pages}{3886}.

\bibitem[{\citenamefont{{del-Castillo-Negrete}}(2000)}]{delcastillo00a}
\bibinfo{author}{\bibnamefont{{del-Castillo-Negrete}}, \bibfnamefont{D.}},
  \bibinfo{year}{2000}, \bibinfo{journal}{Chaos} \textbf{\bibinfo{volume}{10}},
  \bibinfo{pages}{75}.

\bibitem[{\citenamefont{{del-Castillo-Negrete} and
  Firpo}(2002)}]{delcastillo2002a}
\bibinfo{author}{\bibnamefont{{del-Castillo-Negrete}}, \bibfnamefont{D.}}, and
  \bibinfo{author}{\bibfnamefont{M.~C.} \bibnamefont{Firpo}},
  \bibinfo{year}{2002}, \bibinfo{journal}{Chaos} \textbf{\bibinfo{volume}{12}},
  \bibinfo{pages}{496}.

\bibitem[{\citenamefont{{del-Castillo-Negrete} and
  Morrison}(1993)}]{delcastillo93a}
\bibinfo{author}{\bibnamefont{{del-Castillo-Negrete}}, \bibfnamefont{D.}}, and
  \bibinfo{author}{\bibfnamefont{P.~J.} \bibnamefont{Morrison}},
  \bibinfo{year}{1993}, \bibinfo{journal}{Phys. Fluids A}
  \textbf{\bibinfo{volume}{5}}, \bibinfo{pages}{948}.

\bibitem[{\citenamefont{Denavit}(1981)}]{denavit81a}
\bibinfo{author}{\bibnamefont{Denavit}, \bibfnamefont{J.}},
  \bibinfo{year}{1981}, \bibinfo{journal}{Phys. Fluids}
  \textbf{\bibinfo{volume}{28}}, \bibinfo{pages}{2773}.

\bibitem[{\citenamefont{Dikki}(1960)}]{dikki60a}
\bibinfo{author}{\bibnamefont{Dikki}, \bibfnamefont{L.~A.}},
  \bibinfo{year}{1960}, \bibinfo{journal}{Sov. Phys. Doklady}
  \textbf{\bibinfo{volume}{135}}, \bibinfo{pages}{1179}.

\bibitem[{\citenamefont{Dirac}(1927)}]{dirac27}
\bibinfo{author}{\bibnamefont{Dirac}, \bibfnamefont{P.}}, \bibinfo{year}{1927},
  \bibinfo{journal}{Z. Physik} \textbf{\bibinfo{volume}{44}},
  \bibinfo{pages}{585}.

\bibitem[{\citenamefont{Doxas and Cary}(1997)}]{cary}
\bibinfo{author}{\bibnamefont{Doxas}, \bibfnamefont{I.}}, and
  \bibinfo{author}{\bibfnamefont{J.}~\bibnamefont{Cary}}, \bibinfo{year}{1997},
  \bibinfo{journal}{Phys. Plasmas} \textbf{\bibinfo{volume}{4}},
  \bibinfo{pages}{2508}.

\bibitem[{\citenamefont{Drazin and Reid}(1981)}]{drazin81a}
\bibinfo{author}{\bibnamefont{Drazin}, \bibfnamefont{P.~G.}}, and
  \bibinfo{author}{\bibfnamefont{W.~H.} \bibnamefont{Reid}},
  \bibinfo{year}{1981}, \emph{\bibinfo{title}{Hydrodynamic Stability}}
  (\bibinfo{publisher}{University Press}, \bibinfo{address}{Cambridge}).

\bibitem[{\citenamefont{Driscoll and Fine}(1990)}]{driscoll90a}
\bibinfo{author}{\bibnamefont{Driscoll}, \bibfnamefont{C.~F.}}, and
  \bibinfo{author}{\bibfnamefont{K.~S.} \bibnamefont{Fine}},
  \bibinfo{year}{1990}, \bibinfo{journal}{Phys. Fluids B}
  \textbf{\bibinfo{volume}{2}}, \bibinfo{pages}{1359}.

\bibitem[{\citenamefont{Drummond} \emph{et~al.}(1970)\citenamefont{Drummond,
  Malmberg, {O'Neil}, and Thompson}}]{drummond70a}
\bibinfo{author}{\bibnamefont{Drummond}, \bibfnamefont{W.~E.}},
  \bibinfo{author}{\bibfnamefont{J.~H.} \bibnamefont{Malmberg}},
  \bibinfo{author}{\bibfnamefont{T.~M.} \bibnamefont{{O'Neil}}}, and
  \bibinfo{author}{\bibfnamefont{J.~R.} \bibnamefont{Thompson}},
  \bibinfo{year}{1970}, \bibinfo{journal}{Phys. Fluids}
  \textbf{\bibinfo{volume}{13}}, \bibinfo{pages}{2422}.

\bibitem[{\citenamefont{Ecker and Hoelling}(1963)}]{Eck63}
\bibinfo{author}{\bibnamefont{Ecker}, \bibfnamefont{G.}}, and
  \bibinfo{author}{\bibfnamefont{J.}~\bibnamefont{Hoelling}},
  \bibinfo{year}{1963}, \bibinfo{journal}{Phys. Fluids}
  \textbf{\bibinfo{volume}{6}}, \bibinfo{pages}{70}.

\bibitem[{\citenamefont{Eliassen} \emph{et~al.}(1953)\citenamefont{Eliassen,
  H{\o}iland, and Riis}}]{eliassen}
\bibinfo{author}{\bibnamefont{Eliassen}, \bibfnamefont{A.}},
  \bibinfo{author}{\bibfnamefont{E.}~\bibnamefont{H{\o}iland}}, and
  \bibinfo{author}{\bibfnamefont{E.}~\bibnamefont{Riis}}, \bibinfo{year}{1953},
  \bibinfo{journal}{Institute for Weather and Climate Res. Oslo, Sweden, Pub.
  No.} \textbf{\bibinfo{volume}{1}}.

\bibitem[{\citenamefont{Eliasson and Shukla}(2007)}]{shukla}
\bibinfo{author}{\bibnamefont{Eliasson}, \bibfnamefont{B.}}, and
  \bibinfo{author}{\bibfnamefont{P.~K.} \bibnamefont{Shukla}},
  \bibinfo{year}{2007}, \bibinfo{journal}{Phys. Reports}
  \textbf{\bibinfo{volume}{442}}, \bibinfo{pages}{225}.

\bibitem[{\citenamefont{Engevik}(1966)}]{engevik}
\bibinfo{author}{\bibnamefont{Engevik}, \bibfnamefont{L.}},
  \bibinfo{year}{1966}, \bibinfo{journal}{Dept. Applied Math. Bergen, Norway,
  Pub. No.} \textbf{\bibinfo{volume}{11}}.

\bibitem[{\citenamefont{Evstatiev} \emph{et~al.}(2003)\citenamefont{Evstatiev,
  Morrison, and Horton}}]{evstatiev1}
\bibinfo{author}{\bibnamefont{Evstatiev}, \bibfnamefont{E.~G.}},
  \bibinfo{author}{\bibfnamefont{P.~J.} \bibnamefont{Morrison}}, and
  \bibinfo{author}{\bibfnamefont{W.}~\bibnamefont{Horton}},
  \bibinfo{year}{2003}, \bibinfo{journal}{Phys. Plasmas}
  \textbf{\bibinfo{volume}{10}}, \bibinfo{pages}{4090}.

\bibitem[{\citenamefont{Evstatiev} \emph{et~al.}(2005)\citenamefont{Evstatiev,
  Morrison, and Horton}}]{evstatiev2}
\bibinfo{author}{\bibnamefont{Evstatiev}, \bibfnamefont{E.~G.}},
  \bibinfo{author}{\bibfnamefont{P.~J.} \bibnamefont{Morrison}}, and
  \bibinfo{author}{\bibfnamefont{W.}~\bibnamefont{Horton}},
  \bibinfo{year}{2005}, \bibinfo{journal}{Phys. Plasmas}
  \textbf{\bibinfo{volume}{12}}, \bibinfo{pages}{072198}.

\bibitem[{\citenamefont{Feix} \emph{et~al.}(1994)\citenamefont{Feix, Bertrand,
  and Ghizzo}}]{feix94a}
\bibinfo{author}{\bibnamefont{Feix}, \bibfnamefont{M.~R.}},
  \bibinfo{author}{\bibfnamefont{P.}~\bibnamefont{Bertrand}}, and
  \bibinfo{author}{\bibfnamefont{A.}~\bibnamefont{Ghizzo}},
  \bibinfo{year}{1994}, in \emph{\bibinfo{booktitle}{Advances in kinetic theory
  and computing}}, edited by
  \bibinfo{editor}{\bibfnamefont{B.}~\bibnamefont{Perthame}}
  (\bibinfo{publisher}{World Scientific}, \bibinfo{address}{Singapore}), pp.
  \bibinfo{pages}{44--81}.

\bibitem[{\citenamefont{Fj{\o}rtoft}(1950)}]{fjortoft50a}
\bibinfo{author}{\bibnamefont{Fj{\o}rtoft}, \bibfnamefont{R.}},
  \bibinfo{year}{1950}, \bibinfo{journal}{Geofys. Publ.}
  \textbf{\bibinfo{volume}{17}}, \bibinfo{pages}{1}.

\bibitem[{\citenamefont{Flierl} \emph{et~al.}(1987)\citenamefont{Flierl,
  Malanotte-Rizzoli, and Zabusky}}]{flierl87a}
\bibinfo{author}{\bibnamefont{Flierl}, \bibfnamefont{G.~R.}},
  \bibinfo{author}{\bibfnamefont{P.}~\bibnamefont{Malanotte-Rizzoli}}, and
  \bibinfo{author}{\bibfnamefont{N.~J.} \bibnamefont{Zabusky}},
  \bibinfo{year}{1987}, \bibinfo{journal}{J. Phys. Ocean.}
  \textbf{\bibinfo{volume}{17}}, \bibinfo{pages}{1408}.

\bibitem[{\citenamefont{Flierl and Morrison}(2011)}]{flierl11}
\bibinfo{author}{\bibnamefont{Flierl}, \bibfnamefont{G.~R.}}, and
  \bibinfo{author}{\bibfnamefont{P.~J.} \bibnamefont{Morrison}},
  \bibinfo{year}{2011}, \bibinfo{journal}{Physica D}
  \textbf{\bibinfo{volume}{240}}, \bibinfo{pages}{212}.

\bibitem[{\citenamefont{Franklin} \emph{et~al.}(1972)\citenamefont{Franklin,
  Hamberger, and Smith}}]{franklin72a}
\bibinfo{author}{\bibnamefont{Franklin}, \bibfnamefont{R.~N.}},
  \bibinfo{author}{\bibfnamefont{S.~M.} \bibnamefont{Hamberger}}, and
  \bibinfo{author}{\bibfnamefont{G.~J.} \bibnamefont{Smith}},
  \bibinfo{year}{1972}, \bibinfo{journal}{Phys. Rev. Lett.}
  \textbf{\bibinfo{volume}{29}}, \bibinfo{pages}{914}.

\bibitem[{\citenamefont{Fried and Conte}(1961)}]{fried}
\bibinfo{author}{\bibnamefont{Fried}, \bibfnamefont{B.~D.}}, and
  \bibinfo{author}{\bibfnamefont{S.~D.} \bibnamefont{Conte}},
  \bibinfo{year}{1961}, \emph{\bibinfo{title}{The plasma dispersion function}}
  (\bibinfo{publisher}{Academic Press}, \bibinfo{address}{London}).

\bibitem[{\citenamefont{Frieman} \emph{et~al.}(1963)\citenamefont{Frieman,
  Bodner, and Rutherford}}]{frieman63a}
\bibinfo{author}{\bibnamefont{Frieman}, \bibfnamefont{E.}},
  \bibinfo{author}{\bibfnamefont{S.}~\bibnamefont{Bodner}}, and
  \bibinfo{author}{\bibfnamefont{P.}~\bibnamefont{Rutherford}},
  \bibinfo{year}{1963}, \bibinfo{journal}{Phys. Fluids}
  \textbf{\bibinfo{volume}{6}}, \bibinfo{pages}{1298}.

\bibitem[{\citenamefont{Fujiwara}(1981)}]{fujiwara}
\bibinfo{author}{\bibnamefont{Fujiwara}, \bibfnamefont{T.}},
  \bibinfo{year}{1981}, \bibinfo{journal}{{Publ. Astron. Soc. Japan}}
  \textbf{\bibinfo{volume}{33}}, \bibinfo{pages}{531}.

\bibitem[{\citenamefont{Gakhov}(1990)}]{gakhov}
\bibinfo{author}{\bibnamefont{Gakhov}, \bibfnamefont{F.}},
  \bibinfo{year}{1990}, \emph{\bibinfo{title}{Boundary Value Problems}}
  (\bibinfo{publisher}{Dover}, \bibinfo{address}{New York}).

\bibitem[{\citenamefont{Goldstein and Hultgren}(1988)}]{goldstein88a}
\bibinfo{author}{\bibnamefont{Goldstein}, \bibfnamefont{M.~E.}}, and
  \bibinfo{author}{\bibfnamefont{L.~S.} \bibnamefont{Hultgren}},
  \bibinfo{year}{1988}, \bibinfo{journal}{J.~Fluid Mech.}
  \textbf{\bibinfo{volume}{197}}, \bibinfo{pages}{259}.

\bibitem[{\citenamefont{Goldstein and Leib}(1988)}]{goldstein88b}
\bibinfo{author}{\bibnamefont{Goldstein}, \bibfnamefont{M.~E.}}, and
  \bibinfo{author}{\bibfnamefont{S.~J.} \bibnamefont{Leib}},
  \bibinfo{year}{1988}, \bibinfo{journal}{J.~Fluid Mech.}
  \textbf{\bibinfo{volume}{191}}, \bibinfo{pages}{481}.

\bibitem[{\citenamefont{Goldstein and Leib}(1989)}]{goldstein89a}
\bibinfo{author}{\bibnamefont{Goldstein}, \bibfnamefont{M.~E.}}, and
  \bibinfo{author}{\bibfnamefont{S.~J.} \bibnamefont{Leib}},
  \bibinfo{year}{1989}, \bibinfo{journal}{J. Fluid Mech.}
  \textbf{\bibinfo{volume}{207}}, \bibinfo{pages}{73}.

\bibitem[{\citenamefont{Haberman}(1973)}]{haberman73a}
\bibinfo{author}{\bibnamefont{Haberman}, \bibfnamefont{R.}},
  \bibinfo{year}{1973}, \bibinfo{journal}{Stud. Appl. Math.}
  \textbf{\bibinfo{volume}{51}}, \bibinfo{pages}{139}.

\bibitem[{\citenamefont{Hagstrom and Morrison}(2011)}]{hagstrom}
\bibinfo{author}{\bibnamefont{Hagstrom}, \bibfnamefont{G.~I.}}, and
  \bibinfo{author}{\bibfnamefont{P.~J.} \bibnamefont{Morrison}},
  \bibinfo{year}{2011}, \bibinfo{journal}{Trans. Theory and Stat. Phys.}
  \textbf{\bibinfo{volume}{39}}, \bibinfo{pages}{466}.

\bibitem[{\citenamefont{Hagstrom and Morrison}(2013)}]{HagMor}
\bibinfo{author}{\bibnamefont{Hagstrom}, \bibfnamefont{G.~I.}}, and
  \bibinfo{author}{\bibfnamefont{P.~J.} \bibnamefont{Morrison}},
  \bibinfo{year}{2013}, in \emph{\bibinfo{booktitle}{Spectral Analysis,
  Stability and Bifurcations in Modern Nonlinear Physical Systems}}, edited by
  \bibinfo{editor}{\bibfnamefont{O.}~\bibnamefont{Kirrilov}}
  (\bibinfo{publisher}{Springer-Verlag}, \bibinfo{address}{Berlin}).

\bibitem[{\citenamefont{Haynes}(1985)}]{haynes85a}
\bibinfo{author}{\bibnamefont{Haynes}, \bibfnamefont{P.~H.}},
  \bibinfo{year}{1985}, \bibinfo{journal}{J.~Fluid Mech.}
  \textbf{\bibinfo{volume}{161}}, \bibinfo{pages}{493}.

\bibitem[{\citenamefont{Heath} \emph{et~al.}(2012)\citenamefont{Heath, Gamba,
  Morrison, and Michler}}]{Heath}
\bibinfo{author}{\bibnamefont{Heath}, \bibfnamefont{R.~E.}},
  \bibinfo{author}{\bibfnamefont{I.~M.} \bibnamefont{Gamba}},
  \bibinfo{author}{\bibfnamefont{P.~J.} \bibnamefont{Morrison}}, and
  \bibinfo{author}{\bibfnamefont{C.}~\bibnamefont{Michler}},
  \bibinfo{year}{2012}, \bibinfo{journal}{J. Comp. Phys.}
  \textbf{\bibinfo{volume}{231}}, \bibinfo{pages}{1140}.

\bibitem[{\citenamefont{Hickernell}(1984)}]{hickernell84a}
\bibinfo{author}{\bibnamefont{Hickernell}, \bibfnamefont{F.~J.}},
  \bibinfo{year}{1984}, \bibinfo{journal}{J.~Fluid Mech.}
  \textbf{\bibinfo{volume}{142}}, \bibinfo{pages}{431}.

\bibitem[{\citenamefont{Howard}(1964)}]{howard64b}
\bibinfo{author}{\bibnamefont{Howard}, \bibfnamefont{L.~N.}},
  \bibinfo{year}{1964}, \bibinfo{journal}{J. de M\'{e}canique}
  \textbf{\bibinfo{volume}{3}}, \bibinfo{pages}{433}.

\bibitem[{\citenamefont{Huerre}(1987)}]{huerre87a}
\bibinfo{author}{\bibnamefont{Huerre}, \bibfnamefont{P.}},
  \bibinfo{year}{1987}, \bibinfo{journal}{Proc. Roy. Soc. A}
  \textbf{\bibinfo{volume}{409}}, \bibinfo{pages}{369}.

\bibitem[{\citenamefont{Illner}(2000)}]{Illner00}
\bibinfo{author}{\bibnamefont{Illner}, \bibfnamefont{R.}},
  \bibinfo{year}{2000}, \bibinfo{journal}{Fields Inst. Comm}
  \textbf{\bibinfo{volume}{27}}, \bibinfo{pages}{98}.

\bibitem[{\citenamefont{Imamura} \emph{et~al.}(1969)\citenamefont{Imamura,
  Sugihara, and Taniuti}}]{imamura69a}
\bibinfo{author}{\bibnamefont{Imamura}, \bibfnamefont{T.}},
  \bibinfo{author}{\bibfnamefont{R.}~\bibnamefont{Sugihara}}, and
  \bibinfo{author}{\bibfnamefont{T.}~\bibnamefont{Taniuti}},
  \bibinfo{year}{1969}, \bibinfo{journal}{J. Phys. Soc. Japan}
  \textbf{\bibinfo{volume}{27}}, \bibinfo{pages}{1623}.

\bibitem[{\citenamefont{Inagaki and Konishi}(1993)}]{inagaki}
\bibinfo{author}{\bibnamefont{Inagaki}, \bibfnamefont{S.}}, and
  \bibinfo{author}{\bibfnamefont{T.}~\bibnamefont{Konishi}},
  \bibinfo{year}{1993}, \bibinfo{journal}{{Publ. Astron. Soc. Japan}}
  \textbf{\bibinfo{volume}{45}}, \bibinfo{pages}{733}.

\bibitem[{\citenamefont{van Kampen}(1955)}]{vankampen55a}
\bibinfo{author}{\bibnamefont{van Kampen}, \bibfnamefont{N.~G.}},
  \bibinfo{year}{1955}, \bibinfo{journal}{Physica}
  \textbf{\bibinfo{volume}{21}}, \bibinfo{pages}{949}.

\bibitem[{\citenamefont{Kato}(1966)}]{kato}
\bibinfo{author}{\bibnamefont{Kato}, \bibfnamefont{T.}}, \bibinfo{year}{1966},
  \emph{\bibinfo{title}{Perturbation Theory for Linear Operators}}
  (\bibinfo{publisher}{Springer-Verlag, Berlin}).

\bibitem[{\citenamefont{Killworth and {McIntyre}}(1985)}]{killworth85a}
\bibinfo{author}{\bibnamefont{Killworth}, \bibfnamefont{P.~D.}}, and
  \bibinfo{author}{\bibfnamefont{M.~E.} \bibnamefont{{McIntyre}}},
  \bibinfo{year}{1985}, \bibinfo{journal}{J.~Fluid Mech.}
  \textbf{\bibinfo{volume}{161}}, \bibinfo{pages}{449}.

\bibitem[{\citenamefont{Krall and Trivelpiece}(1973)}]{krall}
\bibinfo{author}{\bibnamefont{Krall}, \bibfnamefont{N.~A.}}, and
  \bibinfo{author}{\bibfnamefont{A.~W.} \bibnamefont{Trivelpiece}},
  \bibinfo{year}{1973}, \emph{\bibinfo{title}{Principles of Plasma Physics}}
  (\bibinfo{publisher}{New York: McGraw-Hill}).

\bibitem[{\citenamefont{Krechetnikov and Marsden}(2007)}]{krech}
\bibinfo{author}{\bibnamefont{Krechetnikov}, \bibfnamefont{R.}}, and
  \bibinfo{author}{\bibfnamefont{J.~E.} \bibnamefont{Marsden}},
  \bibinfo{year}{2007}, \bibinfo{journal}{Rev. Mod. Phys.}
  \textbf{\bibinfo{volume}{79}}, \bibinfo{pages}{519}.

\bibitem[{\citenamefont{Kre\u{i}n}(1950)}]{krein}
\bibinfo{author}{\bibnamefont{Kre\u{i}n}, \bibfnamefont{M.}},
  \bibinfo{year}{1950}, \bibinfo{journal}{Dokl. Akad. Nauk SSSR A}
  \textbf{\bibinfo{volume}{73}}, \bibinfo{pages}{445}.

\bibitem[{\citenamefont{Kueny and Morrison}(1995)}]{kueny}
\bibinfo{author}{\bibnamefont{Kueny}, \bibfnamefont{C.~S.}}, and
  \bibinfo{author}{\bibfnamefont{P.~J.} \bibnamefont{Morrison}},
  \bibinfo{year}{1995}, \bibinfo{journal}{Phys. Plasmas}
  \textbf{\bibinfo{volume}{2}}, \bibinfo{pages}{1926}.

\bibitem[{\citenamefont{Kuramoto}(1975)}]{kuramoto75a}
\bibinfo{author}{\bibnamefont{Kuramoto}, \bibfnamefont{Y.}},
  \bibinfo{year}{1975}, in \emph{\bibinfo{booktitle}{International symposium on
  mathematical problems in theoretical physics}}, edited by
  \bibinfo{editor}{\bibfnamefont{H.}~\bibnamefont{Araki}}
  (\bibinfo{publisher}{Springer-Verlag}, \bibinfo{address}{New York}),
  number~\bibinfo{number}{39} in \bibinfo{series}{Lecture Notes in Physics},
  pp. \bibinfo{pages}{420--422}.

\bibitem[{\citenamefont{Lancelotti and Dorning}(2003)}]{lancelotti}
\bibinfo{author}{\bibnamefont{Lancelotti}, \bibfnamefont{C.}}, and
  \bibinfo{author}{\bibfnamefont{J.~J.} \bibnamefont{Dorning}},
  \bibinfo{year}{2003}, \bibinfo{journal}{{Phys. Rev. E}}
  \textbf{\bibinfo{volume}{68}}, \bibinfo{pages}{026406}.

\bibitem[{\citenamefont{Landahl}(1962)}]{landahl62a}
\bibinfo{author}{\bibnamefont{Landahl}, \bibfnamefont{M.}},
  \bibinfo{year}{1962}, \bibinfo{journal}{J. Fluid Mech.}
  \textbf{\bibinfo{volume}{13}}, \bibinfo{pages}{607}.

\bibitem[{\citenamefont{Landahl}(1986)}]{landahl86}
\bibinfo{author}{\bibnamefont{Landahl}, \bibfnamefont{M.}},
  \bibinfo{year}{1986}, in \emph{\bibinfo{booktitle}{1986 Summer Study Program
  in Geophysical Fluid Dynamics at Woods Hole: Order and Disorder in Turbulent
  Shear Flow}}, edited by
  \bibinfo{editor}{\bibfnamefont{M.}~\bibnamefont{Stern}}
  (\bibinfo{publisher}{Woods Hole, MA: Woods Hole Oceanographic Institution}),
  volume \bibinfo{volume}{WHOI-86-45}, pp. \bibinfo{pages}{44--50}.

\bibitem[{\citenamefont{Landau}(1946)}]{Landau1946}
\bibinfo{author}{\bibnamefont{Landau}, \bibfnamefont{L.~D.}},
  \bibinfo{year}{1946}, \bibinfo{journal}{J. Phys. (Moscow)}
  \textbf{\bibinfo{volume}{10}}, \bibinfo{pages}{25}.

\bibitem[{\citenamefont{Leib and Goldstein}(1989)}]{leib89a}
\bibinfo{author}{\bibnamefont{Leib}, \bibfnamefont{S.~J.}}, and
  \bibinfo{author}{\bibfnamefont{M.~E.} \bibnamefont{Goldstein}},
  \bibinfo{year}{1989}, \bibinfo{journal}{Phys. Fluids A}
  \textbf{\bibinfo{volume}{1}}, \bibinfo{pages}{513}.

\bibitem[{\citenamefont{Lin}(1945)}]{lin45a}
\bibinfo{author}{\bibnamefont{Lin}, \bibfnamefont{C.~C.}},
  \bibinfo{year}{1945}, \bibinfo{journal}{Quart. App. Maths.}
  \textbf{\bibinfo{volume}{3}}, \bibinfo{pages}{117}.

\bibitem[{\citenamefont{Lynden-Bell}(1967)}]{lyndenbell}
\bibinfo{author}{\bibnamefont{Lynden-Bell}, \bibfnamefont{D.}},
  \bibinfo{year}{1967}, \bibinfo{journal}{Mon. Not. R. astr. Soc.}
  \textbf{\bibinfo{volume}{136}}, \bibinfo{pages}{101}.

\bibitem[{\citenamefont{MacKay and Saffman}(1986)}]{mackay}
\bibinfo{author}{\bibnamefont{MacKay}, \bibfnamefont{R.~S.}}, and
  \bibinfo{author}{\bibfnamefont{P.~G.} \bibnamefont{Saffman}},
  \bibinfo{year}{1986}, \bibinfo{journal}{Proc. R. Soc. Lond. A}
  \textbf{\bibinfo{volume}{406}}, \bibinfo{pages}{115}.

\bibitem[{\citenamefont{Malkus and Veronis}(1958)}]{malkus58a}
\bibinfo{author}{\bibnamefont{Malkus}, \bibfnamefont{W.~V.~R.}}, and
  \bibinfo{author}{\bibfnamefont{G.}~\bibnamefont{Veronis}},
  \bibinfo{year}{1958}, \bibinfo{journal}{J.~Fluid Mech.}
  \textbf{\bibinfo{volume}{4}}, \bibinfo{pages}{225}.

\bibitem[{\citenamefont{Malmberg and Wharton}(1964)}]{Mal64}
\bibinfo{author}{\bibnamefont{Malmberg}, \bibfnamefont{J.~H.}}, and
  \bibinfo{author}{\bibfnamefont{C.~B.} \bibnamefont{Wharton}},
  \bibinfo{year}{1964}, \bibinfo{journal}{Phys. Rev. Lett.}
  \textbf{\bibinfo{volume}{17}}.

\bibitem[{\citenamefont{Malmberg} \emph{et~al.}(1966)\citenamefont{Malmberg,
  Wharton, and Drummond}}]{Mal66}
\bibinfo{author}{\bibnamefont{Malmberg}, \bibfnamefont{J.~H.}},
  \bibinfo{author}{\bibfnamefont{C.~B.} \bibnamefont{Wharton}}, and
  \bibinfo{author}{\bibfnamefont{W.~E.} \bibnamefont{Drummond}},
  \bibinfo{year}{1966}, in \emph{\bibinfo{booktitle}{Plasma Physics and
  Controlled Nuclear Fusion Research}} (\bibinfo{publisher}{International
  Atomic Energy Agency, Vienna}), volume~\bibinfo{volume}{1}.

\bibitem[{\citenamefont{Manfredi}(1997)}]{manfredi97a}
\bibinfo{author}{\bibnamefont{Manfredi}, \bibfnamefont{G.}},
  \bibinfo{year}{1997}, \bibinfo{journal}{Phys. Rev. Lett.}
  \textbf{\bibinfo{volume}{79}}, \bibinfo{pages}{2815}.

\bibitem[{\citenamefont{Marsden and Ratiu}(1999)}]{marsden}
\bibinfo{author}{\bibnamefont{Marsden}, \bibfnamefont{J.~E.}}, and
  \bibinfo{author}{\bibfnamefont{T.~S.} \bibnamefont{Ratiu}},
  \bibinfo{year}{1999}, \emph{\bibinfo{title}{{I}ntroduction to {M}echanics and
  {S}ymmetry}} (\bibinfo{publisher}{Springer-Verlag}, \bibinfo{address}{New
  York, NY}).

\bibitem[{\citenamefont{Mattor and Mitchell}(1996)}]{mattor96}
\bibinfo{author}{\bibnamefont{Mattor}, \bibfnamefont{N.}}, and
  \bibinfo{author}{\bibfnamefont{T.~B.} \bibnamefont{Mitchell}},
  \bibinfo{year}{1996}, \bibinfo{journal}{{Strophys. J.}}
  \textbf{\bibinfo{volume}{472}}, \bibinfo{pages}{532}.

\bibitem[{\citenamefont{van~der Meer}(1985)}]{meer}
\bibinfo{author}{\bibnamefont{van~der Meer}, \bibfnamefont{J.~C.}},
  \bibinfo{year}{1985}, \emph{\bibinfo{title}{The Hamiltonian Hopf
  bifurcation}}, Lecture Notes in Mathematics 1160
  (\bibinfo{publisher}{Springer-Verlag}, \bibinfo{address}{Berlin}).

\bibitem[{\citenamefont{Mitchell and Driscoll}(1994)}]{mitchell94}
\bibinfo{author}{\bibnamefont{Mitchell}, \bibfnamefont{T.~B.}}, and
  \bibinfo{author}{\bibfnamefont{C.~F.} \bibnamefont{Driscoll}},
  \bibinfo{year}{1994}, \bibinfo{journal}{{Phys. Rev.Lett.}}
  \textbf{\bibinfo{volume}{73}}, \bibinfo{pages}{2196}.

\bibitem[{\citenamefont{M\"{o}ller and Montgomery}(1999)}]{montgomery99}
\bibinfo{author}{\bibnamefont{M\"{o}ller}, \bibfnamefont{J.~D.}}, and
  \bibinfo{author}{\bibfnamefont{M.~T.} \bibnamefont{Montgomery}},
  \bibinfo{year}{1999}, \bibinfo{journal}{J. Atmos. Sci.}
  \textbf{\bibinfo{volume}{56}}, \bibinfo{pages}{1674}.

\bibitem[{\citenamefont{Morrison}(1980)}]{morrison80}
\bibinfo{author}{\bibnamefont{Morrison}, \bibfnamefont{P.~J.}},
  \bibinfo{year}{1980}, \bibinfo{journal}{Phys. Lett. A}
  \textbf{\bibinfo{volume}{80}}, \bibinfo{pages}{383}.

\bibitem[{\citenamefont{Morrison}(1982)}]{morrison82}
\bibinfo{author}{\bibnamefont{Morrison}, \bibfnamefont{P.~J.}},
  \bibinfo{year}{1982}, in \emph{\bibinfo{booktitle}{{M}athematical {M}ethods
  in {H}ydrodynamics and {I}ntegrability in {D}ynamical {S}ystems}}, edited by
  \bibinfo{editor}{\bibfnamefont{M.}~\bibnamefont{Tabor}} and
  \bibinfo{editor}{\bibfnamefont{Y.}~\bibnamefont{Treve}}
  (\bibinfo{publisher}{American Inst. Phys.}), volume~\bibinfo{volume}{88}, pp.
  \bibinfo{pages}{13--46}.

\bibitem[{\citenamefont{Morrison}(1994)}]{morrison94}
\bibinfo{author}{\bibnamefont{Morrison}, \bibfnamefont{P.~J.}},
  \bibinfo{year}{1994}, \bibinfo{journal}{Phys. Plasmas}
  \textbf{\bibinfo{volume}{1}}, \bibinfo{pages}{1447}.

\bibitem[{\citenamefont{Morrison}(1998)}]{morrison98}
\bibinfo{author}{\bibnamefont{Morrison}, \bibfnamefont{P.~J.}},
  \bibinfo{year}{1998}, \bibinfo{journal}{Rev. Mod. Phys.}
  \textbf{\bibinfo{volume}{70}}, \bibinfo{pages}{467}.

\bibitem[{\citenamefont{Morrison}(2000)}]{morrison00a}
\bibinfo{author}{\bibnamefont{Morrison}, \bibfnamefont{P.~J.}},
  \bibinfo{year}{2000}, \bibinfo{journal}{Trans. Theory and Stat. Phys.}
  \textbf{\bibinfo{volume}{29}}, \bibinfo{pages}{397}.

\bibitem[{\citenamefont{Morrison}(2003)}]{morrison03}
\bibinfo{author}{\bibnamefont{Morrison}, \bibfnamefont{P.~J.}},
  \bibinfo{year}{2003}, in \emph{\bibinfo{booktitle}{Nonlinear Processes in
  Geophysical Fluid Dynamics}}, edited by \bibinfo{editor}{\bibfnamefont{O.~U.}
  \bibnamefont{Velasco~Fuentes}},
  \bibinfo{editor}{\bibfnamefont{J.}~\bibnamefont{Sheinbaum}}, and
  \bibinfo{editor}{\bibfnamefont{J.}~\bibnamefont{Ochoa}}
  (\bibinfo{publisher}{Kluwer, Dordrecht}), pp. \bibinfo{pages}{53--69}.

\bibitem[{\citenamefont{Morrison}(2013)}]{morrison13}
\bibinfo{author}{\bibnamefont{Morrison}, \bibfnamefont{P.~J.}},
  \bibinfo{year}{2013}, \bibinfo{journal}{Phys. Plasmas}
  \textbf{\bibinfo{volume}{20}}, \bibinfo{pages}{012104}.

\bibitem[{\citenamefont{Morrison and Greene}(1980)}]{MG80}
\bibinfo{author}{\bibnamefont{Morrison}, \bibfnamefont{P.~J.}}, and
  \bibinfo{author}{\bibfnamefont{J.~M.} \bibnamefont{Greene}},
  \bibinfo{year}{1980}, \bibinfo{journal}{Phys. Rev. Lett.}
  \textbf{\bibinfo{volume}{45}}, \bibinfo{pages}{790}.

\bibitem[{\citenamefont{Morrison and Kotschenreuther}(1990)}]{morrison90}
\bibinfo{author}{\bibnamefont{Morrison}, \bibfnamefont{P.~J.}}, and
  \bibinfo{author}{\bibfnamefont{M.}~\bibnamefont{Kotschenreuther}},
  \bibinfo{year}{1990}, in \emph{\bibinfo{booktitle}{Nonlinear World: IV
  International Workshop on Nonlinear and Turbulent Processes in Physics}},
  edited by \bibinfo{editor}{\bibfnamefont{V.~G.} \bibnamefont{BarÕyakhtar}},
  \bibinfo{editor}{\bibfnamefont{V.~M.} \bibnamefont{Chernousenko}},
  \bibinfo{editor}{\bibfnamefont{N.~S.} \bibnamefont{Erokhin}},
  \bibinfo{editor}{\bibfnamefont{A.~B.} \bibnamefont{Sitenko}}, and
  \bibinfo{editor}{\bibfnamefont{V.~E.} \bibnamefont{Zakharov}}
  (\bibinfo{publisher}{World Scientific}, \bibinfo{address}{Singapore}), pp.
  \bibinfo{pages}{910--932}.

\bibitem[{\citenamefont{Morrison and Pfirsch}(1989)}]{mp89}
\bibinfo{author}{\bibnamefont{Morrison}, \bibfnamefont{P.~J.}}, and
  \bibinfo{author}{\bibfnamefont{D.}~\bibnamefont{Pfirsch}},
  \bibinfo{year}{1989}, \bibinfo{journal}{Phys. Rev. A}
  \textbf{\bibinfo{volume}{40}}, \bibinfo{pages}{3898}.

\bibitem[{\citenamefont{Morrison and Pfirsch}(1990)}]{mp90}
\bibinfo{author}{\bibnamefont{Morrison}, \bibfnamefont{P.~J.}}, and
  \bibinfo{author}{\bibfnamefont{D.}~\bibnamefont{Pfirsch}},
  \bibinfo{year}{1990}, \bibinfo{journal}{Phys. Fluids B}
  \textbf{\bibinfo{volume}{2}}, \bibinfo{pages}{1105}.

\bibitem[{\citenamefont{Morrison and Pfirsch}(1992)}]{MP92}
\bibinfo{author}{\bibnamefont{Morrison}, \bibfnamefont{P.~J.}}, and
  \bibinfo{author}{\bibfnamefont{D.}~\bibnamefont{Pfirsch}},
  \bibinfo{year}{1992}, \bibinfo{journal}{Phys.~Fluids}
  \textbf{\bibinfo{volume}{4B}}, \bibinfo{pages}{3038}.

\bibitem[{\citenamefont{Moser}(1975)}]{Moser75}
\bibinfo{author}{\bibnamefont{Moser}, \bibfnamefont{J.}}, \bibinfo{year}{1975},
  \bibinfo{journal}{Adv. Math.} \textbf{\bibinfo{volume}{16}},
  \bibinfo{pages}{197}.

\bibitem[{\citenamefont{Moser}(1976)}]{moser76}
\bibinfo{author}{\bibnamefont{Moser}, \bibfnamefont{J.}}, \bibinfo{year}{1976},
  \bibinfo{journal}{Comm. Pure App. Math.} \textbf{\bibinfo{volume}{29}},
  \bibinfo{pages}{727}.

\bibitem[{\citenamefont{Mouhot and Villani}(2011)}]{cedric}
\bibinfo{author}{\bibnamefont{Mouhot}, \bibfnamefont{C.}}, and
  \bibinfo{author}{\bibfnamefont{C.}~\bibnamefont{Villani}},
  \bibinfo{year}{2011}, \bibinfo{journal}{Acta Math.}
  \textbf{\bibinfo{volume}{207}}, \bibinfo{pages}{9}.

\bibitem[{\citenamefont{Mynick and Kaufman}(1978)}]{mynick}
\bibinfo{author}{\bibnamefont{Mynick}, \bibfnamefont{H.~E.}}, and
  \bibinfo{author}{\bibfnamefont{A.}~\bibnamefont{Kaufman}},
  \bibinfo{year}{1978}, \bibinfo{journal}{Phys. Fluids}
  \textbf{\bibinfo{volume}{21}}, \bibinfo{pages}{653}.

\bibitem[{\citenamefont{O'Neil}(1965)}]{oneil65a}
\bibinfo{author}{\bibnamefont{O'Neil}, \bibfnamefont{T.~M.}},
  \bibinfo{year}{1965}, \bibinfo{journal}{Phys. Fluids}
  \textbf{\bibinfo{volume}{8}}, \bibinfo{pages}{2255}.

\bibitem[{\citenamefont{O'Neil} \emph{et~al.}(1971)\citenamefont{O'Neil,
  Winfrey, and Malmberg}}]{oneil71a}
\bibinfo{author}{\bibnamefont{O'Neil}, \bibfnamefont{T.~M.}},
  \bibinfo{author}{\bibfnamefont{J.~H.} \bibnamefont{Winfrey}}, and
  \bibinfo{author}{\bibfnamefont{J.~H.} \bibnamefont{Malmberg}},
  \bibinfo{year}{1971}, \bibinfo{journal}{Phys. Fluids}
  \textbf{\bibinfo{volume}{14}}, \bibinfo{pages}{1204}.

\bibitem[{\citenamefont{Onishchenko}
  \emph{et~al.}(1971)\citenamefont{Onishchenko, Linetskii, Matsiborko, Shapiro,
  and Shevchenko}}]{onishchenko71a}
\bibinfo{author}{\bibnamefont{Onishchenko}, \bibfnamefont{I.~N.}},
  \bibinfo{author}{\bibfnamefont{A.~R.} \bibnamefont{Linetskii}},
  \bibinfo{author}{\bibfnamefont{N.~G.} \bibnamefont{Matsiborko}},
  \bibinfo{author}{\bibfnamefont{V.~D.} \bibnamefont{Shapiro}}, and
  \bibinfo{author}{\bibfnamefont{V.~I.} \bibnamefont{Shevchenko}},
  \bibinfo{year}{1971}, \bibinfo{journal}{Soviet Phys. JETP.}
  \textbf{\bibinfo{volume}{11}}, \bibinfo{pages}{281}.

\bibitem[{\citenamefont{Padhye and Horton}(1999)}]{padhye}
\bibinfo{author}{\bibnamefont{Padhye}, \bibfnamefont{N.}}, and
  \bibinfo{author}{\bibfnamefont{W.}~\bibnamefont{Horton}},
  \bibinfo{year}{1999}, \bibinfo{journal}{Phys. Plasmas}
  \textbf{\bibinfo{volume}{6}}.

\bibitem[{\citenamefont{Pedlosky}(1987)}]{pedlosky}
\bibinfo{author}{\bibnamefont{Pedlosky}, \bibfnamefont{J.}},
  \bibinfo{year}{1987}, \emph{\bibinfo{title}{Geophysical Fluid Dynamics}}
  (\bibinfo{publisher}{Springer-Verlag}).

\bibitem[{\citenamefont{Penrose}(1960)}]{penrose}
\bibinfo{author}{\bibnamefont{Penrose}, \bibfnamefont{O.}},
  \bibinfo{year}{1960}, \bibinfo{journal}{Phys. Fluids}
  \textbf{\bibinfo{volume}{3}}, \bibinfo{pages}{258}.

\bibitem[{\citenamefont{Redekopp}(1977)}]{redekopp}
\bibinfo{author}{\bibnamefont{Redekopp}, \bibfnamefont{L.~G.}},
  \bibinfo{year}{1977}, \bibinfo{journal}{J. Fluid Mech.}
  \textbf{\bibinfo{volume}{82}}, \bibinfo{pages}{725}.

\bibitem[{\citenamefont{Reutov}(1980)}]{reutov}
\bibinfo{author}{\bibnamefont{Reutov}, \bibfnamefont{V.~P.}},
  \bibinfo{year}{1980}, \bibinfo{journal}{Izvestiya, Atmosph. Oceanic Phys.}
  \textbf{\bibinfo{volume}{16}}, \bibinfo{pages}{938}.

\bibitem[{\citenamefont{Rosenbluth and Simon}(1964)}]{rosenbluth64a}
\bibinfo{author}{\bibnamefont{Rosenbluth}, \bibfnamefont{M.~N.}}, and
  \bibinfo{author}{\bibfnamefont{A.}~\bibnamefont{Simon}},
  \bibinfo{year}{1964}, \bibinfo{journal}{Phys. Fluids}
  \textbf{\bibinfo{volume}{7}}, \bibinfo{pages}{557}.

\bibitem[{\citenamefont{Rosencrans and Sattinger}(1966)}]{rosencrans66a}
\bibinfo{author}{\bibnamefont{Rosencrans}, \bibfnamefont{S.~I.}}, and
  \bibinfo{author}{\bibfnamefont{D.~H.} \bibnamefont{Sattinger}},
  \bibinfo{year}{1966}, \bibinfo{journal}{J. Math. and Phys.}
  \textbf{\bibinfo{volume}{45}}, \bibinfo{pages}{289}.

\bibitem[{\citenamefont{Rossi} \emph{et~al.}(1997)\citenamefont{Rossi,
  Lingevitch, and Bernoff}}]{rossi97a}
\bibinfo{author}{\bibnamefont{Rossi}, \bibfnamefont{L.~F.}},
  \bibinfo{author}{\bibfnamefont{J.~F.} \bibnamefont{Lingevitch}}, and
  \bibinfo{author}{\bibfnamefont{A.~J.} \bibnamefont{Bernoff}},
  \bibinfo{year}{1997}, \bibinfo{journal}{Phys. Fluids}
  \textbf{\bibinfo{volume}{9}}, \bibinfo{pages}{2329}.

\bibitem[{\citenamefont{Schade}(1964)}]{schade64a}
\bibinfo{author}{\bibnamefont{Schade}, \bibfnamefont{H.}},
  \bibinfo{year}{1964}, \bibinfo{journal}{Phys. Fluids}
  \textbf{\bibinfo{volume}{7}}, \bibinfo{pages}{623}.

\bibitem[{\citenamefont{Schecter} \emph{et~al.}(2000)\citenamefont{Schecter,
  Dubin, Cass, Driscoll, Lansky, and {O'Neil}}}]{schecter2000a}
\bibinfo{author}{\bibnamefont{Schecter}, \bibfnamefont{D.~A.}},
  \bibinfo{author}{\bibfnamefont{D.~H.~E.} \bibnamefont{Dubin}},
  \bibinfo{author}{\bibfnamefont{A.~C.} \bibnamefont{Cass}},
  \bibinfo{author}{\bibfnamefont{C.~F.} \bibnamefont{Driscoll}},
  \bibinfo{author}{\bibfnamefont{I.~M.} \bibnamefont{Lansky}}, and
  \bibinfo{author}{\bibfnamefont{T.~M.} \bibnamefont{{O'Neil}}},
  \bibinfo{year}{2000}, \bibinfo{journal}{Phys. Fluids}
  \textbf{\bibinfo{volume}{12}}, \bibinfo{pages}{2397}.

\bibitem[{\citenamefont{Shagalov} \emph{et~al.}(2009)\citenamefont{Shagalov,
  Reutov, and Rybushkina}}]{shagalov09}
\bibinfo{author}{\bibnamefont{Shagalov}, \bibfnamefont{S.~V.}},
  \bibinfo{author}{\bibfnamefont{V.~P.} \bibnamefont{Reutov}}, and
  \bibinfo{author}{\bibfnamefont{G.~V.} \bibnamefont{Rybushkina}},
  \bibinfo{year}{2009}, \bibinfo{journal}{{Izvestiya Atmospheric and Oceanic
  Physics}} \textbf{\bibinfo{volume}{45}}, \bibinfo{pages}{629}.

\bibitem[{\citenamefont{Shagalov} \emph{et~al.}(2010)\citenamefont{Shagalov,
  Reutov, and Rybushkina}}]{shagalov10}
\bibinfo{author}{\bibnamefont{Shagalov}, \bibfnamefont{S.~V.}},
  \bibinfo{author}{\bibfnamefont{V.~P.} \bibnamefont{Reutov}}, and
  \bibinfo{author}{\bibfnamefont{G.~V.} \bibnamefont{Rybushkina}},
  \bibinfo{year}{2010}, \bibinfo{journal}{{Izvestiya Atmospheric and Oceanic
  Physics}} \textbf{\bibinfo{volume}{46}}, \bibinfo{pages}{95}.

\bibitem[{\citenamefont{Shukhman}(1989)}]{shukhman89a}
\bibinfo{author}{\bibnamefont{Shukhman}, \bibfnamefont{I.~G.}},
  \bibinfo{year}{1989}, \bibinfo{journal}{J.~Fluid Mech.}
  \textbf{\bibinfo{volume}{200}}, \bibinfo{pages}{425}.

\bibitem[{\citenamefont{Simon and Rosenbluth}(1976)}]{rosenbluth76a}
\bibinfo{author}{\bibnamefont{Simon}, \bibfnamefont{A.}}, and
  \bibinfo{author}{\bibfnamefont{M.}~\bibnamefont{Rosenbluth}},
  \bibinfo{year}{1976}, \bibinfo{journal}{Phys. Fluids}
  \textbf{\bibinfo{volume}{19}}, \bibinfo{pages}{1567}.

\bibitem[{\citenamefont{Smereka}(1998)}]{smerk}
\bibinfo{author}{\bibnamefont{Smereka}, \bibfnamefont{P.}},
  \bibinfo{year}{1998}, \bibinfo{journal}{Physica}
  \textbf{\bibinfo{volume}{124D}}, \bibinfo{pages}{104}.

\bibitem[{\citenamefont{Smith and Pereira}(1978)}]{smith}
\bibinfo{author}{\bibnamefont{Smith}, \bibfnamefont{G.~R.}}, and
  \bibinfo{author}{\bibfnamefont{N.}~\bibnamefont{Pereira}},
  \bibinfo{year}{1978}, \bibinfo{journal}{Phys. Fluids}
  \textbf{\bibinfo{volume}{21}}, \bibinfo{pages}{2253}.

\bibitem[{\citenamefont{Stein and Weiss}(1971)}]{stein}
\bibinfo{author}{\bibnamefont{Stein}, \bibfnamefont{E.~M.}}, and
  \bibinfo{author}{\bibfnamefont{G.}~\bibnamefont{Weiss}},
  \bibinfo{year}{1971}, \emph{\bibinfo{title}{Introduction to Fourier analysis
  on Euclidean spaces}} (\bibinfo{publisher}{Princeton University Press},
  \bibinfo{address}{Princeton, New Jersey}).

\bibitem[{\citenamefont{Stewartson}(1978)}]{stewartson78a}
\bibinfo{author}{\bibnamefont{Stewartson}, \bibfnamefont{K.}},
  \bibinfo{year}{1978}, \bibinfo{journal}{Geophys. Astrophys. Fluid Dyn.}
  \textbf{\bibinfo{volume}{9}}, \bibinfo{pages}{185}.

\bibitem[{\citenamefont{Stewartson}(1981)}]{stewartson81a}
\bibinfo{author}{\bibnamefont{Stewartson}, \bibfnamefont{K.}},
  \bibinfo{year}{1981}, \bibinfo{journal}{IMA J.~Applied Math.}
  \textbf{\bibinfo{volume}{27}}, \bibinfo{pages}{133}.

\bibitem[{\citenamefont{Strogatz}(2000)}]{strogatz}
\bibinfo{author}{\bibnamefont{Strogatz}, \bibfnamefont{S.~H.}},
  \bibinfo{year}{2000}, \bibinfo{journal}{Physica D}
  \textbf{\bibinfo{volume}{143}}, \bibinfo{pages}{1}.

\bibitem[{\citenamefont{Strogatz} \emph{et~al.}(1992)\citenamefont{Strogatz,
  Mirollo, and Matthews}}]{strogatz92a}
\bibinfo{author}{\bibnamefont{Strogatz}, \bibfnamefont{S.~H.}},
  \bibinfo{author}{\bibfnamefont{R.~E.} \bibnamefont{Mirollo}}, and
  \bibinfo{author}{\bibfnamefont{P.~C.} \bibnamefont{Matthews}},
  \bibinfo{year}{1992}, \bibinfo{journal}{Phys. Rev. Lett.}
  \textbf{\bibinfo{volume}{68}}, \bibinfo{pages}{2730}.

\bibitem[{\citenamefont{Stuart}(1960)}]{stuart60a}
\bibinfo{author}{\bibnamefont{Stuart}, \bibfnamefont{J.~T.}},
  \bibinfo{year}{1960}, \bibinfo{journal}{J.~Fluid Mech.}
  \textbf{\bibinfo{volume}{9}}, \bibinfo{pages}{353}.

\bibitem[{\citenamefont{Sturrock}(1958)}]{sturrock}
\bibinfo{author}{\bibnamefont{Sturrock}, \bibfnamefont{J.~A.}},
  \bibinfo{year}{1958}, \bibinfo{journal}{Phys. Rev.}
  \textbf{\bibinfo{volume}{112}}, \bibinfo{pages}{1488}.

\bibitem[{\citenamefont{Sugihara and Kamimura}(1972)}]{sugihara72a}
\bibinfo{author}{\bibnamefont{Sugihara}, \bibfnamefont{R.}}, and
  \bibinfo{author}{\bibfnamefont{T.}~\bibnamefont{Kamimura}},
  \bibinfo{year}{1972}, \bibinfo{journal}{J. Phys. Soc. Japan}
  \textbf{\bibinfo{volume}{33}}, \bibinfo{pages}{206}.

\bibitem[{\citenamefont{Tassi} \emph{et~al.}(2009)\citenamefont{Tassi, Chandre,
  and Morrison}}]{TaChaMo09}
\bibinfo{author}{\bibnamefont{Tassi}, \bibfnamefont{E.}},
  \bibinfo{author}{\bibfnamefont{C.}~\bibnamefont{Chandre}}, and
  \bibinfo{author}{\bibfnamefont{P.~J.} \bibnamefont{Morrison}},
  \bibinfo{year}{2009}, \bibinfo{journal}{Phys. Plasmas}
  \textbf{\bibinfo{volume}{16}}, \bibinfo{pages}{082301}.

\bibitem[{\citenamefont{Tassi and Morrison}(2011)}]{tassi1}
\bibinfo{author}{\bibnamefont{Tassi}, \bibfnamefont{E.}}, and
  \bibinfo{author}{\bibfnamefont{P.~J.} \bibnamefont{Morrison}},
  \bibinfo{year}{2011}, \bibinfo{journal}{Phys. Plasmas}
  \textbf{\bibinfo{volume}{18}}, \bibinfo{pages}{032115}.

\bibitem[{\citenamefont{Tassi} \emph{et~al.}(2008)\citenamefont{Tassi,
  Morrison, Waelbroeck, and Grasso}}]{tassi2}
\bibinfo{author}{\bibnamefont{Tassi}, \bibfnamefont{E.}},
  \bibinfo{author}{\bibfnamefont{P.~J.} \bibnamefont{Morrison}},
  \bibinfo{author}{\bibfnamefont{F.~L.} \bibnamefont{Waelbroeck}}, and
  \bibinfo{author}{\bibfnamefont{D.}~\bibnamefont{Grasso}},
  \bibinfo{year}{2008}, \bibinfo{journal}{Plasma Phys Cont. Fusion}
  \textbf{\bibinfo{volume}{50}}, \bibinfo{pages}{085014}.

\bibitem[{\citenamefont{Tennyson} \emph{et~al.}(1994)\citenamefont{Tennyson,
  Meiss, and Morrison}}]{tennyson94a}
\bibinfo{author}{\bibnamefont{Tennyson}, \bibfnamefont{J.~L.}},
  \bibinfo{author}{\bibfnamefont{J.~D.} \bibnamefont{Meiss}}, and
  \bibinfo{author}{\bibfnamefont{P.~J.} \bibnamefont{Morrison}},
  \bibinfo{year}{1994}, \bibinfo{journal}{Physica D}
  \textbf{\bibinfo{volume}{71}}, \bibinfo{pages}{1}.

\bibitem[{\citenamefont{Turner and Gilbert}(2007)}]{turner07}
\bibinfo{author}{\bibnamefont{Turner}, \bibfnamefont{M.~R.}}, and
  \bibinfo{author}{\bibfnamefont{A.~D.} \bibnamefont{Gilbert}},
  \bibinfo{year}{2007}, \bibinfo{journal}{J. Fluid Mech.}
  \textbf{\bibinfo{volume}{593}}, \bibinfo{pages}{255}.

\bibitem[{\citenamefont{Turner and Gilbert}(2008)}]{turner08}
\bibinfo{author}{\bibnamefont{Turner}, \bibfnamefont{M.~R.}}, and
  \bibinfo{author}{\bibfnamefont{A.~D.} \bibnamefont{Gilbert}},
  \bibinfo{year}{2008}, \bibinfo{journal}{J. Fluid Mech.}
  \textbf{\bibinfo{volume}{614}}, \bibinfo{pages}{381}.

\bibitem[{\citenamefont{Vanneste}(1996)}]{vanneste}
\bibinfo{author}{\bibnamefont{Vanneste}, \bibfnamefont{J.}},
  \bibinfo{year}{1996}, \bibinfo{journal}{{J. Fluid Mech.}}
  \textbf{\bibinfo{volume}{323}}, \bibinfo{pages}{317}.

\bibitem[{\citenamefont{Warn and Gauthier}(1989)}]{wg}
\bibinfo{author}{\bibnamefont{Warn}, \bibfnamefont{T.}}, and
  \bibinfo{author}{\bibfnamefont{P.}~\bibnamefont{Gauthier}},
  \bibinfo{year}{1989}, \bibinfo{journal}{Tellus}
  \textbf{\bibinfo{volume}{41A}}, \bibinfo{pages}{115}.

\bibitem[{\citenamefont{Warn and Warn}(1978)}]{warn78a}
\bibinfo{author}{\bibnamefont{Warn}, \bibfnamefont{T.}}, and
  \bibinfo{author}{\bibfnamefont{H.}~\bibnamefont{Warn}}, \bibinfo{year}{1978},
  \bibinfo{journal}{Stud. Appl. Math.} \textbf{\bibinfo{volume}{59}},
  \bibinfo{pages}{37}.

\bibitem[{\citenamefont{Watson}(1960)}]{watson60a}
\bibinfo{author}{\bibnamefont{Watson}, \bibfnamefont{J.}},
  \bibinfo{year}{1960}, \bibinfo{journal}{J.~Fluid Mech.}
  \textbf{\bibinfo{volume}{9}}, \bibinfo{pages}{371}.

\bibitem[{\citenamefont{Weitzner}(1963)}]{weitzner63a}
\bibinfo{author}{\bibnamefont{Weitzner}, \bibfnamefont{H.}},
  \bibinfo{year}{1963}, \bibinfo{journal}{Phys. Fluids}
  \textbf{\bibinfo{volume}{6}}, \bibinfo{pages}{1123}.

\bibitem[{\citenamefont{Yamaguchi} \emph{et~al.}(2004)\citenamefont{Yamaguchi,
  Barr\'e, Bouchet, Dauxois, and Ruffo}}]{yamaguchi04}
\bibinfo{author}{\bibnamefont{Yamaguchi}, \bibfnamefont{Y.~Y.}},
  \bibinfo{author}{\bibfnamefont{J.}~\bibnamefont{Barr\'e}},
  \bibinfo{author}{\bibfnamefont{F.}~\bibnamefont{Bouchet}},
  \bibinfo{author}{\bibfnamefont{T.}~\bibnamefont{Dauxois}}, and
  \bibinfo{author}{\bibfnamefont{S.}~\bibnamefont{Ruffo}},
  \bibinfo{year}{2004}, \bibinfo{journal}{Physica A}
  \textbf{\bibinfo{volume}{337}}, \bibinfo{pages}{36}.

\end{thebibliography}
\newpage

\end{document}